\documentclass[prd,nofootinbib,preprint,superscriptaddress]{revtex4}

\usepackage{amsmath, amssymb, amsthm, graphicx, epsfig, fancyhdr,epsfig, slashed, mathrsfs}

\usepackage{tikzsymbols}
\usepackage{tikz,xcolor
}
\usepackage{natbib}
\usepackage{float}

\usepackage{mathtools}

\definecolor{cadmiumred}{rgb}{0.89, 0.0, 0.13}
\definecolor{candyapplered}{rgb}{1.0, 0.03, 0.0}
\definecolor{copper}{rgb}{0.72, 0.45, 0.2}
\definecolor{darkmagenta}{rgb}{0.55, 0.0, 0.55}

\definecolor{ao(english)}{rgb}{0.0, 0.5, 0.0}
\usepackage{hyperref}
\hypersetup{colorlinks,linkcolor={green!50!black},citecolor={cyan},urlcolor=ao(english)
}

\definecolor{lime}{HTML}{A6CE39}
\DeclareRobustCommand{\orcidicon}{
	\begin{tikzpicture}
	\draw[lime, fill=lime] (0,0) 
	circle [radius=0.2] 
	node[white] {{\fontfamily{qag}\selectfont \tiny ID}};
	\draw[white, fill=white] (-0.0625,0.095) 
	circle [radius=0.007];
	\end{tikzpicture}
	\hspace{-2mm}
}

\foreach \x in {Anish,Maxim, Shila,Lalak}{\expandafter\xdef\csname orcid\x\endcsname{\noexpand\href{https://orcid.org/\csname orcidauthor\x\endcsname}
			{\noexpand\orcidicon}}
}

\newcommand{\be}{\begin{equation}}
\newcommand{\ee}{\end{equation}}
\newcommand{\bea}{\begin{eqnarray}}
\newcommand{\eea}{\end{eqnarray}}

\definecolor{dukeblue}{rgb}{0.0, 0.0, 0.61}

\def\hds#1{\href{https://doi.org/#1}}

\newcommand{\spp}[1]{\textcolor{red}{[Shila: #1]}}

 \usepackage{
 ulem,cancel,
 multirow,
 slashed,
 upgreek, 
 physics, cases}
 \usepackage{easyReview}
\usepackage{bbold}

\newcommand{\eq}[1]{\begin{align}#1\end{align}}

\definecolor{amber(sae/ece)}{rgb}{1.0, 0.49, 0.0}

\definecolor{blue(ncs)}{rgb}{0.0, 0.53, 0.74}

\definecolor{darkviolet}{rgb}{0.58, 0.0, 0.83}

\def\mg#1{\color{magenta}{\bf #1}}

\definecolor{americanrose}{rgb}{1.0, 0.01, 0.24}

\definecolor{jazzberryjam}{rgb}{0.65, 0.04, 0.37}
\definecolor{jonquil}{rgb}{0.98, 0.85, 0.37}
\definecolor{mikadoyellow}{rgb}{1.0, 0.77, 0.05}
\definecolor{schoolbusyellow}{rgb}{1.0, 0.85, 0.0}

\newcommand{\ba}{\begin{eqnarray}}
\newcommand{\ea}{\end{eqnarray}}

\newcommand{\unit}[1]{\, \text{#1}}

\newcommand{\lt}{\left }
\newcommand{\rt}{\right }
\newcommand{\seq}{\simeq}
\newcommand{\gsim}{\gtrsim}
\newcommand{\lsim}{\lesssim}

\def\gs{g_{\star}}
\def\gss{g_{\star, s}}
\def\mpl{M_P}
\def\GeV{\unit{GeV}}
\def\eV{\unit{eV}}
\def\keV{\unit{keV}}

\def\cmb{\text{CMB}}

\def\ncmb{N_{\cmb}}

\newcommand{\Planck}{\textit{Planck}}

\newcommand{\KeckArray}{\textit{Keck Array}}
\newcommand{\BICEP}{\textsc{Bicep}}

\newcommand{\cmbsfour}{{CMB-S4}}
\newcommand{\LB}{\textit{LiteBIRD}}

\def\Trh{T_{\rm rh}}
\def\rh{{\rm rh}}

\def\Br{\text{Br}}
\def\cc{\bar{\chi}\chi}
\def\ccp{\chi\chi}

\def\sm{\text{SM}}
\def\bsm{\text{BSM}}
\def\dm{\text{DM}}
\def\cdm{CDM}
\def\wdm{\text{WDM}}
\def\bbn{BBN}

\def\ncmb{{\cal N}_{\cmb}}

\def\td{\text{d}}
\def\pd{\partial}

\def\g{\gamma}

\def\r{\rho}
\def\a{\alpha}
\def\b{\beta}
\def\m{\mu}
\def\G{\Gamma}
\def\L{\Lambda}
\def\c{\chi}
\def\vp{\varphi}

\def\mNo{`(m)NM-N-1'}
\def\mNt{`(m)NM-N-2'}
\def\mNr{`(m)NM-N-3'}
\def\mNPo{`(m)NMP-N-1'}
\def\mNPt{`(m)NMP-N-2'}
\def\mNPr{`(m)NMP-N-3'}
\def\mNPf{`(m)NMP-N-4'}
\def\mCWo{`(m)NM-CW-1'}
\def\mCWt{`(m)NM-CW-2'}
\def\nmn{\text{NM-N}}
\def\nmpn{\text{NMP-N}}
\def\nmcw{\text{NM-CW}}

\def\lop{{\rm 1-loop}}

\def\cs{\boldsymbol{a}}
\def\hubble{{\cal H}}

\def\cc{\bar{\chi}\chi}
\def\lH{\lambda_H}

\def\mH{m_H}
\def\mc{m_{\chi}}
\def\yc{y_\c}

\def\Yco{Y_{\c,0}}

\def\lO{\lambda_{12}}
\def\ld{\lambda_{12}}
\def\lT{\lambda_{22}}
\def\ls{\lambda_{22}}
\def\HH{H^\dagger H}

\def\Trh{T_{\rm rh}}
\def\Tmax{T_{max}}

\def\Br{\text{Br}}

\def\one{2-to-2 scattering of non-relativistic inflaton with graviton as the mediator}
\def\two{2-to-2 scattering of \sm~particles with graviton as mediator with $\mc\ll\Trh$}
\def\three{2-to-2 scattering of \sm~particles with graviton as mediator with $\Tmax\gg \mc\gg\Trh$}
\def\four{2-to-2 scattering of \sm~particles with inflaton as mediator with $\Trh \ll m_\phi$}

\def\cq{{\cal Q}}

\def\jframe{\text{Jordan frame}}
\def\eframe{\text{Einstein frame}}

\usepackage{xstring}
\usepackage[capitalise]{cleveref}
\Crefname{figure}{Fig.}{Figs.}
\Crefname{section}{Sec.}{Secs.}
\newcommand{\Ccite}[1]{%
\IfSubStr{#1}{,}{Refs.~}{Ref.~}\cite{#1}%
}

\crefname{equation}{eq.}{eqs.}
\Crefname{equation}{Eq.}{Eqs.}

\def\wdth{0.6\linewidth}

\begin{document}


\title{Post-inflationary production of particle Dark Matter: \\ \it{non-minimal Natural and Coleman--Weinberg inflationary scenarios}}

\author{Anish Ghoshal\orcidAnish{}}
\email{anish.ghoshal@fuw.edu.pl}
\affiliation{Institute of Theoretical Physics, Faculty of Physics, University of Warsaw, ul. Pasteura 5, 02-093 Warsaw, Poland}

\author{Maxim~Yu.~Khlopov\orcidMaxim{}}
\email{khlopov@apc.in2p3.fr}
\affiliation{Virtual Institute of Astroparticle Physics, 75018 Paris, France}
\affiliation{Institute of Physics, Southern Federal University, 344090 Rostov on Don, Russia}
\affiliation{National Research Nuclear University “MEPHI”, 115409 Moscow, Russia}

\author{Zygmunt Lalak\orcidLalak{}}
\email{zygmunt.lalak@fuw.edu.pl}
\affiliation{Institute of Theoretical Physics, Faculty of Physics, University of Warsaw, ul. Pasteura 5, 02-093 Warsaw, Poland}

\author{Shiladitya Porey\orcidShila{}}
\email{shiladityamailbox@gmail.com}
\affiliation{Department of Physics, Novosibirsk State University, Pirogova 2, 630090 Novosibirsk, Russia}

\begin{abstract}

\textit{
We investigate the production of non-thermal fermionic dark matter particles during the reheating era following slow roll inflation, driven by inflaton $\varphi$ non-minimally coupled to the curvature scalar, ${\cal R}$. Two types of non-minimal couplings are considered: $\xi \varphi^2 {\cal R}$ for both natural (referred to as NM-N) and for Coleman-Weinberg (referred to as NM-CW) inflation, and $\alpha\left( 1+ \cos\left(\frac{\varphi}{f_a}\right)\right)$ only for natural inflation (referred to as NMP-N), where $\alpha$ and $\xi$ are dimensionless parameters and $f_a$ is an energy scale. 
We determine benchmark values for slow roll inflationary scenarios satisfying current bounds from Cosmic Microwave Background (CMB) radiation measurement and find the mass of inflaton to be $m_\phi\sim {\cal O}\qty(10^{12}) \GeV$ for all three inflationary scenarios and tensor-to-scalar ratio, $r\sim 0.0177$ (for NM-N), $\sim 0.0097$ (for NMP-N), and $r\sim 0.0157$ (for NM-CW) which fall inside $1-\sigma$ contour on scalar spectral index versus $r$ plane of Planck2018+BICEP3+KeckArray2018 joint analysis, and can be probed by future CMB~observations e.g. Simons Observatory. 
 We then show that dark matter particles produced from the decay of inflaton can fully match the present-day cold dark matter (CDM) yield, as well as other cosmological constraints, if the coupling value between inflaton and dark matter, $y_\chi$, and the dark matter mass, $\mc$, are within the range $10^{-1}\gtrsim y_\chi\gtrsim 10^{-20}$ for NM-N and NMP-N ($10^{-4}\gtrsim y_\chi\gtrsim 10^{-20}$  for NM-CW) and ${\cal O}\qty(\text{keV})\lesssim m_\chi\lesssim m_\phi/2$ (for NM-N, NMP-N, and NM-CW). The exact range of $y_\chi$ and $m_\chi$  varies with different benchmark values as well as parameters of inflation, like energy scale of inflation and $r$, some of which are within reach of next-generation CMB experiments.  
}

\end{abstract}

\maketitle

\section{Introduction}

Big bang cosmological model has been widely accepted model describing the origin of the universe, with evidence such as the Hubble expansion of distant galaxies and the Cosmic Microwave Background (\cmb) radiation supporting its validity. However, precise analysis of the CMB data suggests that our universe is nearly spatially flat, homogenous, and isotropic on large scales, with nearly uniform temperature of CMB photons ($2.73\unit{K}$) which are difficult to reconcile with the Big Bang Model. On the other hand, Big Bang model suggests that if we trace back in time from the current state of the expanding and cold universe, it would contract into a point with extremely high energy during its origin, and the universe is expanding ever since,  with varying expansion rates during different cosmological epochs. Cosmological inflation is one of the primeval epochs that occurs for a very short time during which the energy density of the universe is dominated by that of inflaton, and consequently,
space-time expands exponentially.  This epoch successfully addresses most of the shortcomings of Big Bang model. During this epoch, the Hubble radius remained almost constant, and due to the finite size of the observable universe, it is expected that quantum relic scalar and tensor perturbations are generated, which at a later stage plays crucial role in the structure formation, generation of relic gravitational waves, and perhaps even in the formation of primordial black holes.  This epoch was first introduced in literature by Refs.~\cite{Starobinsky:1980te, Guth:1980zm, Linde:1981mu, Albrecht:1982wi}. Since then, this epoch has become so popular that numerous inflationary models have been proposed in the literature. However, analyses of the latest data from Planck~and~BICEP~observations~\cite{Planck:2018jri, BICEP2:2018kqh} have resulted in a reassessment of the preferred feasible inflation models.
 Multi-field inflationary scenarios that involve non-negligible coupling between the light scalars during inflation are in tension due to the absence of large non-Gaussianities in the CMB ~data~\cite{Folkerts:2013tua}. On the other hand, several single field inflation scenarios, including slow roll chaotic inflation with potential of monomial form ($V(\phi)\propto \phi^p$, specifically with $p=1,2,4$), large field inflation with field value greater than Planck mass are ruled out even by $2-\sigma$ on the $(n_s,r)$ plane (see e.g., Ref.~\cite{Martin:2013tda} for a comprehensive review). However, with non-minimal coupling to gravity (Ricci scalar $\mathcal{R}$) several inflationary models can be revived with suitable choice of the coupling \cite{Hertzberg:2010dc}.

Among other unresolved issues of the Big Bang model is the understanding of the origin and composition of dark matter (\dm) and it remains one of the biggest mysteries. \cmb~data manifests that $26.5\%$ of the total mass-energy density of the universe is in the form of \cdm. Despite the existence of numerous gravitational interactions of DM in addition to the CMB data, there has not been any direct evidence of non-gravitational signatures of DM to date. As a result, the identity of \dm~and how it came to exist in the universe remain unknown \cite{Bertone:2004pz}.
Several hypotheses have been proposed to explain the mechanism of DM production during different cosmological epochs~\cite{Bernal:2018hjm} (see also~\cite{Choudhury:2023hvf,Choudhury:2023rks,Choudhury:2023jlt,Choudhury:2023vuj,Choudhury:2014hua,Choudhury:2014sxa,Choudhury:2013jya,Choudhury:2013iaa,Choudhury:2011jt,Choudhury:2011rz}). The interaction channels through which DM particles are produced, together with its cross-section with other particles, control the decoupling time from the Standard Model (\sm) relativistic plasma of the universe. In this study, we focus on the production of \dm~during one of the earliest cosmological epochs, specifically, during the reheating era \footnote{One of the widely popular models of thermal \dm~is the Weakly Interacting Massive Particles (WIMP) scenario. However, the credibility of the theory about WIMP is currently in doubt state since  the attempts to produce them at LHC have been unsuccessful, or the scattering off nuclei by them or their annihilation has not been detected in various direct and indirect DM detection experiments.}. Among non-thermal \dm~production scenarios, Feebly Interacting Massive Particles (FIMP)~\cite{McDonald:2001vt, Choi:2005vq, Kusenko:2006rh, Petraki:2007gq, Hall:2009bx, Bernal:2017kxu,Haque:2021mab,Haque:2022kez,Haque:2023yra} may be a viable option, owing to their significantly weaker interactions, which prevent them from attaining thermal equilibrium with relativistic plasma in the early universe. In contrast to WIMP, the number density of FIMP  depends on initial conditions. FIMP DM particles can be formed mainly in three ways:   first,  their number density freezes as soon as the temperature of the relativistic plasma drops below the mass of the DM particles. Second, they can be produced from the decay of heavy particles such as moduli fields, the curvaton (see, e.g. Ref.~\cite{Baer:2014eja}), or the inflaton which is considered in this work. Third, they can be generated through the scattering of \sm~particles mediated by inflaton or graviton~\cite{Garny:2015sjg, Tang:2016vch, Tang:2017hvq, Garny:2017kha, Bernal:2018qlk}. Recently several scenarios involving testability of FIMP DM involving long-lived particles at laboratories \cite{Barman:2023ktz,Barman:2022scg,Barman:2022njh,Barman:2021yaz,Barman:2021lot,Ghosh:2023tyz,Elor:2021swj,Bhattiprolu:2022sdd,Chakraborty:2023ocr} and involving primordial Gravitational Waves of cosmological origin have been proposed \cite{Ghoshal:2022ruy,Berbig:2023yyy,Paul:2018njm,Ghoshal:2022jdt}.

{In this paper since we motivate particle nature of dark matter and its formation from inflaton, we choose to work with two well-known particle physics motivated re-normalizable Coleman-Weinberg potential which is RGE-improved Higgs-like scalar potential in weakly coupled theories and natural inflation which is non-perturbatively generated ALP scalar potential which is present in strongly-coupled theories.}
In this work, alongside Ref.~\cite{Ghoshal:2023jhh}, we present a scenario where the \dm~particles are non-thermal and produced from the inflaton during the reheating era which is preceded by single-field natural inflation where inflaton is non-minimally coupled to Ricci scalar. For the inflationary part, we have considered natural inflation motivated by cosine potential of axion-like particles in general and Coleman--Weinberg inflation. We derive the conditions on DM parameter space from the inflationary constraint analysis and DM phenomenology constraints on inflaton DM couplings and inflaton and DM masses.

	\textit{The paper is organized as follows}: we present the action of our model in~\cref{Sec:Inflationary models}. Next, in~\cref{Sec:Non-minimal natural,Sec:Non-minimal-periodic natural,Sec:Nonminimal CW}, we explore three inflationary scenarios: non-minimal natural inflation with both $\xi \vp^2{\cal R}$ and periodic non-minimal coupling, and non-minimal natural inflation with both $\xi \vp^2{\cal R}$ non-minimal coupling. For each inflationary scenario, we study the slow-roll inflation, reheating, and production of dark matter (\dm). In ~\cref{Sec:Conclusion}, we summarize our findings.

We use $\hbar=c=k_B=1$ unit throughout this paper, such that the value of reduced Planck mass is $\mpl=2.4\times 10^{18}\GeV$. We also 
consider that the space-time metric is diagonal with signature $(+,-,-,-)$.

\section{Lagrangian density}
\label{Sec:Inflationary models}
In this analysis, we assume that the inflationary epoch in the early universe is driven by an axion or axion-like pseudo-Nambu–Goldstone boson, $\vp$, which is singlet under \sm~gauge transformations, and the inflaton is non-minimally coupled to gravity. Therefore, the action of inflaton in Jordan frame is given by~\cite{Nozari:2007eq,Cheong:2021kyc,Kodama:2021yrm} 
\eq{\label{Eq:Jordan frame action}
{\cal S}^{JF}\supset \int \td^4 x  \sqrt{-g^{JF}} \lt(\frac{1}{2}M_P^2 \, \Omega^2(\varphi)\, {\cal R}^{JF}   + \frac{1}{2} \lt( \partial \vp\rt)^2  -  V^{JF}(\vp)\rt)\,,
}
where the notion $JF$ in superscript denotes that the associated parameter is defined in Jordan frame, for example, $g^{JF}$ and ${\cal R}^{JF} $ represent the determinant of space-time metric and Ricci scalar, respectively, in Jordan frame. Here, $\vp$ is the inflaton in Jordan frame, $\Omega^2(\vp)\equiv \Omega^2\neq 0$, , and $\Omega^2(v_{\rm inf})=\mpl^2$, where $v_{\rm inf}$ is the vacuum expectation value of inflation at the end of inflation. 
The last condition ensures ordinary Einstein gravity at low energy scale~\cite{Watanabe:2006ku} (also~\cite{Kannike:2015fom}). 

Now, by conformally transforming the space-time metric in Jordan frame, we can derive the space-time metric in Einstein frame as
\begin{align}
g^E_{\mu\nu} = \Omega^{2} \, g^{JF}_{\mu\nu}\,.
\end{align}
Hereafter, we use superscript $E$, which implies that the associated parameter is defined in Einstein frame. In this fame, the gravity sector is canonical. Therefore, 
the action in Einstein frame is
\eq{\label{Eq:Einstein action first step}
{\cal S}^E\supset \int \td^4 x  \sqrt{-g^E} \lt(\frac{1}{2}M_P^2 \,  {\cal R}^E   + \frac{1}{2} \Pi(\vp)  \lt( \partial \vp\rt)^2  -  \Omega^{-4}\, V^{JF}(\vp)\rt)\,,
} 
where~\cite{Shaposhnikov:2020fdv}
\begin{align}
\Omega^2(\vp)= 1+ \frac{F(\vp)}{\mpl^2}\,, 
\qquad 
\Pi(\vp)  \equiv \frac{1}{ \Omega^{2}}  + \kappa \,  \frac{3}{2 M_{P}^{2}} \frac{1}{ \Omega^{4} }
\lt(\frac{\td F (\vp)}{\td \vp}\rt)^2\,,
 \label{eq:2eq}
\end{align}
along with $\kappa = 1$ (metric formalism).  The first term on the right-hand side of~\cref{eq:2eq} arises from conformal rescaling while the second term arises from transformation of Ricci scalar (see~\Ccite{Park:2008hz}). To express the kinetic energy of the inflaton in canonical form, we need to redefine the inflaton as
\begin{align} \label{Eq:JordantoEinstein}
\frac{d \phi}{d \varphi} = \sqrt{\Pi(\vp)}\,.
\end{align}
Hence, $\phi$ is the inflaton in Einstein frame. The value of $\phi$ as a function of $\vp$ can be found using~\cref{Eq:JordantoEinstein}. 
The action of~\cref{Eq:Einstein action first step} then transforms as
\ba
{\cal S}^E\supset \int \td^4 x  \sqrt{-g^E} \lt(\frac{1}{2}M_P^2 \,  {\cal R}^E   + \frac{1}{2}   \lt( \partial \phi\rt)^2  -  V^E(\phi)\rt)\,,
\ea 
where the potential of the inflaton in Einstein frame is 
\be 
V^E(\phi)= \frac{V^{JF}\lt(\vp(\phi)\rt)}{\Omega^4 \lt(\vp(\phi)\rt)}\,.
\ee

For slow roll inflation in \eframe, the first two potential slow roll parameters are defined as~\cite{Oda:2017zul,Jarv:2016sow,Kodama:2021yrm}
\eq{
&\epsilon_{V}^E = \frac{\mpl^2}{2} \qty(\frac{\td \ln [V^E]}{\td \vp} \frac{\td \vp}{\td \phi})^2 \,,\\
&\eta_{V}^E = \mpl^2 \qty(\frac{1}{V^E} \frac{\td^2 V^E}{\td \vp^2} \lt(\frac{\td \vp}{\td \phi} \rt)^2 - \frac{\td \ln [V^E]}{\td \vp} \frac{\td^2 \phi}{\td \vp^2} \lt(\frac{\td \vp}{\td \phi} \rt)^3 )\,.
}
Both of these $\epsilon_{V}^E$ and $\lt|\eta_{V}^E\rt|$ should be $<1$ to continue the slow roll phase, and this ensures that the inflaton is over damped and Hubble parameter, $\hubble$, remains nearly constant during inflation~\cite{Lyth:1993eu}. The slow roll phase ceases whenever at least one of these two parameters reaches $\sim 1$. 
During slow roll inflation, nearly constant value of $\hubble$ (quasi-de Sitter) and exponential growth of cosmological scale factor $\cs$ implies that comoving Hubble radius ($\qty(\cs\hubble)^{-1}$) shrinks exponentially, resulting in comoving length scales possibly leaving the Hubble horizon. Thus, if there are two length scale $k_*$ and $k_{\rm end}$ that exit the casual horizon just at the beginning and end of inflation, then the number of e-folds by which $\cs$ is increased during this time is~\cite{Oda:2017zul}
\begin{eqnarray}
\ncmb=\frac{1}{\mpl}\int_{\vp_{{\rm end}}}^{\vp_{*}} \td \vp \; \frac{1}{\sqrt{2 \,\epsilon_{V}^E} }
\lt(\frac{\td \phi}{\td \vp}\rt) \,,
 \label{EFold}
\end{eqnarray} 
where $\vp_{*}$ and $\vp_{{\rm end}}$ are the values of inflaton corresponding to when $k_\star$ and $k_{\rm end}$ leave the horizon. If the length scale $k_*$ reenters within the Hubble radius today and Cosmic Microwave Background (\cmb)~observations are measured, then at least $\ncmb\sim 60$~\cite{Baumann:2022mni} ($k_*\sim 0.05 \unit{Mpc}^{-1}$) (or $\ncmb\gsim 66-70$, depending on inflationary model~\cite{Bambi:2015mba}) is required to solve the horizon problem.  

Because of the finite size of $\hubble$, scalar and tensor quantum perturbations are expected to be generated during slow roll inflationary phase. The exponential expansion during inflation stretches the wavelengths of the quantum perturbations even beyond the Hubble horizon. Those perturbations upon reentering the horizon, is measured in terms of two point correlation function i.e. power spectrum which are defined as  
\eq{
&\mathcal{P}_s \left( k \right) = A_s \left(  \frac{k}{k_*} \right)^{n_s -1 + (1/2) \alpha_s \ln(k/k_*) + (1/6)\beta_s (\ln(k/k_*))^2 }  \label{eq:define scalar power spectrum}\,,\\
& \mathcal{P}_h \left( k \right) = A_t \left(  \frac{k}{k_*} \right)^{n_t + (1/2) d n_t/d \ln k \ln(k/k_*) + \cdots } \,,\label{eq:define tensor power spectrum}
}
where $k_*$ is the pivot scale at which bounds on inflation are measured from the analysis of \cmb~data, and 
\begin{itemize}
    \item $\mathcal{P}_s$  and $\mathcal{P}_h $ : scalar and tensor power spectrum of primordial perturbation, 
    \item $A_s$ and $A_t$: amplitude of scalar and tensor power spectrum,
    \item $n_s$ and $n_t$: scalar and tensor spectral index, 
    \item $\a_s$ and $\b_s$: running and running of running of scalar spectral index.
\end{itemize}
For the given single-field inflationary potential $A_s$ and $n_s$ can be measured as 
\eq{
A_s\approx  \frac{1}{24 \pi^2}\frac{1}{\mpl^4}\left. \frac{V^E}{ \epsilon_{V}^E } \right|_{k=k_\star}\,,
&& n_s(\vp_*) = 1-6\,\epsilon_{V}^E (\vp_*)\, +\, 2\, \eta_{V}^E(\vp_*)\,.
 \label{PSpec}
}
For $n_s<1$, $\mathcal{P}_s$ is red tilted i.e. more power at larger length scales. Now, if we define tensor-to-scalar ratio $r= A_t/A_s$, then~\cite{Planck:2018jri} 
\eq{
V^E (\varphi_*)= \frac32 \pi^2 \, r(\varphi_*) \, A_s\, \mpl^4,
}
where we have used slow roll approximation $ r(\vp_*) \approx 16 \, \epsilon_{V}^E(\vp_*)$. 
Current \cmb~constraints on $A_s$, $n_s$, and $r$ are mentioned in~\cref{Table:PlanckData}. 
\begin{table}[H]
\begin{center}
\caption{ \centering\it Constraints on inflationary parameters from \Planck, \Planck+\BICEP+\KeckArray~and other \cmb~experiments (T and E for temperature and E-mode polarization of \cmb).} \label{Table:PlanckData}
\begin{tabular}{ |c|| c| c||c| }
\hline
$\ln(10^{10} A_s)$ & $
3.044\pm 0.014$ & $68\%$, TT,TE,EE+lowE+lensing+BAO & 
\cite{Aghanim:2018eyx,ParticleDataGroup:2022pth}  \\
 \hline
 $n_s$ & $0.9647\pm 0.0043$ & $68\%$, TT,TE,EE+lowE+lensing+BAO & 
 \cite{Aghanim:2018eyx} \\ 
 \hline 
 $r$ & $0.014^{+0.010}_{-0.011}\, \text{and}$ &  $ 95 \%  \,, \text{BK18, \textsc{Bicep}3, \textit{Keck Array}~2020,}$& \cite{Aghanim:2018eyx, BICEPKeck:2022mhb,BICEP:2021xfz}\\
   & $ <0.036 $ & and \textsl{WMAP} and \textit{Planck}~CMB polarization & (see also \cite{Campeti:2022vom}) \\
 \hline 
\end{tabular}
\end{center}
\end{table}%
%
%
%
%

In addition to the slow roll inflationary epoch, in this article, we also consider the reheating era that comes after inflation and also make the assumption that 
during that reheating era production of non-thermal \dm~particles and \sm~Higgs proceeds concomitantly.  
Therefore, the total action in which we are interested in can be expressed as follows~\cite{Bernal:2021qrl,Ghoshal:2022jeo,Ghoshal:2023noe,Ghoshal:2022aoh,Ghoshal:2024ycp}
\be\label{Eq:Action}
{\cal S}_{\rm tot}\supset {\cal S}^E + 
\int \td x^4 \sqrt{-g^E}\, \lt(  {\cal L}_\chi^E + {\cal L}_H^E + {\cal L}_{\rh}^E \rt)\,.
\ee 
Here, the Lagrangian density of the beyond the standard model (\bsm) vector-like fermionic field $\chi$ and \sm~Higgs doublet $H$ are ${\cal L}_\chi$ and ${\cal L}_H$, respectively, and those can be written in the following form: 
\eq{
{\cal L}_\chi^E &= i \bar{\c}\slashed{\partial} \chi - m_\c \cc \,, \label{Eq:Lagrangian-for-chi}\\
{\cal L}_H^E &= \lt|\partial H\rt|^2 +\mH^2 \HH  - \lH \lt( \HH \rt)^2 \,,\label{Eq:Lagrangian-for-H}
}
where 
both the parameters $\mc$ and $\mH$ have the mass dimension whereas the parameter $\lH$ is dimensionless. Moreover, ${\cal L}_\rh^E$ in~\cref{Eq:Action} comprises interaction between the inflaton and the dark sector as well as between the inflaton and \sm~Higgs.   
Hence,%
\be\label{Eq:reheating lagrangian}
\mathcal{L}_{\rh}^E = - \yc \phi \cc - \lO \phi \HH - \lT \phi^2 \HH +{\cal L}_{\rm scatter} + \text{h.c.} \,.
\ee 
where ${\cal L}_{\rm scatter}$ includes higher order terms that account for the scattering of $\c$ by inflaton, or \sm~particles (including $H$). The couplings $\yc$ and $\ls$ are also dimensionless, but $\ld$ has mass dimension. 
{Much like the coupling $\lO$, coupling $\lambda_{22}$ contributes to the dynamics of reheating. However, the stability analysis imposes a very small permissible upper value for $\lambda_{22}$~\cite{Drees:2021wgd}. Therefore, in this article we mainly discuss the permissible range of $\lO$  since we are interested in perturbative approach of reheating.} 
In the two succeeding sections (\cref{Sec:Non-minimal natural,Sec:Non-minimal-periodic natural}), we consider two different forms of $\Omega^2(\vp)$ and study the slow roll single field inflationary scenarios, and determine benchmark values that satisfy the current \cmb~bounds. We also investigate  the parameter space, involving $\yc$ and $\mc$ under the assumption that the total \cdm~density of the present universe is completely comprised of $\c$ particles produced during the reheating era.

\section{Natural inflation with $\xi \vp^2 {\cal R}$ non-minimal coupling}
\label{Sec:Non-minimal natural}
Natural inflation with canonical kinetic energy term is no longer the favorite candidate for single field inflationary scenario since this model predicts large values of $r$ relative to $2-\sigma$ best-fit contour from \Planck~2018+\BICEP~data on $n_s-r$ plane~\cite{Planck:2018jri,Ghoshal:2023phi}. However, this is one of the inflationary models which has strong motivation based on particle physics theory. In this theory, inflaton is an axion or axionic field which is a pseudo-Nambu-Goldstone boson, arising from the spontaneous breaking of global symmetry and an additional subsequent explicit symmetry breaking. This boson is introduced in the theory to solve strong CP problem~\cite{Peccei:2006as,Hook:2018dlk,Kim:2008hd}. The axion potential which arises non-perturbatively due to the spontaneous breaking of global $U(1)$ symmetry can be written as
\eq{\label{E:potential for natural inflation in Jordan frame}
V_{\rm Natural}(\vp)=\L^4 \lt( 1+ \cos\lt(\frac{\vp}{f_a}\rt)\rt)\,,
}
where $\vp$ represents the canonically normalized axion or axionic field. Here $\L$ is the $U(1)$ explicit symmetry-breaking energy scale of the continuous shift symmetry, 
while $f_a$ determines the spontaneous U(1) symmetry-breaking energy scale. Additionally, $f_a$ is accountable for the small value of self-coupling as $f_a$ reduces its value by $1/f_a$ and, hence, solves the problem of fine-tuning. For this reason, this inflationary model is called natural inflation. 
Furthermore, this potential has a shift symmetry ($V_{\rm Natural} (\vp)=V_{\rm Natural} (\vp+\text{const.})$) and thus protects the flatness of the potential.
On the other hand, the shift symmetry protects the flatness of the potential of axion-inflaton \cite{Kim:2004rp,Freese:1990rb}.

\begin{figure}[H]
    \centering    
    \includegraphics[height=6cm,width=\wdth]{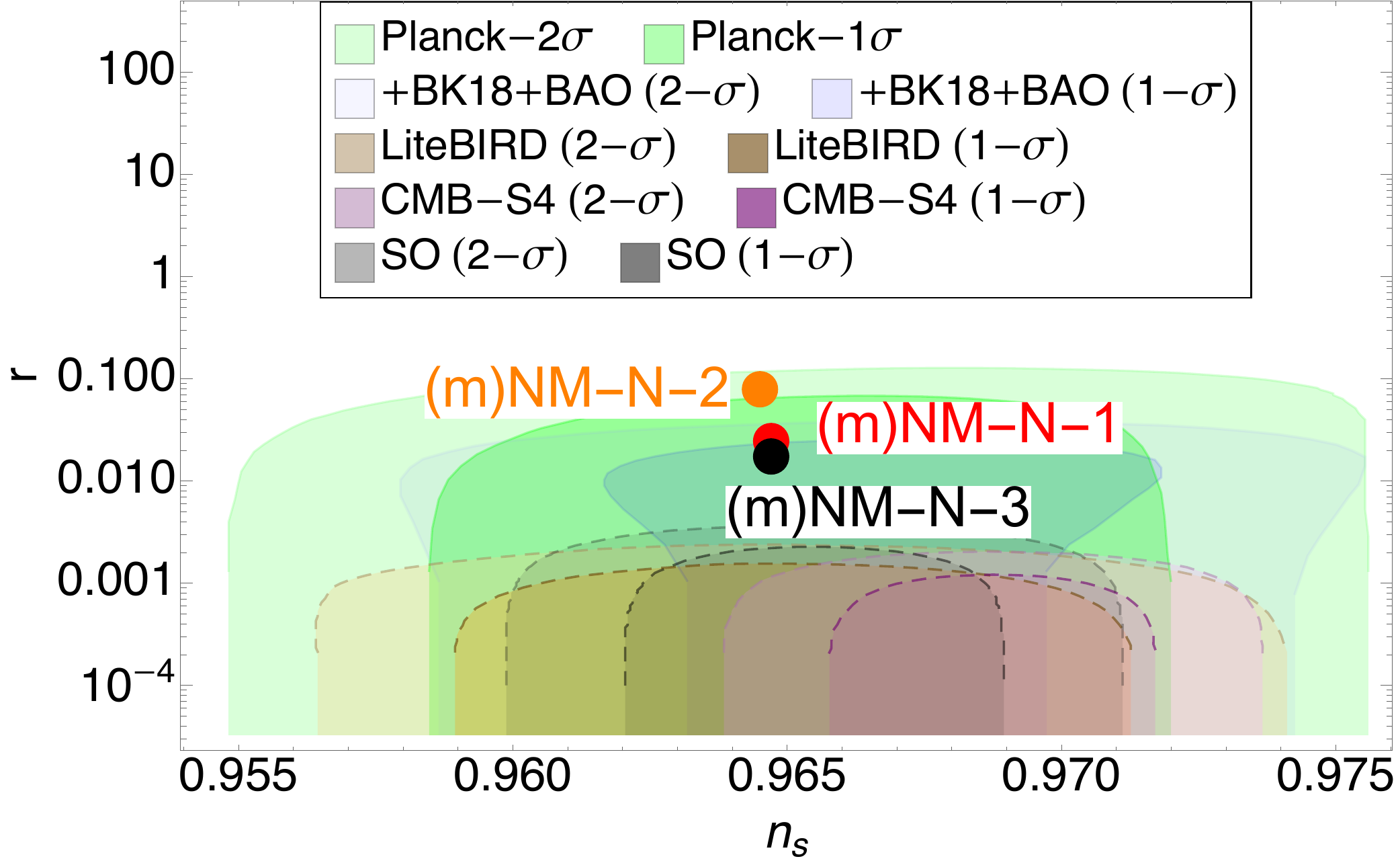} 
    \caption{\it \raggedright \label{Fig:NonminimalNatural-ns-r-bound} 
    The colored dots represent the predicted values of $(n_s,r)$ for benchmark values from~\cref{Table:Benchmark-NonminimalNatural}. Alongside, this figure also displays $95\%$ C.L. ($1-\sigma$) contours highlighted by a deep-colored regions located inside and $68\%$ C.L. ($2-\sigma$) $(n_s, r)$ best-fit contours as light-colored regions located outside. 
   The green-colored region is for best-fit bound from \Planck~2018 {
   (Planck TT,TE,EE+lowE+lensing)} while the blue shaded area presents bound from \Planck~2018+\BICEP3 (2022)+\KeckArray2018 joint analysis. Other colored regions, with dashed lines as periphery, show contours from different forthcoming \cmb~observations (\LB+\cmbsfour+Simons Observatory (SO))~\cite{LiteBIRD:2020khw,CMB-S4:2016ple,SimonsObservatory:2018koc}.
    }
\end{figure}

Being such a theoretically well-motivated model~\cite{Savage:2006tr}, several attempts have been made to modify the theory slightly so that the predicted value of $r$ remains within at least $2-\sigma$ contour of \Planck~data. Additionally, another objective has been to reduce the value of $f_a$ (the range of $f_a$ used in the paper of Planck collaboration is $10\lsim f_a/\mpl\lsim 1585$).
Some of these attempts include~\cite{Cheng:2021qmc} which considers generalized versions of natural inflation,~\cite{Daido:2017wwb} which explores the incorporation of multiple axion-like fields. 
However, in this section, we consider the inflationary model presented in Ref.~\cite{Reyimuaji:2020goi} where a non-minimal coupling between the inflaton and gravity sector has been considered, and non-minimal coupling has the following form :
\eq{ 
\Omega^2(\vp)= 1+ \xi \frac{\vp^2}{\mpl^2}\,,
} 
where $\xi$ is  dimensionless non-minimal coupling between $\vp$ and gravity. This coupling can be arisen at 1-loop order from quantum corrections in curved spacetime~\cite{Birrell:1982ix,Faraoni:1996rf}, and for the theoretical motivation of such form of non-minimal coupling see~Refs.~\cite{Hertzberg:2010dc,Hrycyna:2015vvs}. 
For this coupling, from~\cref{Eq:JordantoEinstein} we get
\be 
\frac{\td \phi}{\td \vp}= \sqrt{\frac{\xi  (6 \xi +1) \varphi ^2 M_P^2+M_P^4}{\left(M_P^2+\xi  \varphi
   ^2\right){}^2}}\,.
\ee 
The potential in \jframe~is given by~\cref{E:potential for natural inflation in Jordan frame} i.e. $V^{JF}(\vp)\equiv V_{\rm Natural}(\vp)$. 
The potential $V^{JF}(\vp)$ has local maxima at $\vp=n f_a\pi$ ($n=0,2,4,\ldots$) and local minima at $\vp=n f_a\pi$ ($n=1,3,5,\ldots$). The potential of non-minimal Natural (abbreviated as NM-N) Inflation in Einstein frame is
\eq{\label{Eq:Pot_EinsteinFrame_NonMinimalNatural}
V^{E}(\phi(\vp))\equiv 
V^{E}_{\nmn}(\phi(\vp))=  \frac{\L^4 \lt( 1+ \cos\lt(\frac{\vp}{f_a}\rt)\rt)}{\qty( 1+\xi \, \vp^2/\mpl^2  )^2 }\,.
}
For this inflationary model in metric formalism, 
the benchmark values 
are presented in~\cref{Table:Benchmark-NonminimalNatural}. The value of $f_a/\mpl\sim {\cal O}\qty(1)$ is required for $n_s\sim 0.9647$ which is much smaller than the value of $f_a/\mpl$ required for minimal natural inflationary scenario. Following Ref.~\cite{Reyimuaji:2020goi}, we choose negative value of $\xi$
 such that $n_s$ remains within $1-\sigma$ best fit contour from \Planck~2018 data. The lower bound on $\xi$ comes from the condition $\lt(1+\xi \qty(\vp_{\rm min}/\mpl)^2 \rt)>0$. 
 The predicted $n_s,r$ values of the three benchmarks from~\cref{Table:Benchmark-NonminimalNatural} are shown in~\cref{Fig:NonminimalNatural-ns-r-bound}, along with bounds from current \cmb~observations (\Planck+\BICEP+\KeckArray) and future prospective reaches from other upcoming \cmb~observations. Predicted value of $r$ of benchmark \mNo~is within $2-\sigma$ contour of \Planck2018+\BICEP3 (2022)+\KeckArray2018 data, while predicted value of $r$ of benchmark \mNt~is within $2-\sigma$ contour \Planck2018 data. On the other hand, benchmark \mNr~shows that for \nmn~inflation and for a slightly larger value of e-folds (e.g. $\ncmb\sim 70$) predicted $r$ value falls even within $1-\sigma$ best-fit contour of \Planck2018+\BICEP3 (2022)+\KeckArray2018 combined analysis.  
\begin{table}[H]
\centering
\caption{\it Benchmark value for \nmn~inflationary model in metric formalism ($\varphi_{\rm min}$ is the location of minimum).
}
\label{Table:Benchmark-NonminimalNatural}
\begin{tabular}{|c ||c | c | c | c | c | c | c| c|c| } 
 \hline
 {\it Benchmark} & $f_a/\mpl$ & $\xi$ & $\varphi_*/\mpl$ & $\varphi_{\rm end}/\mpl$ &  $n_s$ & $r\times 10^2$ & $\L^4/\mpl^4$ & $\varphi_{\rm min}/\mpl $ & $\ncmb$\\
 \hline
 \hline
(m)NM-N-1 &$3.01$ & $-10^{-2}$ & $3.5140$ & $9.1132$ &  $0.9647$ & $2.4841$ & $4.2633\times 10^{-10} $ & $9.4430$& $60$\\ 
\hline 
 (m)NM-N-2 & $8.00$ & $2\times 10^{-6}$ & $10.7614$ & $23.7212$ & $0.9645$ & $7.9452$ & $2.0184\times 10^{-9}$ & $25.1327$ & $60$\\
 \hline 
 (m)NM-N-3 & $3.00$ & $- 10^{-2}$ & $2.9562$ & $9.0878$ & $0.9658$ & $1.7702$ & $2.9508\times 10^{-10}$ & $9.4248$ & $70$\\
 %
 %
 \hline
\end{tabular}
\end{table}
%
%

%
%

\subsection{Reheating}
\label{Sec:Non-minimal-Natural-reheating}
%

\begin{figure}[H]
    \centering    
   \includegraphics[width=\wdth]{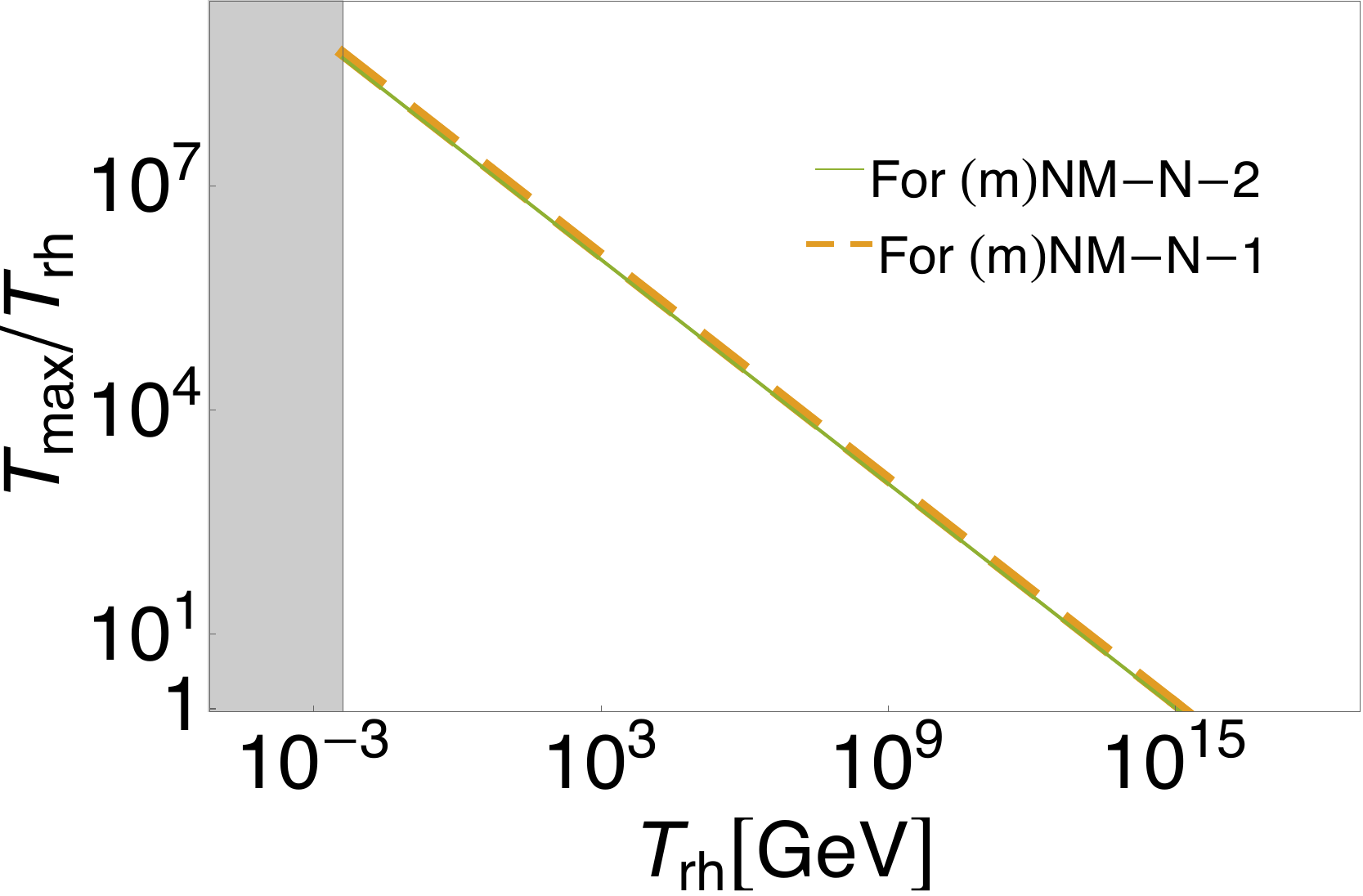}
    \caption{\it \raggedright \label{Fig:Tmax/Trh-Non-minimal_natural}
    Variation of $T_{max}/\Trh$ against $\Trh$ for two benchmark values \mNo~and \mNt~from~\cref{Table:Benchmark-NonminimalNatural}. The gray-colored vertical stripe on the left of the plot of lower bound on $\Trh$ which is $\Trh\nless 4 \unit{MeV}$.
    }
\end{figure}

The reheating phase commences once the slow roll phase ends. During this time, the inflaton field reaches the minimum of its potential, which is also its vacuum expectation value. The inflaton then proceeds to oscillate quasi-periodically around this minimum. 
In this work, we are assuming that potential about the minimum during the reheating era can be approximated as $V^E\sim \phi^2$ such that energy density of oscillating inflaton and pressure averaging over an oscillating cycle during reheating, behaves as~\cite{Enqvist:2012qc} (see also~\cite{Garcia:2020eof}) 
\begin{align}
\label{Eq:arXiv:1201.6164v1}
\rho_{\phi} \propto \cs^{-3}\,, \qquad  &\lt<p\rt>=0\,.
\end{align}
But in addition to the expansion of the universe, the production of \sm~and \bsm~particles from the inflaton also causes the energy density of inflaton to decrease. Reheating epoch is profoundly adiabatic and transforms the universe from being cold and dominated by the energy density of inflaton into being hot and dominated by \sm~radiation. Therefore, we can safely assume
\begin{align}\label{Eq:decay-width-of-inflaton}
 \G_\phi=\Gamma_{\phi\to {h h}}+ \Gamma_{\phi \to \chi\chi} \approx \Gamma_{\phi\to {h h}}\,,
\end{align}	
where $\Gamma_\phi,\Gamma_{\phi\to {h h}},$ and $\Gamma_{\phi \to \chi\chi}$ are respectively the total decay width of inflaton, decay width of inflaton to \sm~Higgs particle $h$, and decay width of inflaton to \dm~particle $\chi$. 
There is no need for an explicit coupling between $\vp$ and those particles to be produced in $f(\varphi){\cal R}$ theories~\cite{Watanabe:2006ku} or \dm~particles can be produced gravitationally (see Refs.~\cite{Kannike:2016jfs,Babichev:2020yeo,Cembranos:2019qlm,Bambi:2015mba}). 
In this work, however, we take advantage of 
the perturbative reheating technique and particle creation in Einstein frame using~\cref{Eq:reheating lagrangian}.
Neglecting the effect of thermal mass~\cite{Kolb:2003ke}, we can find  $\Gamma_{\phi\to {h h}},$ and $\Gamma_{\phi \to \chi\chi}$ from~\cref{Eq:reheating lagrangian} as
\footnote{
A nonminimal coupling between the $\vp$ and the gravity sector is sufficient, for the gravitational-aided production of $\chi$ and $h$ particles during reheating era, without requiring any direct coupling between $\vp$ and $\chi$ or $h$~\cite{Cata:2016epa,Cata:2016dsg}. However, decay width for the production channel ${\vp\to \chi\chi}$ scales with $\mpl^{-2}$ ($\Gamma_{\vp\to \chi\chi}\sim {\xi^2\, m_\phi \, \mc^2 }{\mpl^{-2}}$, where $m_\phi$ is the mass of inflaton), even when $\Omega^2\approx\lt(1+\xi \phi/\mpl\rt)$~\cite{Barman:2023opy}. Consequently, we expect a significantly lower decay width for gravitational production of such decay process compared to the direct coupling scenario mentioned in~\cref{Eq:reheating lagrangian}. This
results in a significantly lower decay width for gravitational production compared to the scenario where inflaton is directly coupled to other fields.
Furthermore, if we consider that the reheating Lagrangian is defined in \jframe, (e.g. $\mathcal{L}_{\rh}^{JF} = - \yc \varphi \cc - \lO \varphi \HH - \lT \varphi^2 \HH$), decay width to two Higgs particle 
$\tilde{\Gamma}_{\phi\to hh}\sim \frac{\xi}{64\pi(1+ 6\xi)}\frac{m_\phi^3}{\mpl^2}\sim 4.97\times 10^{-27}\mpl$ for $\xi\sim 10^{-6}$ and $\lt(m_\phi/\mpl\rt)\sim 10^{-6}$~\cite{Kannike:2015kda}. Conversely, from~\cref{eq:decay width einstein} we get $\Gamma_{\phi\to hh}\simeq\frac{\lO^2}{8\pi m_\phi}\sim 3.98\times 10^{-24}\mpl$ for $\lt(m_\phi/\mpl\rt)\sim 10^{-6}$ and $\qty(\lO/\mpl)\sim10^{-14}$. This leads to the expectation that $\tilde{\Gamma}_{\phi\to hh}< \Gamma_{\phi\to hh}$.  
}.
\eq{\label{eq:decay width einstein}
\Gamma_{\phi\to {h h}}  \simeq \frac{\lambda_{12}^2}{8\pi\, m_{\phi}}\,, 
 \qquad 
 \Gamma_{\phi \to \chi\chi}  \simeq \frac{y_\chi^2\, m_{\phi}}{8\pi}\,,
}
where the physical mass of the inflaton in Einstein frame is 
\be \label{Eq:mass-of-inflaton-NonMinimalNatural}
\frac{m_{\phi}}{M_P} = \left(  \frac{1}{M_P^{2}}   \lt.\frac{\td^2 V^{E}(\phi)}{\td \phi^2} \rt|_{\phi=\phi_{\rm min}} \right)^{1/2}\,.
\ee 
Here, $\phi_{\rm min}$ is the location of the minimum of the potential of~\cref{Eq:Pot_EinsteinFrame_NonMinimalNatural}. Now, let us define the maximum possible temperature during reheating, $\Tmax$, and reheating temperature, $\Trh$, as~\cite{Bernal:2021qrl,Giudice:2000ex,Chung:1998rq} \footnote{(P)reheating production for such tiny couplings is not large in the parameter space of fermionic DM we explore in this paper, for details see \cite{Drewes:2017fmn,Drewes:2019rxn}.}
\eq{ 
\Tmax = 
&\G_\phi^{1/4} \lt( \frac{60}{\gs\, \pi^2} \rt)^{1/4} \lt( \frac{3}{8}\rt)^{2/5} {\cal H}_I^{1/4} \mpl^{1/2} \,,\label{Eq:TMAX}\\
\Trh 
= &\sqrt{\frac{2}{\pi}} \left(\frac{10}{\gs}\right)^{1/4} \sqrt{\mpl} \sqrt{\Gamma_\phi}\,.   \label{Eq:definition of reheating temperature}
} 
The production of \sm~particles by the inflaton causes the temperature of the universe to increase once the reheating epoch starts. This temperature reaches its maximum value, $\Tmax$, and then decreases as a result of the expansion of the universe. When the temperature of the universe drops to $T\sim \Trh$, it then occurs that $\G_\phi\sim \hubble(\Trh)$. At this moment, $\r_{\rm rad}= \r_\phi$, where $\r_{\rm rad}$ is the energy density of \sm~relativistic particles i.e. radiation. The universe then shifts to being dominated by radiation, and the remaining energy density of inflaton rapidly and quickly transforms mostly to $\r_{\rm rad}$. In~\cref{Eq:TMAX,Eq:definition of reheating temperature} $\gs$ is the {\it effective number of relativistic degrees of freedom} which is taken as $\gs=106.75$ (being feebly interacting, $\chi$ might not have the same temperature as that of \sm~radiation), and in~\cref{Eq:TMAX}, ${\cal H}_I$ is the value of Hubble parameter at the end of inflation
(i.e. when $\rho_{\rm rad}=0$~\cite{Giudice:2000ex})
\be\label{Eq:HI-natural}
{\cal H}_I= \sqrt{\frac{V^E\qty(\phi \qty(\vp_{\rm end}))}{3\mpl^2}}\,.
\ee
Following~\cref{Eq:definition of reheating temperature,Eq:decay-width-of-inflaton,Eq:HI-natural}, we can calculate $\G_\phi$, $\Trh$, ${\cal H}_I$, and those values are mentioned in~\cref{Table:reheating_values-NonminimalNatural}. The value of $m_\phi$ is highest for the benchmark \mNo~and lowest for \mNt, while the value of ${\cal H}_I$ is the highest for \mNo. 
\begin{table}[H]
\centering
\caption{\it $m_\phi, \G_\phi, \Trh$, and $\hubble_I$ for the benchmark values from~\cref{Table:Benchmark-NonminimalNatural}.}
\label{Table:reheating_values-NonminimalNatural}
\begin{tabular}{| c | | c | c|  c | c|} 
 \hline
{\it Benchmark} & $m_\phi/\mpl$ & $\G_\phi \,\mpl$ & $\Trh/\lambda_{12}$ & ${\cal H}_I/\mpl$ \\
&   (Eq.~\eqref{Eq:mass-of-inflaton-NonMinimalNatural}) & (Eq.~\eqref{Eq:decay-width-of-inflaton}) & (Eq.~\eqref{Eq:definition of reheating temperature}) &  (\cref{Eq:HI-natural})  \\
[0.5ex]
 \hline\hline
(m)NM-N-1 &   $1.7078 \times 10^{-5}$ & $2329.89\,\lambda_{12}^2$  &$21.3067$ & $5.4539\times 10^{-6}$ \\ 
 \hline
(m)NM-N-2 &   $5.6123\times 10^{-6}$ & $7089.55  \,\lambda_{12}^2$ & $37.1670$ & $3.2284\times 10^{-6}$\\
\hline 
(m)NM-N-3 & $1.4095\times 10^{-5}$
& $2822.91 \, \lO^2$ & $23.4529$ &$4.5216\times 10^{-6}$\\
 \hline
\end{tabular}
\end{table}
From~\cref{Eq:TMAX,Eq:definition of reheating temperature} we can derive 
\be\label{Eq:TmaxTrh-ratio}
\frac{\Tmax}{\Trh} = \lt(\frac{3}{8}\rt)^{2/5}  \lt( \frac{{\cal H}_I}{{\cal H}(\Trh)} \rt)^{1/4}\,,
\ee
where we have used $\hubble(\Trh)= (2/3)\G_\phi$~\cite{Bernal:2021qrl} and 
\be 
\hubble^2 (\Trh)=\frac{1}{3\mpl^2}\lt(\frac{\pi^2}{30} \gs \, \Trh^4\rt)\,,
\ee 
since the universe is radiation dominated at $T=\Trh$. Therefore, $\Tmax/\Trh$ should decrease with increasing values of $\Trh$. 
The variation of $\Tmax/\Trh$ against $\Trh$ for the benchmark values `(m)NM-N-1' and `(m)NM-N-2' are displayed in~\cref{Fig:Tmax/Trh-Non-minimal_natural} - the dashed line is for `(m)NM-N-1' and solid line is for `(m)NM-N-2'. The lower bound (gray-colored vertical stripe on the left) is from the fact that $\Trh$ must be greater than $\sim 4 \unit{MeV}$ for successful Big Bang Nucleosynthesis (\bbn)~\cite{Giudice:2000ex}.%
\footnote{This bound may alter to $5-1.8\unit{MeV}$ if we consider neutrino oscillations or decay of massive \bsm~particles before \bbn, e.g. Refs.~\cite{Hasegawa:2019jsa,Kawasaki:1999na,Kawasaki:2000en}}
To keep $\Tmax/\Trh\gsim 1$, leads to an upper limit on $\Trh$: $1.3826\times 10^{15}\GeV$ for `(m)NM-N-1', and $1.0637\times 10^{15}\GeV$ for `(m)NM-N-2'. Additionally, the maximum value of $\Tmax/\Trh\sim 5.8791\times 10^8$ for \mNo~and $\sim 5.1568\times 10^8 $ for \mNt~for $\Trh\sim 4 \unit{MeV}$.
These large values of $\Tmax/\Trh$ imply that production of any particle with a mass higher than $\Trh$ is still possible during the reheating process.%
\footnote{The production of gravitino tightens the bound on $\Trh$. We ignore this constraint since we are not considering supersymmetry in our study.
}
~Furthermore, lower bound on $\Trh$ leads to establishing the minimum permissible value of $\lO$ which are $\lO/\mpl\gsim 7.8223\times 10^{-23}$ for \mNo, $\lO/\mpl\gsim 4.4843\times 10^{-23}$ for \mNt, and $\lO/\mpl\gsim 7.1064\times 10^{-23}$ for \mNr.
%
%
%
%
%
%

\subsection{Dark Matter Production and Relic Density}
In this section, we focus on the \dm~production during reheating for \nmn~inflationary scenario. 
We are assuming that the \dm~particles are so feebly interacting that they are unable to reach thermal equilibrium with the \sm~relativistic plasma of the universe. 
Thus, if we define $n_\chi$ as number density and $N_\c \equiv n_\c \, \cs^3$ as comoving number density of $\c$-particles, then the equation that governs the time evolution of $N_\c$ is
\eq{\label{Eq:Boltzaman equation for comoving number density}%
\frac{\td N_\c}{\td t} = \cs^3 \gamma  \,,
}
%
where $t$ is the physical time and $\g \equiv \g(t)$ is the rate of production of $n_\c$.
When the universe is dominated by the energy density of the oscillating inflaton i.e. when the temperature of the universe $T$ is within the range $\Tmax>T>\Trh$, $\hubble$ and $\r_\phi$  are given by~\cite{Bernal:2021qrl}
\eq{
\label{Eq:Hubble parameter during reheating+Eq:rho_phi}
{\cal H} = \frac{\pi}{3} \sqrt{\frac{\gs}{10}} \frac{T^4}{M_P\, \Trh^2}\,,
\qquad 
&\r_{\phi} = \frac{ \pi^2  \gs }{30 } \frac{T^8}{ \Trh^4} \,.
}
By applying~\cref{Eq:arXiv:1201.6164v1,Eq:Hubble parameter during reheating+Eq:rho_phi} to~\cref{Eq:Boltzaman equation for comoving number density}, we obtain
\eq{\label{Eq:dNchidT}
\frac{\td N_\c}{\td T} = - \frac{8 \mpl }{\pi} \lt(\frac{10}{  \gs}\rt)^{1/2} \frac{\Trh^{10}}{T^{13}} 
\, \cs^3 (\Trh) \, \g  \,.
}
With the assumption that entropy density of the universe, $s(T)$ (which is expressed as $s(T)=\lt({2 \pi^2}/{45}\rt) \gss \,T^3 $), is conserved after the completion of reheating era, we can derive \dm~yield $Y_\c(T)$ which is specified as $Y_\c(T)=n_\c(T)/s(T)$. 
Here $\gss$ is the {\it effective number of degrees of freedom in entropy}. The energy density of $\chi$ continues to drop over time and eventually becomes non-relativistic, contributing to the \cdm~energy density of the universe. The current CDM yield may be computed as 
\be\label{Eq:present day CDM yield}
 Y_{{\rm CDM},0}  = \frac{\Omega_{\rm CDM}\, \r_{\rm crit} }{s_0 \, m_\c}=\frac{4.3. \times 10^{-10}}{\mc} \,.
\ee 
Here $m_\c$ is in $\GeV$ and \cdm~density of the universe $\Omega_{\rm CDM}=0.12\, h_{\cmb}^{-2}$ with scaling factor for Hubble expansion rate $h_{\cmb}\approx0.674$,  present day entropy  density $s_0=2891.2 \unit{cm}^{-3}$, and critical density of the universe $\r_{\rm crit}=1.053\times 10^{-5}\, h_{\cmb}^{2}\, \GeV  \unit{cm}^{-3}$ (in $k_B=c=1$ unit)~\cite{ParticleDataGroup:2022pth}. In the next two subsections, we look into the production of $\c$ from the decay of inflaton and via scattering mechanisms.

\subsection{DM from decay}
Here, we look at the creation of $\c$ through the decay of inflaton. 
Assuming that the production of $\c$ takes place only via the decay of inflaton ($\phi\to \cc$), 
%
\be\label{Eq:gamma for decay}
\g = 2 \, \Br \, \G_\phi \,  \frac{\r_{\phi}}{m_{\phi}} \,.
\ee 
where $\Br$ is the branching fraction for the production of $\c$ from the decay-channel which is defined as
\be\label{Eq:Br}
\Br = \frac{\G_{\phi \to \ccp} }{\G_{\phi \to \ccp} + \G_{\phi \to hh} } \simeq  \frac{\G_{\phi \to \ccp} }{ \G_{\phi \to hh} }   = m_{\phi}^2  \lt( \frac{\yc}{\lO}\rt)^2
\ee
Substituting~\cref{Eq:gamma for decay} in~\cref{Eq:dNchidT} and integrating, we obtain the \dm~Yield $\Yco$, from the decay of inflaton as 
%
\eq{
\Yco= \frac{N_\c (\Trh)}{s(\Trh)\, \cs^3(\Trh)}
&\simeq \frac{3}{\pi} \frac{\gs}{\gss} \sqrt{\frac{10}{\gs}} \frac{\mpl \, \Gamma}{m_{\phi}\, \Trh} \text{Br}%
 = 1.163\times 10^{-2} \mpl \frac{\yc^2 }{\Trh}\,. \label{Eq:ychi0-decay-total}
}
Use of $s(\Trh)$, the value of entropy density at $T=\Trh$, and $\cs(\Trh)$, the value of cosmological scale factor at $T=\Trh$, in~\cref{Eq:ychi0-decay-total} is based on the premise that yield of \dm~does not change from $T=\Trh$ up to  present day. Moreover, in this work, we are considering $\gss(\Trh)=106.75$  
Equating~\cref{Eq:ychi0-decay-total} with~\cref{Eq:present day CDM yield}, we get 
\be\label{Eq:eq to plot 2}
\Trh  \seq  
6.49 \times 10^{25}\, \yc^2 \, \mc \,.
\ee 

\begin{figure}[H]
    \centering    
    \includegraphics[width=\wdth]{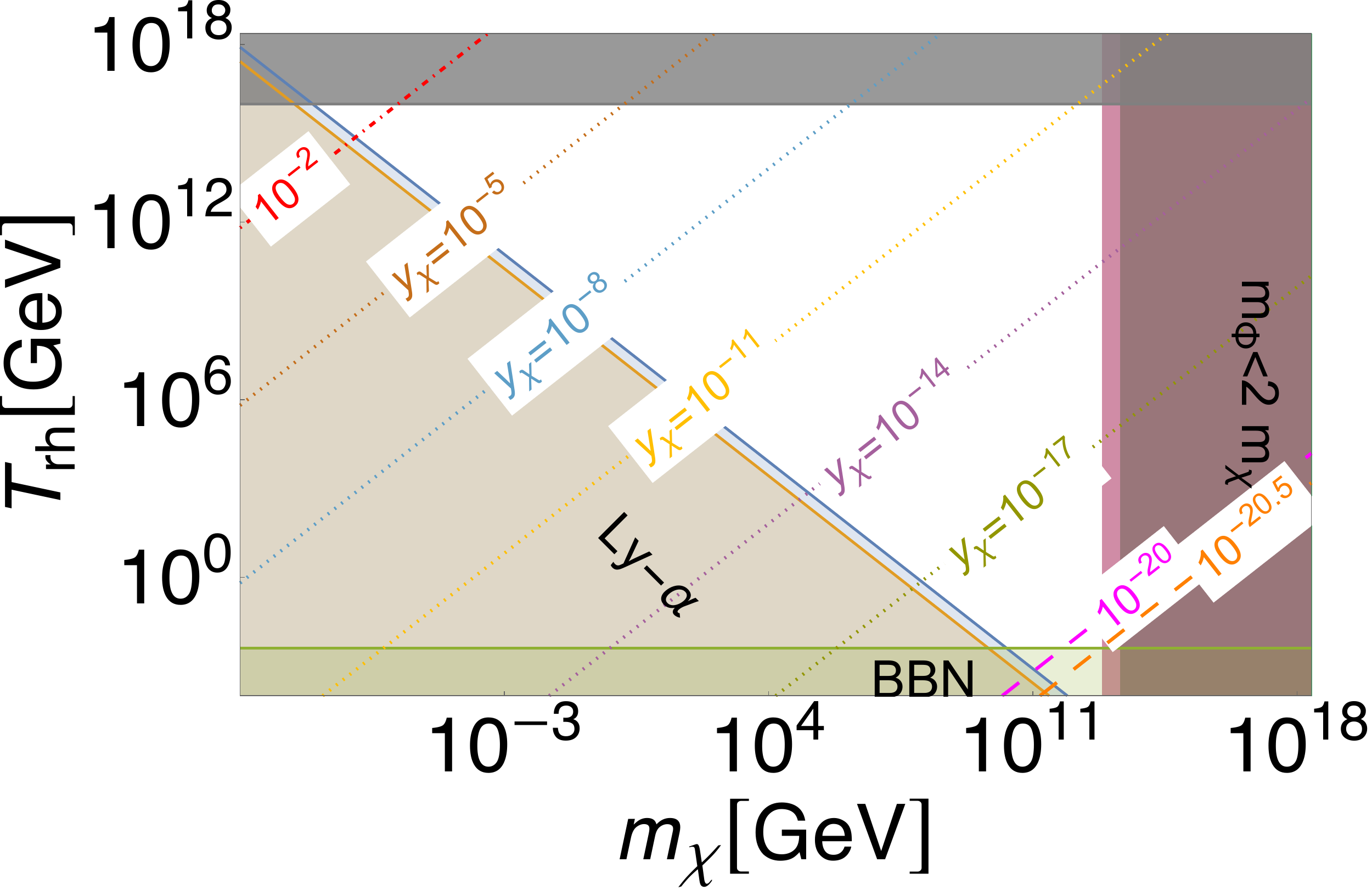}
    \caption{\it \raggedright \label{Fig:CDM_Yield_Decay-NonminimalNatural}
    The white region is allowed on $(\Trh, m_\chi)$ plane for the benchmark values \mNPo~and \mNPt~from~\cref{Table:Benchmark-NonminimalNatural} on log-log scale. The dashed or dashed-dotted lines passing through the white region correspond to~\cref{Eq:eq to plot 2} for different values of $y_\chi$ satisfying present-day CDM density. {
    The red-colored dotted-dashed line in the top-left corner of the plot corresponds to $\yc=10^{-2}$, while magenta and orange colored dashed  lines in the right-bottom corner correspond to $\yc=10^{-20}$ and corresponds to $\yc=10^{-20.5}$, respectively.}  Bounds on this plane are shown as color-shaded regions, and they are coming from: Ly-$\alpha$ bound on the mass of \dm~\cref{Eq:Lyman-alpha-bound}: blue-shaded region, and $\mc\lsim m_\phi/2$: copper-rose colored vertical stripe on the right for `(m)NM-N-1'. For `(m)NM-N-2' same bounds are depicted as peach-colored region and light-copper-rose colored region. Light green-colored stripe at the bottom depicts that $\Trh$ should be $\gsim 4\unit{MeV}$. The gray colored horizontal stripe at the top of the figure indicates that 
    we have restricted ourselves to the highest $\Trh$ possible in the universe to be $10^{16}\GeV$.}  
\end{figure}

If \dm~particles are produced solely from the decay of inflaton in NM-N inflationary scenario and satisfy present-day \cdm~yield, then it must adhere to~\cref{Eq:eq to plot 2} 
which indicates inclined lines on $(\Trh,\mc)$ plane for fixed values of $\yc$. These lines for \nmn~ inflationary scenario are demonstrated as discontinuous lines in~\cref{Fig:CDM_Yield_Decay-NonminimalNatural}. The white region on~\cref{Fig:CDM_Yield_Decay-NonminimalNatural} is allowed.~\cref{Fig:CDM_Yield_Decay-NonminimalNatural} actually shows a comparative view of allowed region for the two benchmark values \mNPo~and \mNPt~which are mentioned in~\cref{Table:Benchmark-NonminimalNatural}. The bounds on this plane are from: Ly-$\alpha$ bound (from~\cref{Eq:Lyman-alpha-bound}) on the mass of \dm~(blue colored region for benchmark `(m)NM-N-1' and peach shaded region for `(m)NM-N-2'), $\Trh\gsim 4\unit{MeV}$: light green shaded horizontal stripe at the bottom, and $\mc\lsim m_\phi/2$ darker and light shaded region with copper-rose color as vertical stripe on the right for `(m)NM-N-1' and `(m)NM-N-2', respectively. Additionally, we assume that the maximum possible value of $\Trh\sim 10^{16}\GeV$ and it is demonstrated by a gray-colored horizontal stripe at the top.
As $m_\phi$ for `(m)NM-N-2' is less than that of `(m)NM-N-1', Ly-$\alpha$ bound which is $\propto m_\phi/\Trh$~\cite{Bernal:2021qrl}, for `(m)NM-N-2' is less than `(m)NM-N-1'. For the same reason, the maximum possible value of $\mc$ is slightly higher in `(m)NM-N-1' than in `(m)NM-N-2'.%
Consequently, the allowed range of $\yc$ passing through the white unshaded region is $10^{-1}\gsim \yc \gsim 10^{-20.5}$ (for $8.20\eV \lsim \mc \lsim 2.05\times 10^{13}\GeV$) for \mNo~and $10^{-1}\gsim  \yc \gsim 10^{-20}$ (for $2.69 \eV \lsim \mc \lsim 6.73\times 10^{12}\GeV$) for \mNt.
The upper bound of $\yc$ and lower bound of $\mc$ alters depending on the assumed maximum value of $\Trh$.  If we assume the maximum value of $\Trh\sim 10^{15}\GeV$, the lower bounds on $\mc$ are $\gsim 81.97\eV$ and $\gsim 26.94\eV$ for \mNo~and \mNt, respectively.
On the other hand, if we assume the maximum value of $\Trh\sim 10^{14}\GeV$, the lower bounds on $\mc$ are $\gsim 819.72\eV$ and $\gsim 269.39\eV$ for \mNo~and \mNt, respectively.
Similarly, for the benchmark \mNr, the lower limit of $\mc$ is $\lsim 1.69\times 10^{13}\GeV$. From this, we conclude that even for the same inflationary scenario, changing the benchmark value can alter the range of $\yc$ and $\mc$ 
required to produce $100\%$ of the total \cdm~relic density of the universe when $\c$ is produced from the decay channel.

%
%
%
%
%
%

This Ly-$\alpha$ bound constrains the parameter space on $(\Trh, \mc)$ plane \footnote{However this bound is not very stringent as long as this is not too close to the Planck scale where the quantum gravity effects may become important. For theories involving Grand Unified Theory (GUT) symmetry breakings, etc. this can be $10^{16}$ GeV. We do not go into details on such issues in this analysis.} such that the produced $\c$ particles behave as \cdm~instead of warm dark matter (\wdm) in the present universe. From the observation of Ly-$\alpha$ absorption lines, the upper bound on present-day velocity of $\c$ particles, $v_\c/c$, and lower bound $m_\c$ are $v_\c/c\lsim 1.8\times 10^{-8}$ and $m_\c\gsim 3.5 \unit{keV}$, respectively, in order to avoid being warm dark matter candidate. 
{
We considered the assumption that the mean initial momentum of DM particles is $m_\phi/2$ at $T=T_{rh}$ (see appendix of~\Ccite{Bernal:2021qrl}).}
Then, the assumption that $\c$ particles are feebly interacting, and because of that the momentum of the particles decreases only due to the expansion of the universe, leading to the bound~\cite{Bernal:2021qrl} 
\be\label{Eq:Lyman-alpha-bound}
\frac{\mc}{\unit{keV}} \gsim 2 \frac{m_\phi}{\Trh}\,,
\ee
where $\mc$ is expressed in $\unit{keV}$. Thus, we conclude that lower permissible range of $m_\chi\gsim {\cal O}\qty(\keV)$ even for $\Trh\sim 10^{16}-10^{15}\GeV$ for \mNo, and \mNt.

\subsection{DM from scattering}
We prefer to choose the maximum allowed value of $\Trh$  as 
\eq{\label{Eq:Nonminimal-natural_max_allowed_Trh}
{\Trh}_{, \text{allowable}} = 
\begin{cases}
    1.3826\times 10^{14} \GeV \, \quad \text{(for `(m)NM-N-1')}\,,\\
    1.0637\times 10^{14}  \GeV \, \quad \text{(for `(m)NM-N-2')} \,,\\
    1.2589 \times 10^{14} \GeV \, \quad \text{(for `(m)NM-N-3')}\,,
\end{cases}
}
such that $\Tmax/\Trh>1$ is maintained. 
For $\Trh\sim {\Trh}_{, \text{allowable}} $, $\Tmax/\Trh\sim 3.16$.
Now, we explore whether \dm~particles generated through the scattering channels for the NM-N inflationary scenario can account fully or partially for the present-day total \cdm~yield or whether there exists a limit on the mass of the $\c$-particles produced via those scattering channels such that it can only contribute completely to the total \cdm~density.
In this study, we consider \dm~yield produced from three 2-to-2 scattering channels
\begin{enumerate}
    \item $Y_{IS,0}$: yield of $\chi$ from scattering of non-relativistic inflaton with graviton as the mediator which is given by
 \eq{
   \label{Eq:yield-DM-scattering-inflaton-graviton}
	    Y_{IS,0} &\simeq \frac{\gs^2}{81920\gss} \sqrt{\frac{10}{\gs}} \left(\frac{\Trh}{\mpl}\right)^3 \left[\left(\frac{\Tmax}{\Trh}\right)^4 - 1 \right] \frac{m_{\chi}^2}{m_{\phi}^2} \left(1-\frac{m_{\chi}^2}{m_{\phi}^2}\right)^{3/2}\,. 
   } 
\item $Y_{\sm g,0}$: yield of $\chi$ from scattering of \sm~particles with graviton as mediator which is given by
\eq{
 \label{Eq:Ysmg0}
 	Y_{\sm g,0} &= 
	\begin{cases}
	\frac{45 \, \alpha_{\dm}}{2\pi^3 \gss} \sqrt{\frac{10}{\gs}} \left(\frac{\Trh}{M_P}\right)^3, \qquad \text{ for } m_\chi \ll \Trh\,,\\
	\frac{45 \,\alpha_{\dm}}{2\pi^3 \gss} \sqrt{\frac{10}{\gs}} \frac{\Trh^7}{M_P^3\, m_\chi^4}\,, \qquad \text{ for } \Tmax \gg m_\chi \gg \Trh\,.
	\end{cases}
}
\item $Y_{\sm i,0}$: yield of $\chi$ from scattering of \sm~particles with inflaton as mediator which is given by
\eq{
Y_{\sm i,0} &\simeq \frac{135\,  y^2_{\chi}\, \lambda_{12}^2}{4 \pi^8\, \gss} \sqrt{\frac{10}{g_\star}}\, \frac{M_P\, \Trh}{m_{\phi}^4}\, , \qquad \text{ for } \Trh \ll m_{\phi}, \Trh > T \,.
}
\end{enumerate}
The value of $\alpha_\dm$ in~\cref{Eq:Ysmg0} is $1.1\times 10^{-3}$~\cite{Bernal:2018qlk} which depends on the coupling of gravitational interaction.

For $\Trh\sim {\Trh}_{, \text{allowable}} $ the condition $ Y_{{\rm CDM},0}\sim Y_{IS,0}$ (from~\cref{Eq:yield-DM-scattering-inflaton-graviton,Eq:present day CDM yield}) leads to the value of $\mc\sim {\mc}^{IS,0}$ as 
\eq{\label{Eq:NMN-large mc is needed-InflatonScattering}
{\mc}^{IS,0} \sim 
\begin{cases}
    4.574 \times 10^{10} \GeV \, \quad \text{(for `(m)NM-N-1')}\,,\\
    2.8312 \times 10^{10}  \GeV \, \quad \text{(for `(m)NM-N-2')} \,,\\
    4.4201 \times 10^{10}  \GeV \, \quad \text{(for `(m)NM-N-3')} \,.
\end{cases}
}
Hence, if $\Trh< {\Trh}_{, \text{allowable}} $, the value of $\mc$ should exceed the values mention in~\cref{Eq:NMN-large mc is needed-InflatonScattering} in order to make the yield of \dm~produced through the scattering process of~\cref{Eq:yield-DM-scattering-inflaton-graviton} comparable to $Y_{{\rm CDM},0}$ mentioned in~\cref{Eq:present day CDM yield}.
Hence, if $\Trh\lsim {\Trh}_{, \text{allowable}}$ and $\mc \gsim {\mc}^{IS,0} $ for \nmn~inflation, then $\c$ produced solely through the 2-to-2 scattering of inflaton with graviton as the mediator can yield $100\%$ of the total \cdm~relic density.%
\footnote{
$\mc$ can have a maximum value $< \sqrt{4\pi} \mpl$, otherwise it would turn into a black hole.
}

When \dm~particles are produced from the scattering of SM particles via graviton
mediation with $\mc \ll \Trh$, then from~\cref{Eq:Ysmg0} with $\Trh\sim {\Trh}_{, \text{allowable}}$,
\eq{
Y_{\sm g,0}=
\begin{cases}
    4.3752 \times 10^{-19} \, \quad \text{(for `(m)NM-N-1')}\,,\\
    1.9925 \times 10^{-19} \, \quad \text{(for `(m)NM-N-2')} \,,\\
    3.3027 \times 10^{-19}  \,  \quad \text{(for `(m)NM-N-3')}\,.
\end{cases}
}
And thus to satisfy $Y_{{\rm CDM},0}$, it is required
\eq{\label{Eq:NMN-large mc is needed}
\mc \gsim 
\begin{cases}
    9.8282 \times 10^{8} \GeV \, \quad \text{(for `(m)NM-N-1')}\,,\\
    2.1581 \times 10^{9} \GeV \, \quad \text{(for `(m)NM-N-2')}\,, \\
    1.302 \times 10^{9} \GeV \, \quad \text{(for `(m)NM-N-3')}\,.
\end{cases}
}

Now, we consider production of $\c$ only via the \three. From~\cref{Eq:Nonminimal-natural_max_allowed_Trh}, the maximum allowed value of $\Trh\sim {\cal O}(10^{14})$ in $\GeV$. Furthermore, for $\Trh\sim 10^{12}\GeV$, $\Tmax/\Trh\sim 30$.

However, for $(\Tmax \gg m_\chi \gg \Trh)$ and $Y_{\sm g,0}=Y_{{\rm CDM},0}$, leads to the relation (in $\GeV$)
\eq{\label{Eq:TERM}
\Trh^{\sm g,0}= 7.1389 \times 10^{12} \, {f}_{\Trh}^{-3/4}\,,
}
where ${f}_{\Trh}=\Trh/\mc$. 
Now, according to~\cref{Eq:TERM}, in order to achieve $\Trh\sim {\cal O}(10^{14}) \GeV$, ${f}_{\Trh}\sim 0.03$ is required.

For the last process, \four, to make $Y_{\sm i,0}\sim f_\c \, Y_{{\rm CDM},0}$, 
where $f_\c$ is the fraction of total \cdm~relic density that can be contributed by $Y_{\sm i,0}$, 
we get  
\eq{\label{Eq:game_changer}
{\mc}^{\sm i,0} = 
4.2164 \times 10^{-5} \frac{m_\phi^4}{\Trh \, \mpl\,\lambda_{12}^2} \, \frac{f_c}{\yc^2}\,\GeV\,.
} 
Then, with $f_\phi= \Trh / m_\phi$ in~\cref{Eq:definition of reheating temperature}
and for $\yc
\sim 10^{-5}$ for all three benchmark values (`(m)NM-N-1', `(m)NM-N-2', and `(m)NM-N-3')
\eq{\label{Eq:NMN-SMScatteringInflatonMediator}
\mc \sim 
3.2689\times 10^3 \, \frac{f_\c}{f_\phi^3} \,\GeV\,.
 %
    %
    %
    %
}
For instance, with $f_\c=1$ and $f_\phi\sim 10^{-2}$, we get $m_\c \sim {\cal O} (10^{9})\GeV$ to satisfy total \cdm~relic density. 

Hence, three of the four scattering processes (\one, \two, and \four) can effectively produce $\c$ and contribute up to $100\%$ of the total \cdm~relic density of the present universe in the context of \nmn~inflation.


\section{Periodic non-minimal coupling and natural inflation}
\label{Sec:Non-minimal-periodic natural}

%
%
The shift symmetry of a system is preserved if the system is invariant under a transformation of its field variables by a constant value. In natural inflation, as mentioned earlier, $\qty(1+\cos(\varphi/f_a))$ is invariant under the transformation of $\vp\to \vp+\text{ constant}$, where the constant is $2\pi f_a$. In this inflationary model, we consider non-minimal coupling to gravity, which respects the same discrete shift symmetry just like the potential of natural inflation.
 Therefore,
 non-minimal coupling is specified as~\cite{Ferreira:2018nav}
\eq{ \label{Eq:periodic-nominimal-coupling}
 \Omega^2(\vp)= 1+ \alpha\lt( 1+ \cos\lt(\frac{\vp}{f_a}\rt)\rt)\,,
}
where $\alpha$ is dimensionless. 
As a result of such choice of $\Omega^2(\vp)$, both $\Omega^2(\vp)$ and $V^{JF}(\vp)$ possess the same shift symmetry $\vp\to \vp + 2\pi f_a$. 
Then the potential in Einstein frame is
\eq{\label{Eq:Nonminimal-periodic-natural-potential in einstein frame}
V^{E}(\phi(\vp)) \equiv V^{E}_{\rm \nmpn}(\phi(\vp))=\frac{\L^4 \lt( 1+ \cos\lt(\frac{\vp}{f_a}\rt)\rt)}{\qty( 1+ \alpha\lt( 1+ \cos\lt(\frac{\vp}{f_a}\rt)\rt)  )^2}
}
The periodicity of the coupling to gravity and potential is both $2\pi f_a$. That's why hereafter we call this inflationary model as non-minimal periodic Natural (\nmpn) inflation. Accordingly, the part of the coupling containing $\alpha\lt( 1+ \cos\lt(\frac{\vp}{f_a}\rt)\rt)$ vanishes at the minimum of the potential, i.e. during reheating. This particular choice ensures that reheating happens in Einstein frame. 
Here, in~\cref{Eq:periodic-nominimal-coupling}, $\alpha>-0.5$ is required so that Planck scale is well-defined. Furthermore, for $\alpha=0.5$, the periodicity of $V^{JF}(\vp)$ as a function of $\vp$ is equal to the periodicity of  $V(\phi)$ as a function of $\phi$, and for $0.5<\alpha<1$, both periodicity scales are almost of the same order~\cite{Ferreira:2018nav}.

$V^E_{\rm \nmpn}\lt(\phi(\vp)\rt)$ has two sets of stationary points: (a) both at $\vp_{a,{\rm even}}/\mpl=2f_a\pi \mathbb{c}_1$, and at $\vp_{a,{\rm odd}}/\mpl=f_a(\pi+ 2\pi \mathbb{c}_1)$, 
and (b) at $\vp_b/\mpl=f_a \, \tan ^{-1}\left(\pm\frac{\sqrt{2 \alpha -1}}{1-\alpha }\right) + 2\pi \mathbb{c}_1, \,\,\, \text{where}\,\mathbb{c}_1\in \mathbb{Z} $, set of integers. For $\alpha<0.5$, there is only one set of stationary points, i.e. $\vp_{a,{\rm even}}$ and $\vp_{a,{\rm odd}}$,
and at those points
\eq{
\lt.\frac{\td^2 V^E\lt(\phi(\vp)\rt)}{\td \vp^2} \rt|_{\vp=\vp_{a,{\rm even}}}
= \frac{(2 \alpha -1) {\Lambda^4}}{(2 \alpha +1)^3 {f_a}^2}
\,,\qquad 
&& 
\lt.\frac{\td^2 V^E\lt(\phi(\vp)\rt)}{\td \vp^2} \rt|_{\vp=\vp_{a,{\rm odd}}}
=\frac{\L^4}{f_a^2}
} 
Hence, $\vp_{a,{\rm odd}}$ is always the location of local minima, whereas, $\vp_{a,{\rm even}}$ can be maxima, or minima, or saddle points depending on whether $\alpha<0.5$, or $>0.5$, or $=0.5$. In contrast, $\vp_{b}$ is the location of maxima, or minima, or saddle points depending on whether $\alpha>0.5$, or $<0.5$, or $=0.5$. This is because
\be 
\lt.\frac{\td^2 V^E\lt(\phi(\vp)\rt)}{\td \vp^2} \rt|_{\vp=\vp_{b}}=
\frac{\qty(1-2 \alpha)\L^4 }{8 \alpha \,  f_a^2}\,.
\ee 

\begin{figure}[H]
    \centering    
\includegraphics[height=6cm,width=\wdth]{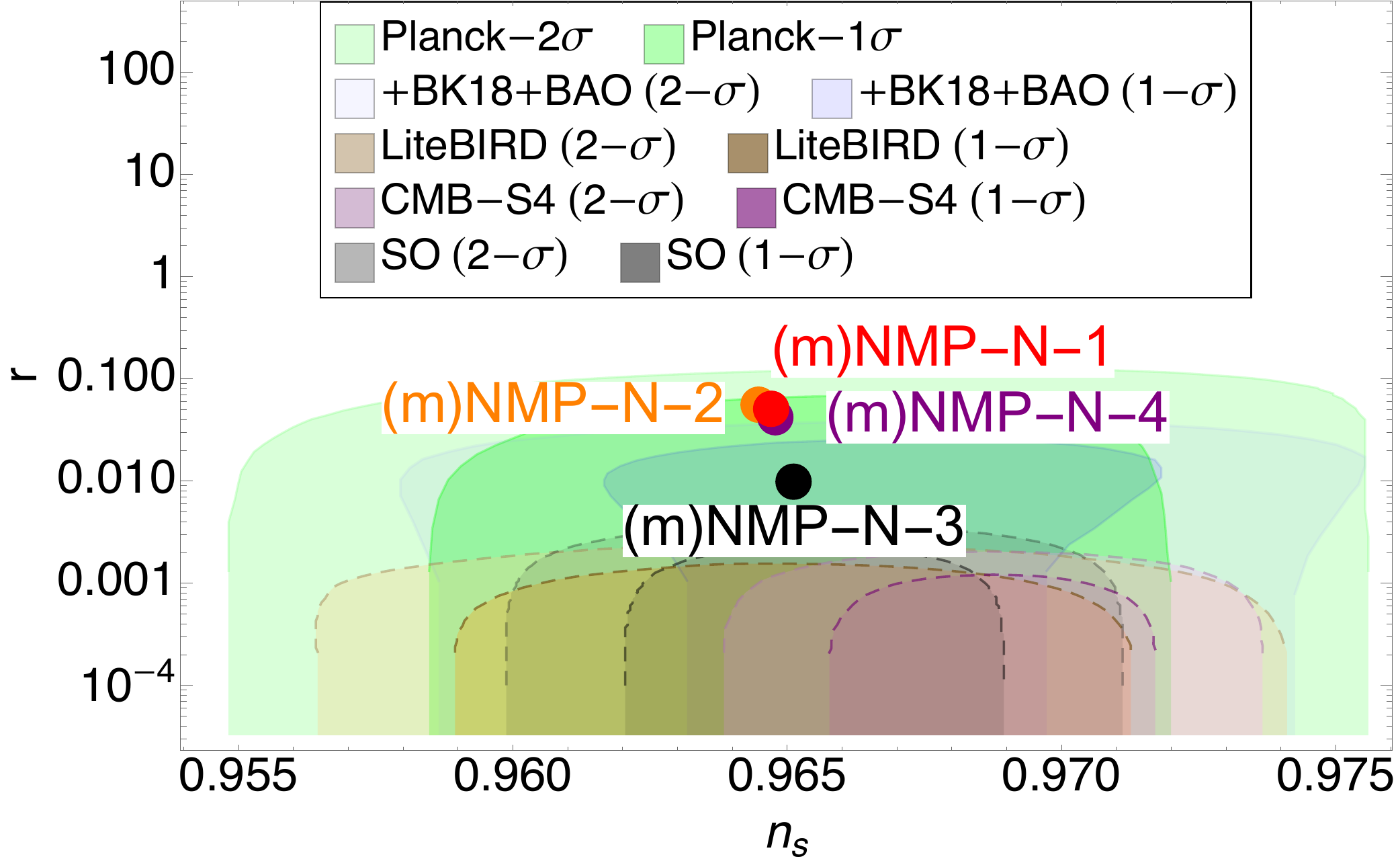}
    \caption{\it \raggedright \label{Fig:NonminimalPeriodicNatural-ns-r-bound}  
    The colored dots represent the predicted value of $(n_s,r)$ for two benchmark values from~\cref{Table:Benchmark-NI-periodic}. Alongside, this figure also displays $68\%$ and $95\%$ C.L. $(n_s, r)$ best-fit contour from the \Planck~2018+\BICEP~and additional forthcoming \cmb~observations mentioned above in~\cref{Fig:NonminimalNatural-ns-r-bound} .
    }
\end{figure}

We consider the slow roll inflationary scenario, where the inflaton begins rolling near $\vp =0$ and continues to roll towards larger values of $\vp$.
When $\alpha<0.5$,
$\vp=0$ is the location of maximum and slow roll inflation stops whenever it becomes $\lt|\eta_V\rt|\sim 1$, and for $f_a\ll\alpha$, inflation may be impossible as $\lt|\eta_V\rt|>1$ even at $\vp=0$.
For $\alpha=0.5$, $\vp=0$ is the location of saddle point, and thus $\eta_V=\epsilon_V=0$ at $\vp=0$.
However, for $\alpha>0.5$ there is a maximum between  $\vp=0$ and the minimum of the potential. In this case, the inflaton may travel across this maximum during its course of evolution.
%
%
%
%
Our chosen benchmark values are listed in~\cref{Table:Benchmark-NI-periodic}. For benchmark `(m)NMP-N-1', $\vp=0$ is a saddle point whereas For benchmark `(m)NMP-N-2', there is a maximum at $\vp/\mpl=9\pi\approx 28.27$. The $n_s,r$ prediction of the benchmark values are displayed in~\cref{Fig:NonminimalPeriodicNatural-ns-r-bound} along with the best-fit contours from current and future \cmb~observations. The predicted value of $r$ for benchmark values \mNPo, \mNPt, and \mNPf~fall within $1-\sigma$ contour of \Planck2018 data. Although the benchmark \mNPr~ is for $\ncmb\approx 70$, the predicted value of $r$ is within $1-\sigma$ contour of \Planck2018+\BICEP3 (2022)+\KeckArray2018 data. This benchmark value indicates that \nmpn~inflation can predict values of $r$ that remain within $1-\sigma$ contour of \Planck+\BICEP~combined data for larger e-folds.~\cref{Table:Benchmark-NI-periodic} also indicates that except for benchmark \mNPf, which we have selected for comparison later in this section, a value of
$f_a/\mpl\sim {\cal O}\qty(10-20)$ is required to predict $n_s\sim 0.9647$. Although these values of $f_a$ are slightly greater than the required value of $f_a$ in~\cref{Table:Benchmark-NonminimalNatural}, these are still smaller than the value required for minimal natural inflation \footnote{The value of $f_a>\mpl$ is associated with the scenario from the point of view of the naturalness of hierarchy of scales may be troublesome but we do not worry about such issues in this analysis. Instead, we focus on the dark matter production from post-inflationary dynamics.}.

 %
%
\begin{table}[H]
\centering
\caption{\it Benchmark values for \nmpn~inflationary model in metric formalism.
}
\label{Table:Benchmark-NI-periodic}
\begin{tabular}{|c||c | c | c | c | c | c | c| c| c|} 
 \hline
{\it Benchmark} &$f_a/\mpl$ & $\alpha$ & $\varphi_*/\mpl$ & $\varphi_{\rm end}/\mpl$ &  $n_s$ & $r\times 10^2$ & $\L^4/\mpl^4$ & $\varphi_{\rm min}/\mpl $&$\ncmb$\\ %
 \hline
\hline 
(m)NMP-N-1 &$11.97$ & $0.5$ & $23.8164$ & $36.1996$ &  $0.9647$ & $5.043$ & $4.4407\times 10^{-9} $ & $ 37.6049 $ & $60$\\ 
\hline 
(m)NMP-N-2 &$18$ & $1$ & $42.4644$ & $55.1417$ &  $0.9645$ & $5.6440$ & $ 1.0048\times 10^{-8}$ & $  56.5487$ &$60$\\ 
\hline 
(m)NMP-N-3 &$6.3$ & $0.48$ & $8.0198$ & $18.408$ &  $0.9651$ & $0.9718$ & $ 6.1337\times 10^{-10}$ & $  19.792$ & $70$ \\ 
\hline 
(m)NMP-N-4 & $164.0$ & $100$ & $500.831$ & $513.815$ &  $0.9648$ & $4.2071$ & $ 6.5157\times 10^{-7}$ & $515.221  $ & $65$ \\ 
\hline
%
%
 %
 %
\end{tabular}
\end{table}
%
%

%
%

\subsection{Reheating and production of \dm~from inflaton decay}
\begin{figure}[H]
    \centering    \includegraphics[width=\wdth]{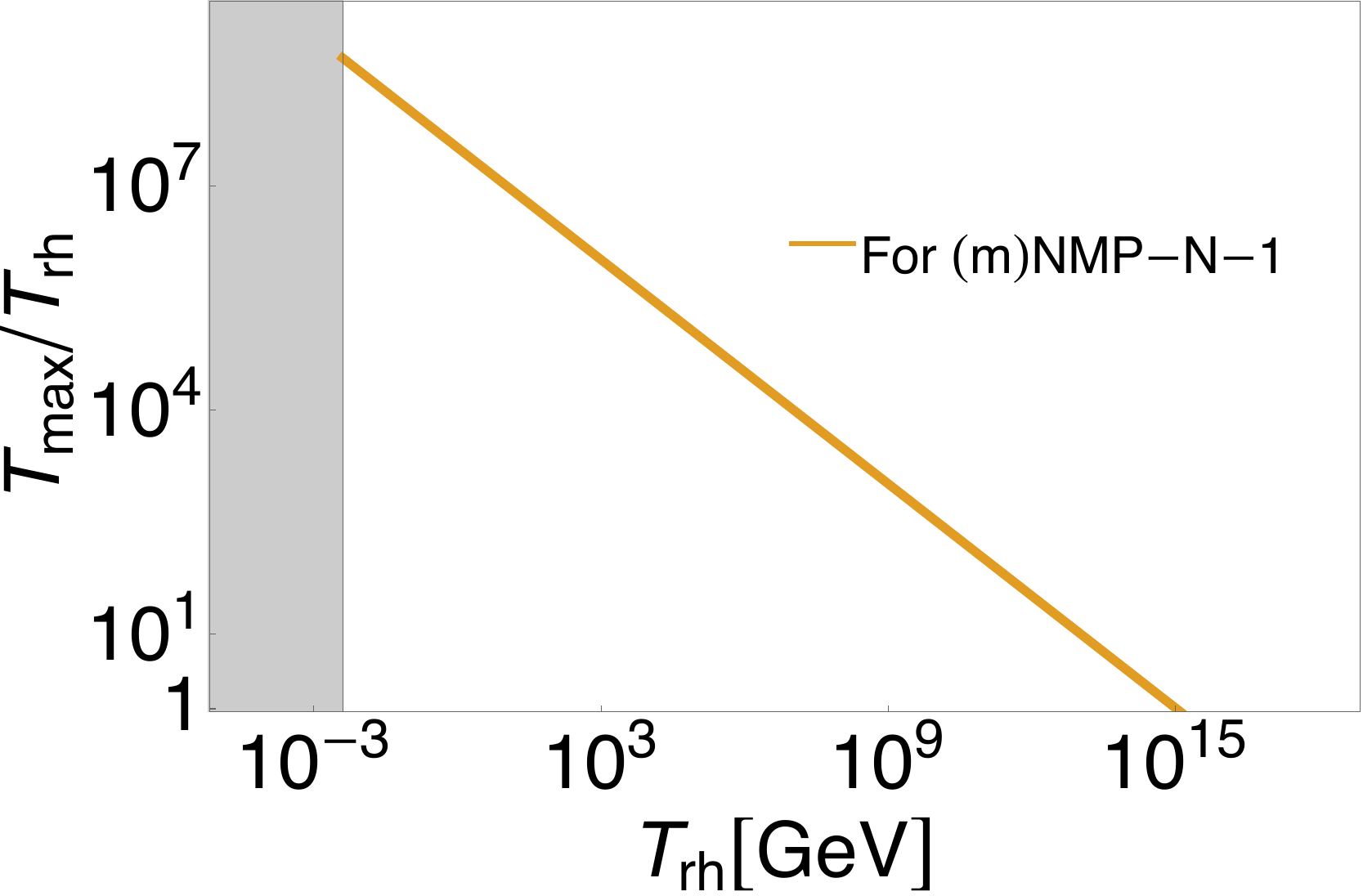}
    \caption{\it \raggedright \label{Fig:Tmax/Trh-Non-minimal-periodic_natural}
    Variation of $T_{max}/\Trh$ against $\Trh$ for benchmark value \mNPo~from~\cref{Table:Benchmark-NI-periodic}. The gray-colored vertical stripe on the left of the plot of lower bound on $\Trh$ which is $\Trh\nless 4 \unit{MeV}$.
    }
\end{figure}

The reheating scenario in this inflationary model is similar to that of NM-N inflation discussed in~\cref{Sec:Non-minimal-Natural-reheating}, although the non-minimal coupling vanishes for \nmpn~inflation at the minimum. The locations of the minima $\vp_{\rm min}/\mpl$ of the potential of~\cref{Eq:Nonminimal-periodic-natural-potential in einstein frame} around which the inflaton oscillates during reheating era for the benchmark values mentioned in~\cref{Table:Benchmark-NI-periodic} and thus, the  mass of the inflaton is given by~\cref{Eq:mass-of-inflaton-NonMinimalNatural}. Following a similar approach, $\G_\varphi, \Trh$, and $\hubble_I$ can be calculated and are listed in~\cref{Table:reheating_values-NonminimalNatural-periodic} for the benchmark values of~\cref{Table:Benchmark-NI-periodic}. 
It can be observed that $m_\phi/\mpl$ is almost of the same order for \mNPo~and \mNPt, and also for \mNPr, \mNPf, and hence, the same conclusion can be drawn for $\Tmax$ as ${\cal H}_I/\mpl$ is almost equal for all benchmark values. For this reason, $\Tmax/\Trh$ versus $\Trh$ is exhibited only for \mNPo~in~\cref{Fig:Tmax/Trh-Non-minimal-periodic_natural}. Gray-colored vertical stripe on the left of the plot represents the lower bound on $\Trh\gsim 4 \unit{MeV}$. From~\cref{Fig:Tmax/Trh-Non-minimal-periodic_natural}, we get the maximum value of $\Tmax/\Trh\sim 5.1378\times 10^8 \GeV$ at $\Trh\sim 4 \unit{MeV}$ and $\Tmax/\Trh\sim 1$ at $\Trh\sim 1.056\times 10^{15}\GeV$. 
Furthermore, lower bound on $\Trh$ leads to establishing the minimum permissible value of $\lO$ which are $\lO/\mpl\gsim 4.4662\times 10^{-23}$ for \mNPo, $\lO/\mpl\gsim 4.4669\times 10^{-23}$ for \mNPt, $\lO/\mpl\gsim 3.753\times 10^{-23}$ for \mNPr, and $\lO/\mpl\gsim 4.1994\times 10^{-23}$ for \mNPf.

\begin{table}[H]
\centering
\caption{\it $m_\varphi, \G_\varphi, \Trh$, and $\hubble_I$ for the benchmark values from~\cref{Table:Benchmark-NI-periodic}.}
\label{Table:reheating_values-NonminimalNatural-periodic}
\begin{tabular}{| c || 
c | c|  c | c|} 
 \hline
{\it Benchmark}& 
$m_\phi/\mpl$ & $\G_\phi \mpl$ & $\Trh/\lambda_{12}$ & ${\cal H}_I/\mpl$ \\
&  
(Eq.~\eqref{Eq:mass-of-inflaton-NonMinimalNatural}) & (Eq.~\eqref{Eq:decay-width-of-inflaton}) & (Eq.~\eqref{Eq:definition of reheating temperature}) &  (\cref{Eq:HI-natural})  \\
[0.5ex]
 \hline\hline
 (m)NMP-N-1 &  
 $5.5671\times 10^{-6}$ & $7147.08 \,\lambda_{12}^2$ & $37.3175 $ & $3.1881\times 10^{-6}$\\ 
 \hline
  (m)NMP-N-2 &  
  $5.5689\times 10^{-6}$ & $ 7144.85\,\lambda_{12}^2$ & $ 37.3117$ & $3.1882\times 10^{-6}$\\ 
 \hline
 (m)NMP-N-3 & $3.9312\times 10^{-6}$ & $10121.40 \, \lO^2$ & $44.4087$ & $2.1915\times 10^{-6}$ \\
 \hline 
(m)NMP-N-4 & $4.9219 \times 10^{-6}$& $ 8083.95 \, \lO^2$ & $ 39.6881$ &  $ 2.8152 \times 10^{-6}$\\
 \hline
\end{tabular}
\end{table}
%
%

%
%
%
%
%
%

%
%
%
\begin{figure}[H]
    \centering    
    \includegraphics[width=\wdth]{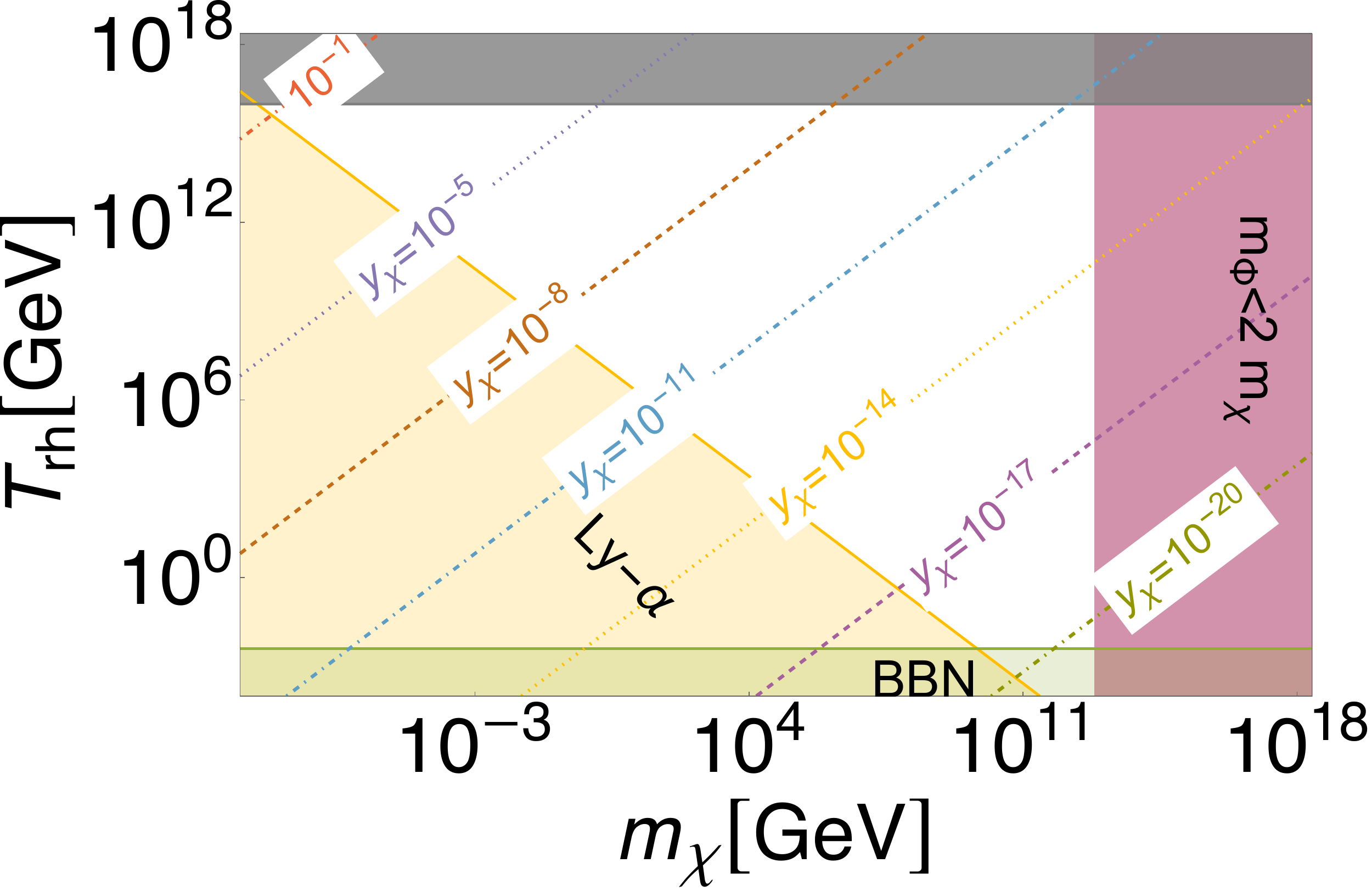}
    \caption{\it \raggedright \label{Fig:CDM_Yield_Decay-NonminimalPeriodicNatural}
    The white region is allowed on $(\Trh, m_\chi)$ plane for the benchmark value \mNPo from~\cref{Table:Benchmark-NI-periodic} on log-log scale. The dashed or dashed-dotted lines passing through the white region correspond to~\cref{Eq:eq to plot 2} for different values of $y_\chi$ satisfying present-day CDM density. {
    The red colored dotted-dashed line in the top-left corner of the plot corresponds to $\yc=10^{-1}$.}
    Bounds on this plane are shown as color-shaded regions, and they are coming from: Ly-$\alpha$ bound on the mass of \dm~(\cref{Eq:Lyman-alpha-bound}): peach-colored region, and $\mc\lsim m_\phi/2$: copper-rose colored vertical stripe on the right. Light green-colored stripe at the bottom depicts that $\Trh$ should be $\gsim 4\unit{MeV}$. The horizontal stripe at the top highlighted in gray is for $\Trh \lsim 10^{16}\GeV$.
    }
\end{figure}%
%
%
%
Similar to~\cref{Fig:CDM_Yield_Decay-NonminimalNatural}, the discontinuous and inclined lines visible on $(\Trh,\mc)$ plane in~\cref{Fig:CDM_Yield_Decay-NonminimalPeriodicNatural} correspond to $\c$ produced from the decay of inflaton during reheating in \nmpn~inflation which can completely account for the total \cdm~density as stated by~\cref{Eq:eq to plot 2}. The bounds on this plane are for Ly-$\alpha$ from~\cref{Eq:Lyman-alpha-bound} (peach-shaded region), from BBN temperature (green-colored horizontal stripe at the bottom), $\mc\lsim m_\phi/2$ (copper-rose colored vertical stripe at the right of the plot), and horizontal stripe at the top highlighted in gray for our assumption that the upper limit for $\Trh$ is $\sim 10^{16}\GeV$. The upper and lower bound of $\yc$ is $\lsim {\cal O}(10^{-1}) $ and $\gsim {\cal O}(10^{-20})$ (for $2.67\eV\lsim  \mc\lsim 6.68\times 10^{12}\GeV$) to account for the total \cdm~yield, produced through decay channel. However Ly-$\alpha$ bound rules out $\mc\lsim {\cal O}\qty(\keV)$ if $\c$ particles contribute to \cdm. Since the allowed region on $(\Trh, \mc)$ plane for \nmpn~inflation varies for all the chosen benchmark values, similar to the scenario explicitly displayed in~\cref{Fig:CDM_Yield_Decay-NonminimalNatural} of \nmn~inflation, it is expected that ranges of $\yc$ and $\mc$ will not be exactly identical for all the benchmark values from~\cref{Table:Benchmark-NI-periodic}.

\subsection{DM from scattering}
Similar to~\cref{Eq:Nonminimal-natural_max_allowed_Trh} in \nmn~inflationary scenario, we can define ${\Trh}_{, \text{allowable}}$ for NMP-N inflationary scenario such that $\Tmax/{\Trh}_{, \text{allowable}}\sim 3.16$ as
\eq{\label{Eq:NonminimalP-natural_max_allowed_Trh}
{\Trh}_{, \text{allowed}} = 
\begin{cases}
    1.0559\times 10^{14} \GeV \, \quad \text{(for `(m)NMP-N-1')}\,,\\
    1.0571\times 10^{14}  \GeV \, \quad \text{(for `(m)NMP-N-2')} \,,\\
    8.7639 \times 10^{13}  \GeV \, \quad \text{(for `(m)NMP-N-3')}\,,\\
    9.9331 \times 10^{13}  \GeV \, \quad \text{(for `(m)NMP-N-4')}\,.
\end{cases}
} 
Now, for $\Trh\lsim {\Trh}_{, \text{ allowable}} $, the condition for $ Y_{{\rm CDM},0}\sim Y_{IS,0}$ (from~\cref{Eq:present day CDM yield,Eq:yield-DM-scattering-inflaton-graviton}) leads to $\mc\sim {\mc}^{IS,0}$ such that
\eq{\label{Eq:NMPN-large mc is needed-InflatonScattering}
{\mc}^{IS,0} \sim 
\begin{cases}
    2.8369 \times 10^{10} \GeV \, \quad \text{(for `(m)NMP-N-1')}\,,\\
    2.8343 \times 10^{10}  \GeV \, \quad \text{(for `(m)NMP-N-2')} \,,\\
    2.7103 \times 10^{10}  \GeV \, \quad \text{(for `(m)NMP-N-3')}\,,\\
    2.7778 \times 10^{10}  \GeV \, \quad \text{(for `(m)NMP-N-4')}\,.
\end{cases}
}
Therefore,  if $\Trh< {\Trh}_{, \text{allowable}} $, the value of $\mc$ should exceed the values mentioned in~\cref{Eq:NMPN-large mc is needed-InflatonScattering} in order to make the yield of \dm~produced through the scattering process of~\cref{Eq:yield-DM-scattering-inflaton-graviton} comparable to $Y_{{\rm CDM},0}$ mentioned in~\cref{Eq:present day CDM yield}.
Hence, if $\Trh\lsim {\Trh}_{, \text{allowed}}$ and $\mc \gsim {\mc}_{,1} $ for \nmn~inflation, then $\c$ produced solely through the 2-to-2 scattering of inflaton with graviton as the mediator can yield $100\%$ of the total \cdm~relic density.

When \dm~particles are produced from the scattering of SM particles via graviton
mediation (\cref{Eq:Ysmg0} with $\mc \ll \Trh$ and $\Trh\sim {\Trh}_{, \text{allowable}}$)
\eq{
Y_{\sm g,0}=
\begin{cases}
    1.9489 \times 10^{-19} \GeV\, \quad \text{(for `(m)NMP-N-1')}\,,\\
    1.9554 \times 10^{-19} \GeV \, \quad \text{(for `(m)NMP-N-2')} \,,\\
    1.1144 \times 10^{-19} \GeV \, \quad \text{(for `(m)NMP-N-3')}\,,\\
    1.6225 \times 10^{-19} \GeV \, \quad \text{(for `(m)NMP-N-4')}\,.
\end{cases}
}
And thus to satisfy $Y_{{\rm CDM},0}$, it is required
\eq{\label{Eq:NMPN-large mc is needed}
\mc \sim 
\begin{cases}
    2.2064 \times 10^{9} \GeV \, \quad \text{(for `(m)NMP-N-1')}\,,\\
    2.1990 \times 10^{9} \GeV \, \quad \text{(for `(m)NMP-N-2')} \,,\\
    3.8586 \times 10^{9} \GeV \, \quad \text{(for `(m)NMP-N-3')} \,,\\
    2.6502 \times 10^{9} \GeV \, \quad \text{(for `(m)NMP-N-4')} \,.
\end{cases}
}

However, since ${\Trh}_{, \text{ allowable}} \sim {\cal O}(10^{14}) \GeV$ (see~\cref{Eq:NonminimalP-natural_max_allowed_Trh}) similar to \nmn~inflation, the conclusion remains the same as that of \nmn, for the production of $\c$ via \three~process.

For the last scattering process (\four), for $Y_{\sm i,0}\sim f_\c \, Y_{{\rm CDM},0}$, and from~\cref{Eq:game_changer}
with $\yc
\sim 10^{-5}$ 
we again get for the four benchmark values
\eq{\label{Eq:NMPN-SMScatteringInflatonMediator}
\mc \sim 
3.2689\times 10^3 \, \frac{f_\c}{f_\phi^3} \GeV\,.
%
%
    %
%
}
If we take $f_\c = 1$ and $f_\phi \sim  10^{-2}$, we can obtain $m_\c \sim {\cal O}(10^9) \GeV$ to satisfy the total \cdm~relic density.




\section{Non-minimal Coleman-Weinberg inflation
}
\label{Sec:Nonminimal CW}
\begin{figure}[H]
    \centering    
    \includegraphics[height=6cm,width=\wdth]{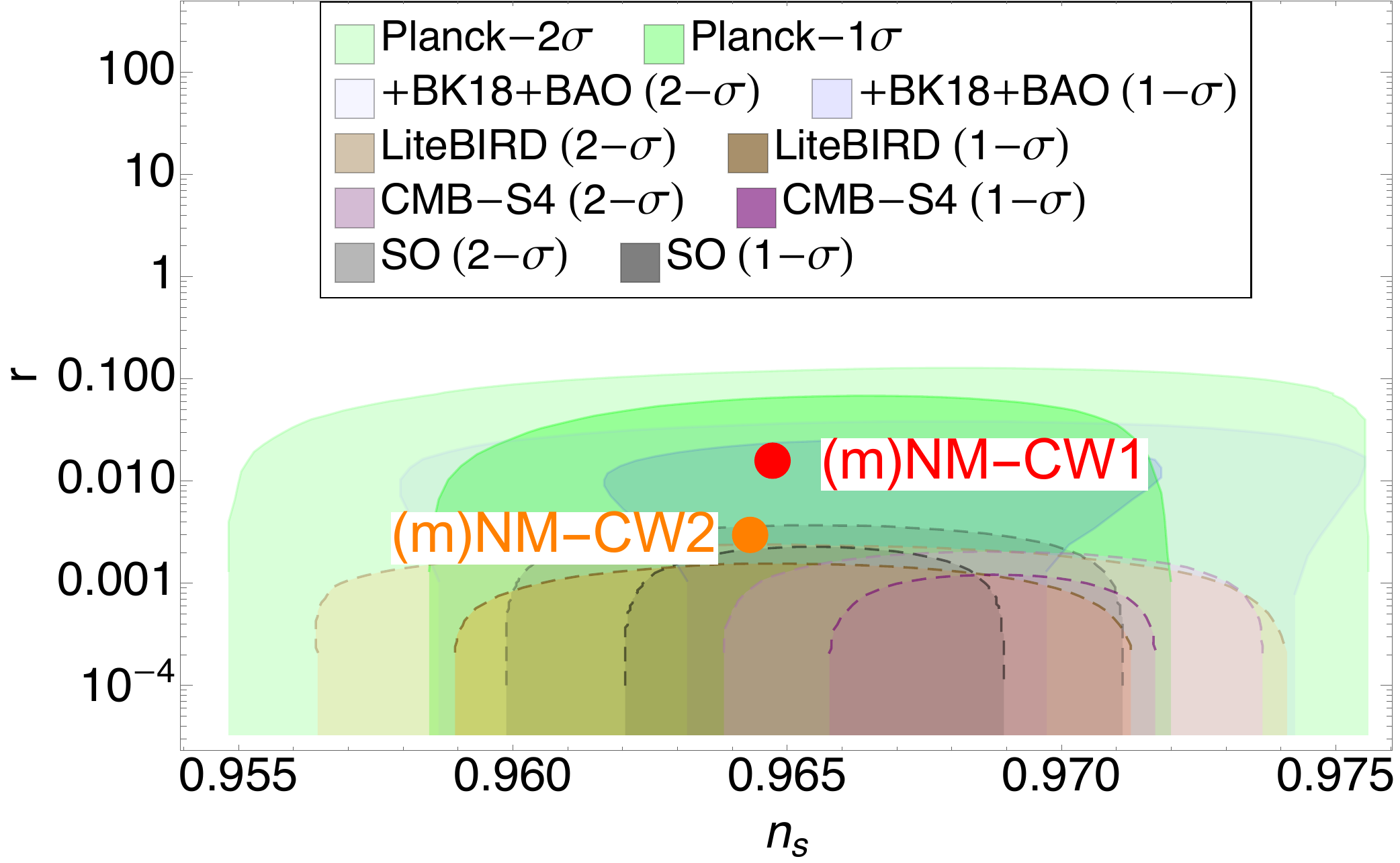} 
    \caption{\it \raggedright \label{Fig:NonminimalCW-ns-r-bound} 
    The predicted values of $n_s, r$ for the two benchmark values from~\cref{Table:Benchmark-CW} are displayed as colored circular dots.
    In addition, this figure also displays $2-\sigma$ and $1-\sigma$ $(n_s, r)$ best-fit contours from the \Planck~2018+\BICEP, as well as additional upcoming \cmb~observations mentioned ealier in~\cref{Fig:NonminimalNatural-ns-r-bound}. 
    }
\end{figure}

%
%
Any fundamental \bsm~scalar particle receives quantum corrections arising either due to self-interactions of gauge or Yukawa interactions present in the theory. Such quantum corrections (mostly coming from dimension-4 operators due to renormalizability conditions) are logarithmic in nature. If such a potential is assumed to drive inflation the renormalization group (RG) improved effective inflationary potential 
which is typically given by~\cite{Racioppi:2017spw, Kannike:2015kda,Kannike:2014mia,Bostan:2019uvv}
\be 
V^{JF}(\vp)=
\frac{\lambda_{\rm eff}(\vp)}{4} \, \vp^4 +\Lambda_{CW}\,,
\ee 
where $\Lambda_{CW}$ is a constant having quartic mass dimension. $\lambda_{\rm eff}(\vp) $ can be written approximately as
\be  
\lambda_{\rm eff}(\vp) =\lambda_{\rm eff}(\cq) + \beta_{\rm eff}(\cq)  \, f(\log[\vp/\cq]) \,.
\ee 
Here, $\cq$ is the renormalization scale such that the beta function $\beta_{\rm eff}(\cq)= \frac{\pd }{\pd \log[\cq]}\lambda_{\rm eff}(\cq)$ and $f(\log[\vp/\cq]) $ is a polynomial function of $\log(\vp/\cq)$.
Considering the 1-loop correction and adjusting $\Lambda_{CW}$ from the condition $V^{JF}(\vp=\cq)=0$, the effective potential in Jordan frame~\cite{Martin:2013tda} 
\be\label{Eq:CW_potential_JordanFrame}
V^{JF}(\vp)= V_{\nmcw}(\varphi)= \frac{A \cq^4}{4}+A \varphi ^4 \left[\log \left(\frac{\varphi }{\cq}\right)-\frac{1}{4}\right]\,.
\ee 
Here $\Lambda_{CW}=A\,\cq^4/4$. $\cq$ has the dimension of mass and $\cq=\lt<\varphi\rt>$ is the renormalization scale with $V(\varphi=\cq)=V'(\varphi=\cq)=0$. $A$ is dimensionless and $A= A(\cq)$ is determined by the beta function of the scalar-quartic-coupling with inflaton. 
Now, we are assuming non-minimal coupling to the gravity of the following form~\cite{Maji:2022jzu}
\be 
\Omega^2(\vp)= 1+ \xi \frac{\vp^2- \cq^2}{\mpl^2}\,,
\ee 
and then the inflaton transforms from Jordan to Einstein as
\eq{\label{Eq:JordantoEinsteinNew}
\frac{\td \phi}{\td \vp}=\sqrt{
\frac{\xi  (6 \xi +1) \varphi^2 M_P^2 - \xi \cq^2\mpl^2+M_P^4}{\left(M_P^2+\xi  (\varphi
   ^2-\cq^2)\right){}^2}\,.
}
}
Thus the potential of this inflationary scenario (non-minimal Coleman Weinberg, abbreviated as \nmcw~inflation) transforms in Einstein frame as
\be\label{Eq:PotCW-Einsteinframe}
V^E(\vp)\equiv V_{\nmcw}^E(\vp)
= \frac{    \frac{A \cq^4}{4}+A \varphi ^4 \left[\log \left(\frac{\varphi }{\cq}\right)-\frac{1}{4}\right]   }{\lt( 1+ \xi \frac{\vp^2-\cq^2}{\mpl^2}\rt)^2}\,.
\ee 
This \nmcw~scenario involves inflation beginning near $\vp=0$ and then the inflaton travels towards $\vp_{\rm min}$, the minimum of the potential of~\cref{Eq:PotCW-Einsteinframe}.
The benchmark values of the above-mentioned inflationary scenario are mentioned in
~\cref{Table:Benchmark-CW} and the $n_s-r$ predictions for the two benchmark values from~\cref{Table:Benchmark-CW} are shown in \cref{Fig:NonminimalCW-ns-r-bound}. The predicted value $r$ for benchmark \mCWo~and \mCWt~fall within $1-\sigma$ and $2-\sigma$ best-fit contour of \Planck2018+\BICEP3 (2022)+\KeckArray2018 data and SO analysis, respectively.

\begin{table}[H]
\centering
\caption{\it Benchmark values for \nmcw~inflationary model in metric formalism (for $\ncmb \approx 60$).}
\label{Table:Benchmark-CW}
\begin{tabular}{|c||c | c | c | c | c | c | c|  } 
 \hline
{\it Benchmark}&$\cq/\mpl$ &$\xi$ &  $\vp_{*}/\mpl$ & $\vp_{\rm end}/\mpl$ &  $n_s$ & $r\times 10^{2}$ & $A$ \\ %
 \hline 
 \hline 
  (m)NM-CW-1  & $52$ & $-0.002$ & $39.0185$ & $ 50.7913$ & $0.964728$ & $1.5708$ & $ \  9.4735\times 10^{-15}$ \\
 \hline
  (m)NM-CW-2 & $ 500$& $-0.003$& $488.912  $& $499.642 $& $0.964318  $& $0.3$ & $ 1.8083 \times 10^{-15} $  \\
 \hline
\end{tabular}
\end{table}
%
%

%
%

\subsection{Stability}
The objective of this section is to determine the upper limit of $\yc$ and $\lO$ in Einstein frame to ensure that the stability of the potential of~\cref{Eq:PotCW-Einsteinframe} is not compromised. 
%
The background-field dependent mass of $H$ and $\chi$ are given by: 
\begin{align}
\widetilde{m}_{\chi}^2 (\phi) = \left( m_\chi + y_\chi \phi \right)^2  \,, 
&& \widetilde{m}_{H}^2 (\phi) = m_H^2 + \lambda_{12} \phi \,.  \label{Eq:Inflaton-dependent-mass-CW}
\end{align}
Next, we consider 
Coleman–Weinberg radiative correction at 1-loop order to the inflaton-potential which is
~\cite{Coleman:1973jx}
\be
V_{\lop}
=\sum_{j
} \frac{n_j}{64\pi^2} (-1)^{2s_j}\widetilde{m}_j^4
\left[ \ln\left( \frac{\widetilde{m}_j^2 
}{\mu^2} \right) - c_j \right]  \,.
\ee
Here, sum is over mainly $j=H$ and $j=\chi$ fields. Moreover, $n_{H,\chi}=4$; $s_H =0$, $s_\chi=1/2$; $c_j=\frac{3}{2}$. Inflaton-dependent mass of $H$ and $\chi$ are already mentioned in~\cref{Eq:Inflaton-dependent-mass-CW}. Then, the first and second derivative of Coleman–Weinberg correction term for $H$ and $\c$ are (with $\mc=m_H=0$)
\eq{
& \lt| V_{{\lop},H}^\prime  
\rt|=\frac{\lO^2 \phi }{8 \pi ^2}  \left(\ln
\left(\frac{\lO \phi }{\mu^2}\right)-1\right)  \, , \label{Eq:stability-H-first-derivative-CW}
%
&&\lt| V_{{\lop},\c}^\prime  
\rt|  = \frac{\Phi ^3 \yc^4 }{4 \pi ^2} \left( 1 - \ln
\left(\frac{\phi ^2 \yc^2}{\mu^2}\right)\right) \, ,\\
%
%
%
%
%
%
& \lt|V_{{\lop},H}^{\prime \prime}
\rt|= \frac{\lO^2 }{8 \pi ^2} \ln
\left(\frac{\lO \phi }{\mu^2}\right) \, .\label{Eq:stability-H-second-derivative-CW}
&& \lt|V_{{\lop},\c}^{\prime \prime} 
\rt|= \frac{1}{8 \pi ^2} \lt(6 \phi ^2 \yc^4  \ln
\left(\frac{\phi ^2 \yc^2}{\mu^2}\right)-2 \phi ^2 \yc^4 \rt)\, .
}
By integrating~\cref{Eq:JordantoEinsteinNew}, and using the resulting function $\phi(\vp)$,~\cref{Eq:stability-H-first-derivative-CW,Eq:stability-H-second-derivative-CW} can be expressed as a function of $\vp$.
However, in \nmcw~Inflation, tree-level potential and its first and second derivative in Einstein frame are 
\eq{
&V_{tree} (\vp)= V_{\nmcw}^E(\vp)= \frac{    \frac{A \cq^4}{4}+A \varphi ^4 \left[\log \left(\frac{\varphi }{\cq}\right)-\frac{1}{4}\right]   }{\lt( 1+ \xi \frac{\vp^2-\cq^2}{\mpl^2}\rt)^2}\,,\\
&V^{\prime}_{tree} (\vp)= \qty(\frac{\td  V_\nmcw^E}{\td \vp} \frac{\td \vp}{\td \phi}) \,,\\
&V^{\prime\prime}_{tree} (\vp)= \frac{\td}{\td \vp}\qty(\frac{\td  V_\nmcw^E}{\td \vp} \frac{\td \vp}{\td \phi}) \frac{\td \vp}{\td \phi}\,.
} 
To obtain the upper limit of $\yc$ and $\lambda_{12}$, we need to satisfy 
\eq{
 \lt| V_{{\lop},H}^\prime  (\vp=\mu)\rt|<V_{\rm tree}^{\prime} (\varphi=\mu) \,,
 \quad
 &&\lt| V_{{\lop},\c}^\prime (\vp=\mu)\rt|< V_{\rm tree}^{\prime} (\varphi=\mu) \,,\\
 \lt|V_{{\lop},H}^{\prime \prime}(\vp=\mu)\rt| <V_{\rm tree}^{\prime\prime} (\varphi=\mu)\,,
 \quad
 &&\lt|V_{{\lop},\c}^{\prime \prime} (\vp=\mu)\rt|<V_{\rm tree}^{\prime\prime} (\varphi=\mu)\,.
}

The permissible upper limit of $\yc$ and $\lO$ for \nmcw~inflationary scenario is listed in~\cref{Table:NM-CW_stability} for $\mu=\vp_*$ and $\m=\vp_{\rm end}$. From this table, we conclude that the permissible upper limit of the couplings $\yc<3.0671\times 10^{-4}$, $\lO/\mpl<2.3304\times 10^{-6}$ for \mCWo, and $\yc<1.1517\times 10^{-4}$, $\lO/\mpl<4.7788 \times 10^{-7}$ for \mCWt.

%
%
\begin{table}[H]
\begin{center}
\caption{\it Allowed range of $\yc$ and $\lO$ for the benchmark values from~\cref{Table:Benchmark-CW}.}
\label{Table:NM-CW_stability}
\vspace{-8pt}
\resizebox{\columnwidth}{!}{
\begin{tabular}{|c|| c| c| c|c|} 
 \hline
 {\it Benchmark} &\multicolumn{2}{|c|}{stability for $\yc$} & \multicolumn{2}{|c|}{stability for $\lO$}\\ [0.5ex] 
  \cline{2-5} 
   & about $\phi_*$ & about $\phi_{\rm end}$ & about $\phi_*$ & about $\phi_{\rm end}$ \\ [0.5ex] 
 \hline\hline
(m)NM-CW-1 & $\yc<3.0956\times 10^{-4}$ & $\yc<3.0671\times 10^{-4}$ & $\lO/\mpl<2.3304\times 10^{-6}$ & $\lO/\mpl<3.3614\times 10^{-6}$ \\ 
 \hline
(m)NM-CW-2 & $\yc< 1.1517\times 10^{-4}$ & $\yc< 2.2136\times 10^{-4}$ & $\lO/\mpl<4.7788 \times 10^{-7}$ & $\lO/\mpl< 2.0868\times 10^{-6}$ \\
 \hline
\end{tabular}
}
\end{center}
\end{table}


\subsection{Reheating and \dm~from decay}
As discussed earlier, the potential of~\cref{Eq:PotCW-Einsteinframe} has a minimum at $\vp=\cq$. 
%
In this work, we assume that $V_{\nmcw}^E(\phi)$ can be approximated around the minimum as $V_{\nmcw}^E(\phi) \sim \phi^2$
 such that during reheating, inflaton oscillates about a quadratic minimum at leading order. 
Then $m_{\phi}$, $\Gamma_\phi$, $\Trh$, and ${\cal H}_I$ for the benchmark values from~\cref{Table:Benchmark-CW} are listed in~\cref{Table:reheating_values-Nonminimal-CW}. 
 The benchmark \mCWt~has a higher value of $m_\phi$ and a smaller value of ${\cal H}_I$
 compared to the benchmark \mCWo. Nevertheless, the values of $m_\phi$, and ${\cal H}_I$ of the two benchmarks are almost identical.

\begin{figure}[H]
    \centering    \includegraphics[width=\wdth]{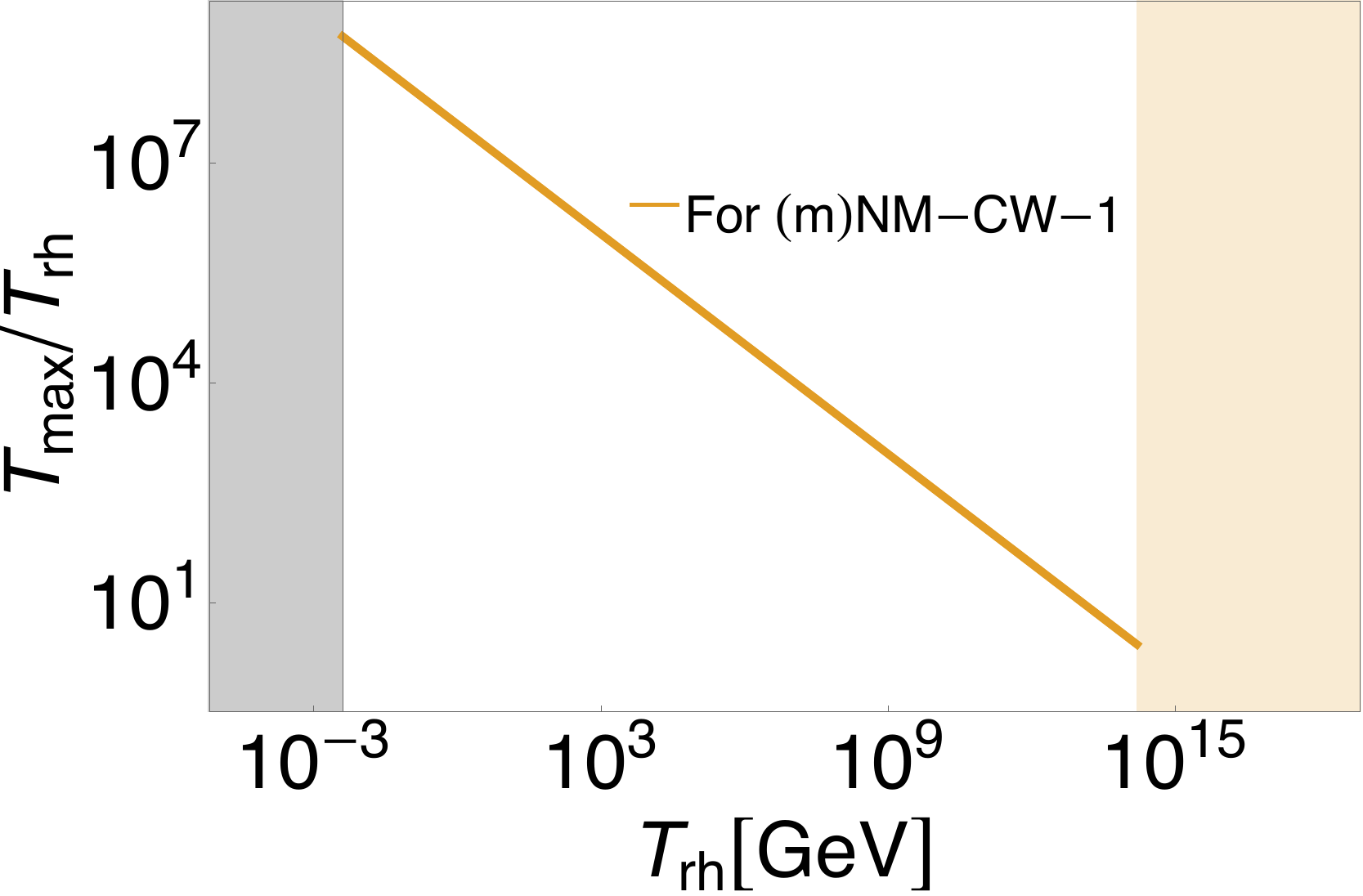}
    \caption{\it \raggedright \label{Fig:Tmax/Trh-Non-minimal-CWOriginal}
    This figure showcases the variation of $T_{max}/\Trh$ against $\Trh$ for benchmark value \mCWo~from~\cref{Table:Benchmark-CW}. 
    The vertical stripe on the left of the plot, highlighted in gray, displays the lower bound on $\Trh$ i.e. $\Trh\gsim 4 \unit{MeV}$. The colorful stripe on the right of the plot shows that there is no permissible value of $\Trh$ based on stability analysis of the corresponding benchmark value from~\cref{Table:NM-CW_stability}.  
    }
\end{figure}

The variation of $\Tmax/\Trh$ vs. $\Trh$ for \nmcw~is illustrated in~\cref{Fig:Tmax/Trh-Non-minimal-CWOriginal}- 
the 
solid line is for \mCWo. 
The gray-colored vertical stripe on the left of the figure shows a limit on $\Trh$ due to the requirement that it must be $\geq 4 \unit{MeV}$.  On the right-hand side of the figure, there is a colored stripe that signifies that those $\Trh$ values are not permitted based on the stability analysis from~\cref{Table:NM-CW_stability}. The maximum value of $\Tmax/\Trh$ is $5.4145 \times 10^{8}$ for \mCWo~(and maximum value of $\Tmax/\Trh \sim 5.0616\times 10^{8}$ for \mCWt~at $\Trh\sim 4 \unit{MeV}$). 
Furthermore, lower bound on $\Trh$ leads to establishing the minimum permissible value of $\lO$ which are $\lO/\mpl\gsim 5.9284\times 10^{-23}$ for \mCWo, and $\lO/\mpl\gsim 6.3255\times 10^{-23}$ for \mCWt.

%
%
%
%
%
%

%
\begin{table}[H]
\centering
\caption{\it $m_\varphi, \G_\varphi, \Trh$, and $\hubble_I$ for the benchmark values from~\cref{Table:Benchmark-CW}.}
\label{Table:reheating_values-Nonminimal-CW}
\begin{tabular}{| c ||  c | c|  c | c|} 
 \hline
{\it Benchmark} & $m_\phi/\mpl$ & $\G_\phi\mpl$ & $\Trh/\lambda_{12}$ & ${\cal H}_I/\mpl$ \\
&   (Eq.~\eqref{Eq:mass-of-inflaton-NonMinimalNatural}) & (Eq.~\eqref{Eq:decay-width-of-inflaton}) & (Eq.~\eqref{Eq:definition of reheating temperature}) & (\cref{Eq:HI-natural})  \\
[0.5ex]
 \hline\hline
 (m)NM-CW-1 &   $9.8092\times 10^{-6}$ & $4056.25 \, \lO^2$  & $28.1132$ & $3.9237\times 10^{-6}$\\ 
 \hline
  (m)NM-CW-2   & $1.1167\times 10^{-5}$ & $3562.97\, \lO^2$  & $ 26.3484$ &  $2.9965\times 10^{-6}$\\ 
 \hline
 %
 %
 %
\end{tabular}
\end{table}

\begin{figure}[H]
    \centering    
   \includegraphics[width=\wdth]{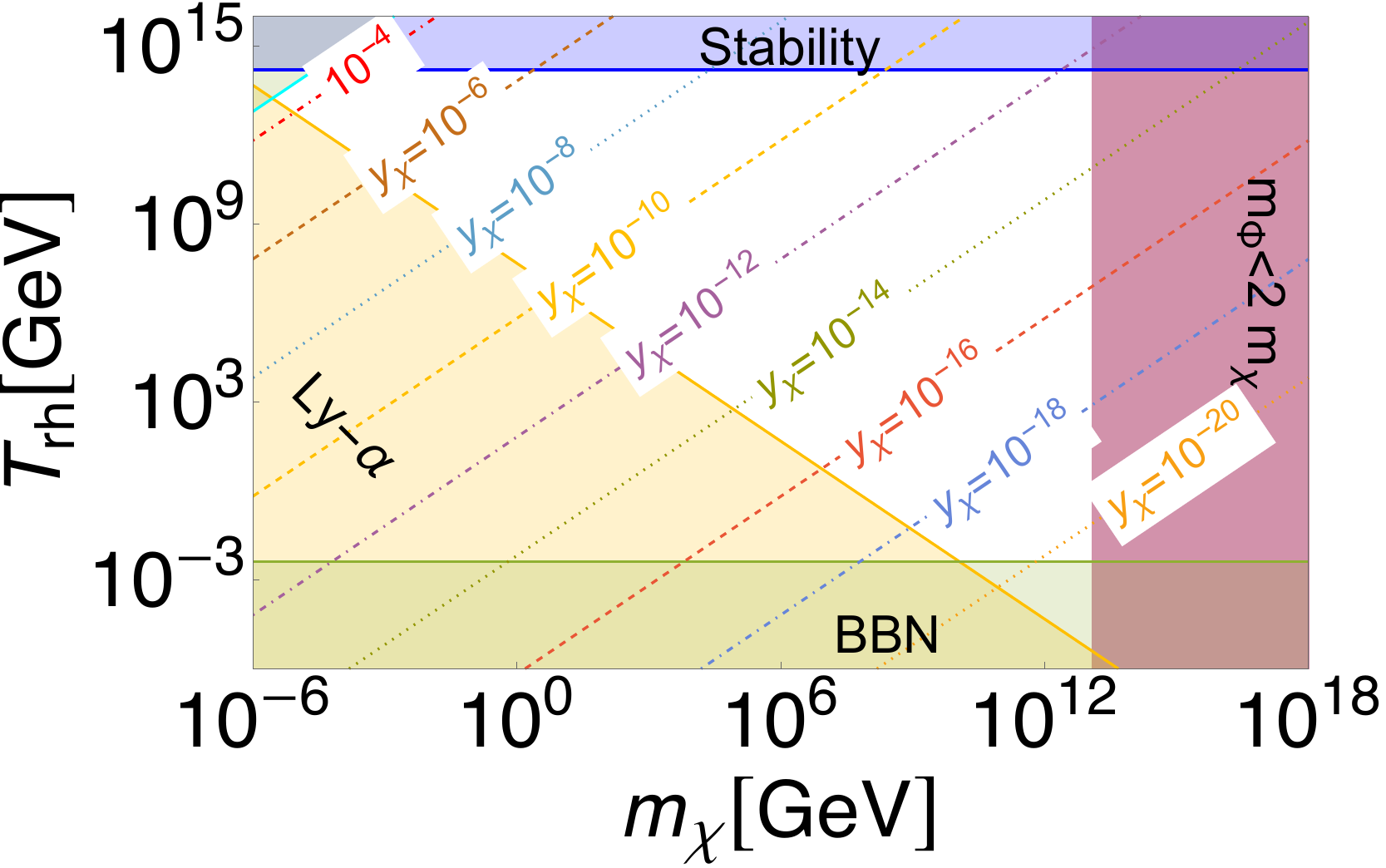}
    \caption{\it \raggedright \label{Fig:CDM_Yield_Decay-NMCW}
    The region without any color on $(\Trh, m_\chi)$ plane, depicted on log-log scale, indicates the allowed region for the benchmark value \mCWo~from~\cref{Table:Benchmark-CW}. The dashed or dashed-dotted lines that intersect the white region corresponds to~\cref{Eq:eq to plot 2} for various values of $y_\chi$ that satisfy present-day \cdm~relic density. {
    The red-colored dotted-dashed line in the top-left corner of the plot corresponds to $\yc=10^{-4}$.}
    The colored regions on this plane represent various bounds – stability analysis: the maximum allowed value of $\Trh$ from the upper limit of $\lO$ from~\cref{Table:NM-CW_stability} (deep blue horizontal stripe at the top) and the maximum allowed value of $\yc$ from~\cref{Table:NM-CW_stability} (cyan-colored wedge-shaped region at the top-left), from Ly-$\alpha$ bound on the mass of \dm~\cref{Eq:Lyman-alpha-bound}: peach-colored region, and $\mc\lsim m_\phi/2$: copper-rose colored vertical stripe on the right. The gray colored stripe at the bottom depicts that $\Trh$ should be $\gsim 4\unit{MeV}$.
    }
\end{figure}

Similarly,~\cref{Fig:CDM_Yield_Decay-NMCW} shows $(\Trh, \mc)$ plane for \nmcw~inflation, particularly for benchmark \mCWo. The discontinuous lines correspond to~\cref{Eq:eq to plot 2}. The largest permissible value of $\lO$ and $\yc$ from stability analysis of~\cref{Table:NM-CW_stability} (orange-colored horizontal stripe at the top of the plot and pink-colored wedge-shaped region at the left-top corner of the plot), BBN temperature: $\Trh\gsim 4 \unit{MeV}$ (gray colored horizontal stripe at the bottom of the plot), and Ly-$\alpha$ bound from~\cref{Eq:Lyman-alpha-bound}: $\Trh\gsim (2 m_\phi)/\mc$ (region shaded with green color) and maximum possible value of $\mc=m_\phi/2$ (vertical stripe at the right, shaded with yellow color). The unshaded region is allowed on $(\Trh, \mc)$ plane and thus if $10^{-4}\gsim\yc\gsim  10^{-20}$ ($2.7743\times 10^{-6}\GeV \lsim \mc \lsim 1.1771\times 10^{13}\GeV$), then $\c$ produced only via the decay of inflaton for the benchmark \mCWo~in \nmcw~inflationary scenario can contribute up to $100\%$ of the entire \cdm~relic density of the present universe. Since $m_\phi$ and ${\cal H}_I$ are almost similar for \mCWo~and \mCWt~(see~\cref{Table:reheating_values-Nonminimal-CW}), Ly$-\alpha$ and $\mc \lsim m_\phi/2$ are almost similar for both benchmarks. However, significant difference in the allowed region on $(\Trh,\mc)$ for the two benchmark values come from the maximum permissible value of $\Trh$ from stability analysis of~\cref{Table:NM-CW_stability}. As a result, the permitted range of $\mc$ for \mCWt~is almost similar - $ 4.5461\times 10^{-6} \GeV \lsim \mc \lsim 1.3401\times 10^{13}\GeV$.

%
%
%

\subsection{DM from scattering}
Similar to GH inflationary scenario, there is an upper limit on $\lO$ from the stability analysis in this case (\nmcw~inflation). Therefore, upper limit on $\Trh$ from stability analysis
\eq{\label{Eq:Nonminimal_CW_max_allowed_Trh}
{\Trh}_{, \text{allowable}}=
\begin{cases}
    1.5724\times 10^{14}\GeV\, \quad \text{for \mCWo}\,, \\
    3.0219\times 10^{13}\GeV\, \quad \text{for \mCWt}\,.
\end{cases}
}
For such maximum possible values of $\Trh$,   $\Tmax/{\Trh}_{, \text{allowable}}>1$ is maintained.
Now, for $\Trh\sim {\Trh}_{, \text{allowable}} $, and the condition $Y_{IS,0}\sim Y_{{\rm CDM},0}$ (from~\cref{Eq:yield-DM-scattering-inflaton-graviton,Eq:eq to plot 2}), the value of $\mc$ is
\eq{\label{Eq:NMCW-large mc is needed-InflatonScattering}
{\mc}^{IS,0}\sim
\begin{cases}
    3.3884 \times 10^{10}\GeV \, \quad \text{for \mCWo}\,,\\
    6.9634\times 10^{10}\GeV \, \quad \text{for \mCWt}\,.
\end{cases}
}
Therefore, for \nmcw, the required condition  to make $Y_{IS,0}\sim Y_{{\rm CDM},0}$ is $\mc\gsim {\mc}^{IS,0}$ for $\Trh\lsim {\Trh}_{, \text{allowable}}$.

In the scenario where \dm~particles are produced through the scattering process involving SM particles and the mediator being a graviton (with $\mc\ll\Trh$) (\cref{Eq:Ysmg0}), assuming $\Trh\sim {\Trh}_{, \text{allowable}}$, we obtain
\eq{
Y_{\sm g,0}=
\begin{cases}
    6.4357 \times 10^{-19} \, \quad \text{(for \mCWo)}\,,\\
    4.5687 \times 10^{-21} \GeV \, \quad \text{(for \mCWt)}\,.
\end{cases}
}
And thus to make $Y_{\sm g,0}\sim Y_{{\rm CDM},0}$ (from~\cref{Eq:present day CDM yield}), we get
\eq{\label{Eq:NMCW-large mc is needed}
\mc \sim 
\begin{cases}
    6.6815 \times 10^{8} \GeV \, \quad \text{(for \mCWo)}\,,\\
    9.4119 \times 10^{10} \GeV \, \quad \text{(for \mCWt)}\,.
\end{cases}
}

However, it appears that the conclusion for \nmcw~is similar to GH inflation for \three~following~\cref{Eq:TERM}. This is because $\Tmax/\Trh\sim 30$ for $\Trh \sim 10^{12}\GeV$ and $\Tmax/{\Trh}_{, \text{allowable}}\sim 3$ for \mCWo~and $\Tmax/{\Trh}_{, \text{allowable}}\sim 10$ for \mCWt~(see~\cref{Eq:Nonminimal_CW_max_allowed_Trh}).

For the production of $\c$ via  2-to-2 scattering of SM particles
with inflaton as mediator (along with the constraint $\Trh\ll m_\phi$), the condition that $Y_{\sm i,0}$ contributes $f_\c$ fraction of total \cdm~relic density (\cref{Eq:NMN-SMScatteringInflatonMediator})
%
along with the maximum permissible values of $\yc$ (from~\cref{Table:NM-CW_stability})
\eq{\label{Eq:CW-con}
{\mc}^{\sm i,0} =
\begin{cases}
    3.4749\, \frac{f_\c}{f_\phi^3}  \GeV \, \quad \text{(for \mCWo)}\,,\\
    24.6447\, \frac{f_\c}{f_\phi^3}  \GeV \, \quad \text{(for \mCWt)}\,.
\end{cases}
}
Therefore, just like in GH inflationary scenario, if $f_\c /f_\phi^3 \sim 1$, then $\mc\sim {\cal O}(10)\GeV$, while for $f_\c /f_\phi^3 \sim 10^8$, $\mc\sim {\cal O}(10^8 -10^9)\GeV$ is required.




\section{Discussion and Conclusion }
\label{Sec:Conclusion}

In this work, we studied single-field slow roll inflationary model with non-minimal coupling between the inflaton and the gravity sector. Two forms of non-minimal coupling - $\xi \vp^2 {\cal R}$ and $ \alpha\lt( 1+ \cos\lt(\frac{\vp}{f_a}\rt)\rt)$ were factored in for non-minimal natural inflation. Additionally, for non-minimal Coleman–Weinberg inflation, we assumed the form of the non-minimal coupling is $\xi \vp^2 {\cal R}$.
For those inflationary scenarios, considering CMB bounds, we explored the parameter space of $\yc$, $\mc$, and $\lO$, for the production of a fermionic \dm~particle $\c$ during post-inflationary reheating era, as a potential candidate for the \cdm~of the present universe. 
Key findings of this investigation are summarized below:
\begin{itemize}

    \item We found that \nmpn~inflationary scenario requires a slightly larger value for $f_a/\mpl$ ($f_a/\mpl\sim {\cal O}(10)$) compared to \nmn~inflationary scenario ($f_a/\mpl\sim {\cal O}(1)$) to satisfy bounds from \cmb~measurements
    (see~\cref{Table:Benchmark-NonminimalNatural,Table:Benchmark-NI-periodic}).  However, 
    both models predict values of $r \sim 1.7231 \times 10^{-2}$ for \nmn, and $r \sim 9.7184\times 10^{-3}$ for \nmpn~
    which fall within $1-\sigma$ contour of joint analysis of \Planck2018+\BICEP3 (2022)+\KeckArray2018 but for $\ncmb> 60$.  These values of $r$ can be verified in the future  by upcoming \cmb~observations, e.g. SO. 
    For instance, predicted value of $r\sim 0.003$ for benchmark `(m)NM-CW-2' (see~\cref{Table:Benchmark-CW,Fig:NonminimalCW-ns-r-bound}) can be validated in the future at $1-\sigma$ CL by SO.
    
   \item From~\cref{Table:reheating_values-NonminimalNatural,Table:reheating_values-NonminimalNatural-periodic,Table:reheating_values-Nonminimal-CW}, we conclude that the value of $m_\phi$ varies depending on benchmarks for three inflationary scenarios, being highest ($\sim 4.0987\times 10^{13}\GeV$) for \mNo~and lowest for \mNPr~($\sim 9.4349\times 10^{12}\GeV$), consequently affecting $\Trh$, Ly-$\alpha$ bound and maximum possible value of $\mc$. However, unlike \nmn~and NM-CW inflationary scenarios, the value of $\hubble_I$ remains nearly the same for various benchmark values ($f_a/\mpl$ was varied from $6.3$ to $164.0$) for \nmpn~inflation due to the vanishing of non-minimal coupling near the minimum.


   \item 
   Variation of $\Tmax/\Trh$ against $\Trh$ for different benchmarks for \nmn, \nmpn, and NM-CW~inflation are shown in~\cref{Fig:Tmax/Trh-Non-minimal_natural,Fig:Tmax/Trh-Non-minimal-periodic_natural,Fig:Tmax/Trh-Non-minimal-CWOriginal}, respectively. $\Tmax/\Trh$ is the highest $(\sim {\cal O}\qty(10^8))$ at $\Trh\sim4\unit{MeV}$, e.g. $\Tmax/\Trh \sim 5.88\times 10^8$ for \mNo,  $\Tmax/\Trh \sim 5.14\times 10^8$ for \mNPo, and $\Tmax/\Trh \sim 5.41\times 10^8$ for NM-CW-1. Hence, inflationary parameters, including $\L$, $f_a$, $\cq$, $A$, and $r$, control $\Tmax/\Trh$. 

   \item 
   From~\cref{Fig:CDM_Yield_Decay-NonminimalNatural,Fig:CDM_Yield_Decay-NonminimalPeriodicNatural,Fig:CDM_Yield_Decay-NMCW} 
   we conclude that $\c$ produced only through inflaton decay can contribute $100\%$ of the total relic density of \cdm~of the present universe if 
   the allowed range of $\mc$ is ${\cal O}(10^{-6}\GeV)\lsim \mc \lsim m_\phi/2$ for \nmn, \nmpn,~and NM-CW inflationary scenarios; and of $\yc$ is ${\cal O}\qty( 10^{-1})\gsim \yc\gsim {\cal O}\qty(10^{-20})$ for \nmn, \nmpn, and ${\cal O}\qty(10^{-4})\gsim \yc\gsim {\cal O}\qty(10^{-20})$ for NM-CW. The exact range of $\yc$ and $\mc$, however, varies depending on our assumed maximum value of $\Trh$ (for NM-N and NMP-N, and maximum permissible value of $\lO$ and $\yc$ from stability analysis for NM-CW), and benchmark values (for example, see~\cref{Fig:CDM_Yield_Decay-NonminimalNatural}), and thus for different values of inflationary parameters and also for different inflationary scenarios. 

   \item If we assume that $\c$ is produced only via scattering of non-relativistic inflaton with graviton as the mediator or via scattering of \sm~particles with graviton as a mediator with $\mc\ll \Trh$, then the \dm~produced through those scattering channels separately for all chosen benchmark values and inflationary scenarios can make $100\%$ of the total \cdm~yield of the present universe, provided $\mc\gsim {\cal O}(10^{10}) \GeV$ (see~\cref{Eq:NMN-large mc is needed-InflatonScattering,Eq:NMPN-large mc is needed-InflatonScattering,Eq:NMCW-large mc is needed-InflatonScattering}) and $\mc \gsim {\cal O}(10^{9}) \GeV$ (see~\cref{Eq:NMN-large mc is needed,Eq:NMPN-large mc is needed,Eq:NMCW-large mc is needed}), respectively. However, the exact value of $\mc$ varies depending on the benchmark as well as the inflationary scenario being considered. However, scattering channel of \sm~particles with graviton as mediator with $\Tmax \gg \mc \gg \Trh$ seems less effective to produce the total \cdm~relic density. Furthermore, for scattering of \sm~particles with inflaton as mediator $m_\chi\sim {\cal O}(10^9)\GeV$ (\cref{Eq:NMN-SMScatteringInflatonMediator,Eq:NMPN-SMScatteringInflatonMediator,Eq:CW-con}), for example, is required for $\Trh\sim 10^{-2} m_\phi$ and $\yc\sim {\cal O}\qty(10^{-5})$
   to contribute completely to the total \cdm~density. 
   


\end{itemize}

{We found the two classes of via particle theory motivated inflationary potentials involving Higgs-like weakly coupled scalar and ALP scalar present in strongly coupled QCD-like theories not only may provide a wide range of CM predictions but also lead to DM formation from inflaton during reheating in suitable regions of the parameter space.}

 In this paper, just by extending the standard model with two degrees of freedom: an axion-like particle (inflaton), and a fermionic DM,  we demonstrated how to obtain the tiny temperature fluctuations as seen in the \cmb~as well as address the \dm~puzzle of the universe in terms of its possible particle origin. 
We showed how future measurements of the CMB from experiments like \cmbsfour, LiteBIRD, and SO~\cite{CMB-S4:2020lpa,LiteBIRD:2022cnt, SimonsObservatory:2018koc} and other such experiments \cite{Hazumi:2019lys,Adak:2021lbu,SPT:2019nip,POLARBEAR:2015ixw,ACT:2020gnv,Harrington:2016jrz,LSPE:2020uos, Mennella:2019cwk,SPIDER:2021ncy}
 will further be able to verify the simple models we have presented if BB-modes are detected and the scale of inflation is measured.  The presence of interaction with inflaton and DM species may lead to non-Gaussianities and may serve as very interesting probes of these models and complementary signatures of non-thermal production of DM. However such a study is beyond the scope of the present draft and will be taken up in future publications.
\medskip

\section*{Acknowledgment}
The authors appreciate the insightful exchanges with Qaisar Shafi.
Work of S.P. is funded by RSF Grant 19-42-02004. The work of M.K. was performed with the financial support provided by the Russian Ministry of Science and Higher Education, project “Fundamental and applied research of cosmic rays”, No. FSWU-2023-0068. Z.L. has been supported by the Polish National Science Center grant 2017/27/B/ ST2/02531.

\begin{figure}[H]
    \centering    
   \includegraphics[width=0.45\linewidth]{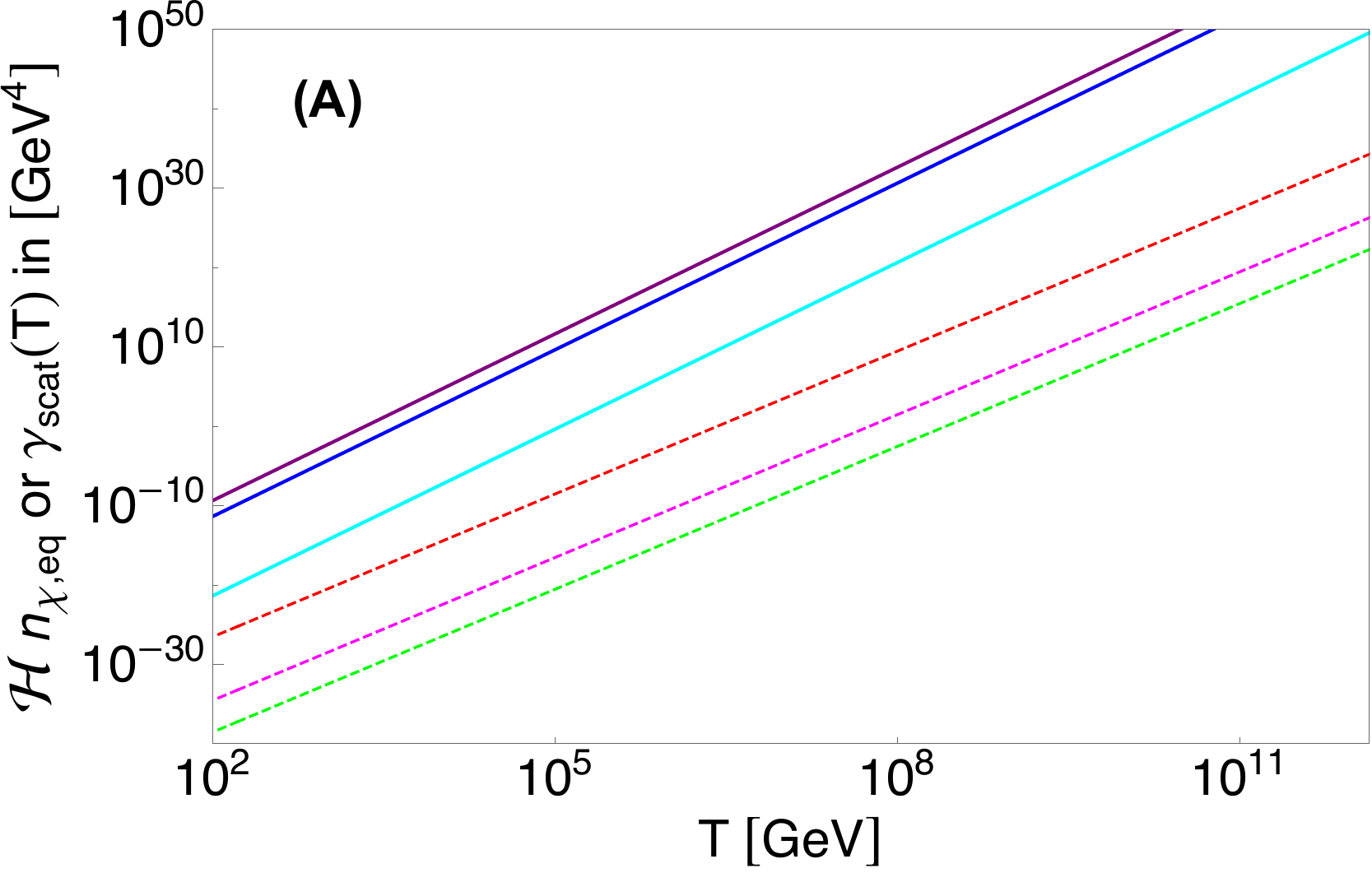}\, \qquad
   \includegraphics[width=0.45\linewidth]{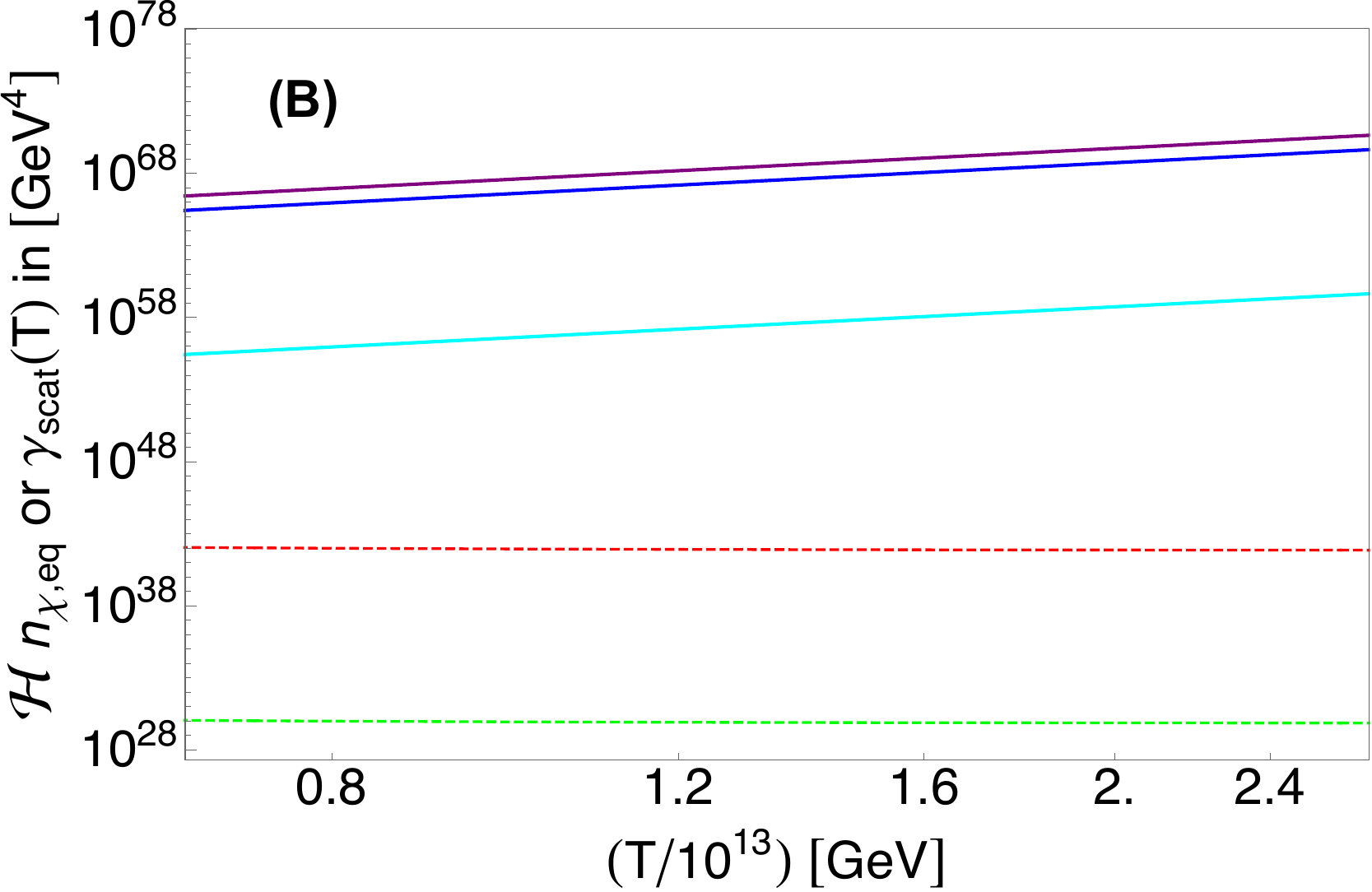}\\
   \vspace{0.6cm}
   \includegraphics[width=0.45\linewidth]{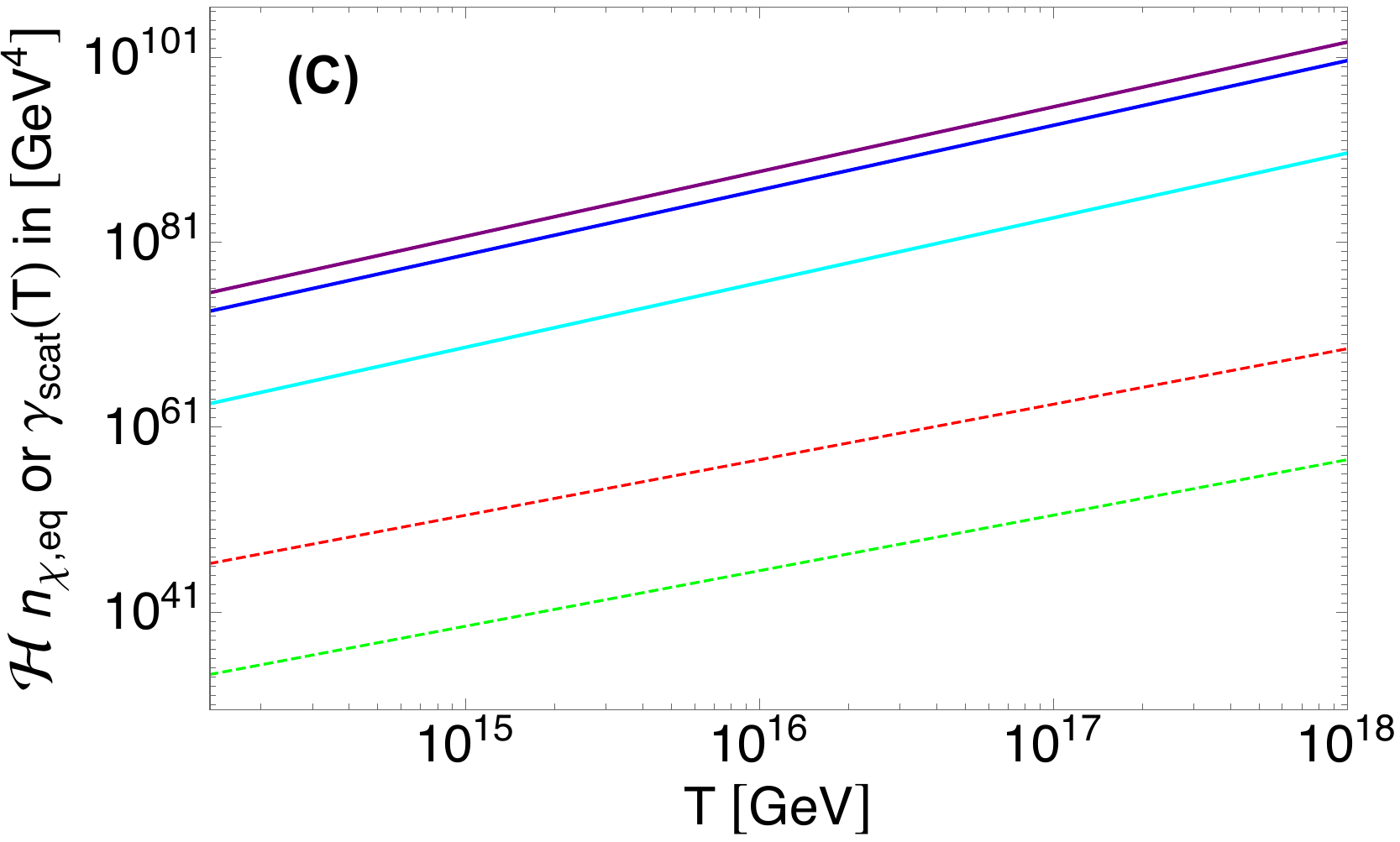}
    \caption{\it \raggedright \label{Fig:thermalization-condition}
    {
    Exploring the the range of $\yc$ and $\lO$ for which $\gamma_{\rm scat}<\hubble n_{\chi,{\rm eq}}$ from~\cref{eq:Riccati type form-scattering source} is maintained such that DM particles can not attain thermal equilibrium with relativistic SM Higgs via inflaton exchange 2-to-2 scattering. Here solid lines are for $\hubble n_\chi$ as a function of $T$: blue-colored curve for $\Trh\sim 100\GeV, \mc\sim 10^3\GeV$, cyan-colored curve for $\Trh\sim 10^7\GeV, \mc\sim 10^3\GeV$, and purple-colored curve for $\Trh\sim 100\GeV, \mc\sim 10\GeV$. Dashed lines are for $\gamma_{\rm scat} (T)$: red-colored curve for $\yc\sim 10^{-4},\lO/\mpl\sim 10^{-6}$, green-colored curve for $\yc\sim 10^{-10},\lO/\mpl\sim 10^{-6}$, and cyan-colored curve for $\yc\sim 10^{-4},\lO/\mpl\sim 10^{-10}$. Top-left panel (labeled as (A)), top-right panel (labeled as (B)), and bottom panel (labeled as (C)) are for `scenario-I',`scenario-II', and `scenario-III', respectively.  
    }
    }
\end{figure}

\appendix
{
\section{Conditions preventing DM–SM thermal equilibrium via inflaton-mediated 2-to-2 scattering}

Here, we explore the conditions, such as the ranges of the couplings $\yc$ and $\lO$ mentioned in~\cref{Eq:reheating lagrangian}, under which $\chi$ particles fail to reach local thermal equilibrium with relativistic SM Higgs during the reheating epoch. During this epoch, inflaton-mediated 2-to-2 scattering between SM particles and DM is inevitable (see~\Ccite{Adshead:2016xxj,Adshead:2019uwj}). The time evolution of number density $\chi$ particles is then governed by
\begin{equation}\label{eq:Riccati type form-scattering source}
    \frac{\dd n_\chi}{\dd t} + 3 \hubble\, n_\chi = \gamma(T) +\gamma_{\rm scat}(T)\,,
\end{equation}
where $\gamma(T)\equiv\gamma$ is defined in ~\Cref{Eq:Boltzaman equation for comoving number density}, and $\gamma_{\rm scat}(T)\equiv \gamma_{\rm scat}$, the production rate of $\chi$ particles from inflaton mediated scattering, is~\cite{Bernal:2021qrl}
\begin{align}\label{eq:gamma-scat}
   \gamma_{\rm scat} (T)  \approx \frac{\yc^2 \lO^2}{64\pi^4} \times
   \begin{cases}
    \frac{32}{\pi} \frac{T^6}{m_\phi^4} \qquad &\text{for }T\ll m_\phi \, \; \text{(scenario-I)},\\
    \frac{m_\phi^2 T}{\Gamma_\phi} K_1 \lt(\frac{m_\phi}{T}\rt) \qquad &\text{for }T\sim m_\phi \, \; \text{(scenario-II)},\\
    \frac{m_\phi T^2}{\Gamma_\phi}\qquad &\text{for }T\gg m_\phi \, \; \text{(scenario-III)}.
   \end{cases}
\end{align}
where $\Gamma_\phi$ is defined in~\cref{Eq:decay-width-of-inflaton}, and $K_1$ is modified Bessel function of order $1$. For `scenario-II' and `scenario-III', $\gamma_{\rm scat}$ is independent of $\lO$, given the assumption used to define $\Gamma_\phi$ in~\cref{Eq:decay-width-of-inflaton}. 
If $\gamma_{\rm scat}\ll H n_\chi$, then thermal equilibrium between DM and SM Higgs cannot be established via inflaton-mediated scattering.~\Cref{Fig:thermalization-condition} exhibits different values of $\yc$ and $\lO$ maintaining $\gamma_{\rm scat}\ll H n_\chi$ for the benchmark (m)NM-N-2 (\Cref{Table:Benchmark-NonminimalNatural}) with $m_\phi\sim {\cal O}\lt(10^{13}\GeV\rt)$. For this estimation, we assume temperature of $\chi$ particles $\sim {\cal O}\lt(T\rt)$, and $n_\chi\sim T^3/m_\chi$, and $\hubble$ during reheating is given by~\Cref{Eq:Hubble parameter during reheating+Eq:rho_phi}. 
Figure at top-left panel of~\Cref{Fig:thermalization-condition} (labeled as (A)) is for `scenario-I' of~\cref{eq:gamma-scat}, and it shows that for for $10\GeV\lsim m_\chi\lsim 100\GeV$, and $10^2 \GeV \lsim \Trh \lsim 10^7\GeV$\footnote{Reheating scenarios with $\Trh\gsim {\cal O}\lt(10^7\rt)\GeV$ may results in an excessive production of gravitino.}, thermal equilibrium is not possible for $100 \lsim T \lsim m_\phi/10$, even with $(\yc \lsim 10^{-4}, \lO\lsim 10^{-6} \mpl)$. Same conclusion about the range of $\yc$ and $\lO$ can  be drawn for `scenario-II' and `scenario-III' from top-right panel and bottom panel of~\Cref{Fig:thermalization-condition}(labeled as (B), and (C), respectively). In~\Cref{Fig:thermalization-condition}.(B) and \Cref{Fig:thermalization-condition}.(C) $T$ is varied in the range: $T$ $m_\phi/10\lsim T\lsim 10 m_\phi$ and $10 m_\phi\lsim T\lsim 10^{18}\GeV$. Similar analysis for other benchmarks from~\cref{Table:Benchmark-NonminimalNatural,Table:Benchmark-NI-periodic,Table:Benchmark-CW} gives the same conclusion as $m_\phi\sim {\cal O}\lt(10^{13}\GeV\rt)$ for our all chosen benchmarks.


}

%
%
%


\bibliographystyle{apsrev4-1}
%


\begin{thebibliography}{125}%
\makeatletter
\providecommand \@ifxundefined [1]{%
 \@ifx{#1\undefined}
}%
\providecommand \@ifnum [1]{%
 \ifnum #1\expandafter \@firstoftwo
 \else \expandafter \@secondoftwo
 \fi
}%
\providecommand \@ifx [1]{%
 \ifx #1\expandafter \@firstoftwo
 \else \expandafter \@secondoftwo
 \fi
}%
\providecommand \natexlab [1]{#1}%
\providecommand \enquote  [1]{``#1''}%
\providecommand \bibnamefont  [1]{#1}%
\providecommand \bibfnamefont [1]{#1}%
\providecommand \citenamefont [1]{#1}%
\providecommand \href@noop [0]{\@secondoftwo}%
\providecommand \href [0]{\begingroup \@sanitize@url \@href}%
\providecommand \@href[1]{\@@startlink{#1}\@@href}%
\providecommand \@@href[1]{\endgroup#1\@@endlink}%
\providecommand \@sanitize@url [0]{\catcode `\\12\catcode `\$12\catcode `\&12\catcode `\#12\catcode `\^12\catcode `\_12\catcode `\%12\relax}%
\providecommand \@@startlink[1]{}%
\providecommand \@@endlink[0]{}%
\providecommand \url  [0]{\begingroup\@sanitize@url \@url }%
\providecommand \@url [1]{\endgroup\@href {#1}{\urlprefix }}%
\providecommand \urlprefix  [0]{URL }%
\providecommand \Eprint [0]{\href }%
\providecommand \doibase [0]{http://dx.doi.org/}%
\providecommand \selectlanguage [0]{\@gobble}%
\providecommand \bibinfo  [0]{\@secondoftwo}%
\providecommand \bibfield  [0]{\@secondoftwo}%
\providecommand \translation [1]{[#1]}%
\providecommand \BibitemOpen [0]{}%
\providecommand \bibitemStop [0]{}%
\providecommand \bibitemNoStop [0]{.\EOS\space}%
\providecommand \EOS [0]{\spacefactor3000\relax}%
\providecommand \BibitemShut  [1]{\csname bibitem#1\endcsname}%
\let\auto@bib@innerbib\@empty
\bibitem [{\citenamefont {Starobinsky}(1980)}]{Starobinsky:1980te}%
  \BibitemOpen
  \bibfield  {author} {\bibinfo {author} {\bibfnamefont {A.~A.}\ \bibnamefont {Starobinsky}},\ }\href {\doibase 10.1016/0370-2693(80)90670-X} {\bibfield  {journal} {\bibinfo  {journal} {Phys. Lett. B}\ }\textbf {\bibinfo {volume} {91}},\ \bibinfo {pages} {99} (\bibinfo {year} {1980})}\BibitemShut {NoStop}%
\bibitem [{\citenamefont {Guth}(1981)}]{Guth:1980zm}%
  \BibitemOpen
  \bibfield  {author} {\bibinfo {author} {\bibfnamefont {A.~H.}\ \bibnamefont {Guth}},\ }\href {\doibase 10.1103/PhysRevD.23.347} {\bibfield  {journal} {\bibinfo  {journal} {Phys. Rev. D}\ }\textbf {\bibinfo {volume} {23}},\ \bibinfo {pages} {347} (\bibinfo {year} {1981})}\BibitemShut {NoStop}%
\bibitem [{\citenamefont {Linde}(1982)}]{Linde:1981mu}%
  \BibitemOpen
  \bibfield  {author} {\bibinfo {author} {\bibfnamefont {A.~D.}\ \bibnamefont {Linde}},\ }\href {\doibase 10.1016/0370-2693(82)91219-9} {\bibfield  {journal} {\bibinfo  {journal} {Phys. Lett. B}\ }\textbf {\bibinfo {volume} {108}},\ \bibinfo {pages} {389} (\bibinfo {year} {1982})}\BibitemShut {NoStop}%
\bibitem [{\citenamefont {Albrecht}\ and\ \citenamefont {Steinhardt}(1982)}]{Albrecht:1982wi}%
  \BibitemOpen
  \bibfield  {author} {\bibinfo {author} {\bibfnamefont {A.}~\bibnamefont {Albrecht}}\ and\ \bibinfo {author} {\bibfnamefont {P.~J.}\ \bibnamefont {Steinhardt}},\ }\href {\doibase 10.1103/PhysRevLett.48.1220} {\bibfield  {journal} {\bibinfo  {journal} {Phys. Rev. Lett.}\ }\textbf {\bibinfo {volume} {48}},\ \bibinfo {pages} {1220} (\bibinfo {year} {1982})}\BibitemShut {NoStop}%
\bibitem [{\citenamefont {Akrami}\ \emph {et~al.}(2020)\citenamefont {Akrami} \emph {et~al.}}]{Planck:2018jri}%
  \BibitemOpen
  \bibfield  {author} {\bibinfo {author} {\bibfnamefont {Y.}~\bibnamefont {Akrami}} \emph {et~al.} (\bibinfo {collaboration} {Planck}),\ }\href {\doibase 10.1051/0004-6361/201833887} {\bibfield  {journal} {\bibinfo  {journal} {Astron. Astrophys.}\ }\textbf {\bibinfo {volume} {641}},\ \bibinfo {pages} {A10} (\bibinfo {year} {2020})},\ \Eprint {http://arxiv.org/abs/1807.06211} {arXiv:1807.06211 [astro-ph.CO]} \BibitemShut {NoStop}%
\bibitem [{\citenamefont {Ade}\ \emph {et~al.}(2018)\citenamefont {Ade} \emph {et~al.}}]{BICEP2:2018kqh}%
  \BibitemOpen
  \bibfield  {author} {\bibinfo {author} {\bibfnamefont {P.~A.~R.}\ \bibnamefont {Ade}} \emph {et~al.} (\bibinfo {collaboration} {BICEP2, Keck Array}),\ }\href {\doibase 10.1103/PhysRevLett.121.221301} {\bibfield  {journal} {\bibinfo  {journal} {Phys. Rev. Lett.}\ }\textbf {\bibinfo {volume} {121}},\ \bibinfo {pages} {221301} (\bibinfo {year} {2018})},\ \Eprint {http://arxiv.org/abs/1810.05216} {arXiv:1810.05216 [astro-ph.CO]} \BibitemShut {NoStop}%
\bibitem [{\citenamefont {Folkerts}\ \emph {et~al.}(2014)\citenamefont {Folkerts}, \citenamefont {Germani},\ and\ \citenamefont {Redondo}}]{Folkerts:2013tua}%
  \BibitemOpen
  \bibfield  {author} {\bibinfo {author} {\bibfnamefont {S.}~\bibnamefont {Folkerts}}, \bibinfo {author} {\bibfnamefont {C.}~\bibnamefont {Germani}}, \ and\ \bibinfo {author} {\bibfnamefont {J.}~\bibnamefont {Redondo}},\ }\href {\doibase 10.1016/j.physletb.2013.12.026} {\bibfield  {journal} {\bibinfo  {journal} {Phys. Lett. B}\ }\textbf {\bibinfo {volume} {728}},\ \bibinfo {pages} {532} (\bibinfo {year} {2014})},\ \Eprint {http://arxiv.org/abs/1304.7270} {arXiv:1304.7270 [hep-ph]} \BibitemShut {NoStop}%
\bibitem [{\citenamefont {Martin}\ \emph {et~al.}(2014)\citenamefont {Martin}, \citenamefont {Ringeval},\ and\ \citenamefont {Vennin}}]{Martin:2013tda}%
  \BibitemOpen
  \bibfield  {author} {\bibinfo {author} {\bibfnamefont {J.}~\bibnamefont {Martin}}, \bibinfo {author} {\bibfnamefont {C.}~\bibnamefont {Ringeval}}, \ and\ \bibinfo {author} {\bibfnamefont {V.}~\bibnamefont {Vennin}},\ }\href {\doibase 10.1016/j.dark.2014.01.003} {\bibfield  {journal} {\bibinfo  {journal} {Phys. Dark Univ.}\ }\textbf {\bibinfo {volume} {5-6}},\ \bibinfo {pages} {75} (\bibinfo {year} {2014})},\ \Eprint {http://arxiv.org/abs/1303.3787} {arXiv:1303.3787 [astro-ph.CO]} \BibitemShut {NoStop}%
\bibitem [{\citenamefont {Hertzberg}(2010)}]{Hertzberg:2010dc}%
  \BibitemOpen
  \bibfield  {author} {\bibinfo {author} {\bibfnamefont {M.~P.}\ \bibnamefont {Hertzberg}},\ }\href {\doibase 10.1007/JHEP11(2010)023} {\bibfield  {journal} {\bibinfo  {journal} {JHEP}\ }\textbf {\bibinfo {volume} {11}},\ \bibinfo {pages} {023} (\bibinfo {year} {2010})},\ \Eprint {http://arxiv.org/abs/1002.2995} {arXiv:1002.2995 [hep-ph]} \BibitemShut {NoStop}%
\bibitem [{\citenamefont {Bertone}\ \emph {et~al.}(2005)\citenamefont {Bertone}, \citenamefont {Hooper},\ and\ \citenamefont {Silk}}]{Bertone:2004pz}%
  \BibitemOpen
  \bibfield  {author} {\bibinfo {author} {\bibfnamefont {G.}~\bibnamefont {Bertone}}, \bibinfo {author} {\bibfnamefont {D.}~\bibnamefont {Hooper}}, \ and\ \bibinfo {author} {\bibfnamefont {J.}~\bibnamefont {Silk}},\ }\href {\doibase 10.1016/j.physrep.2004.08.031} {\bibfield  {journal} {\bibinfo  {journal} {Phys. Rept.}\ }\textbf {\bibinfo {volume} {405}},\ \bibinfo {pages} {279} (\bibinfo {year} {2005})},\ \Eprint {http://arxiv.org/abs/hep-ph/0404175} {arXiv:hep-ph/0404175} \BibitemShut {NoStop}%
\bibitem [{\citenamefont {Bernal}\ \emph {et~al.}(2018{\natexlab{a}})\citenamefont {Bernal}, \citenamefont {Chatterjee},\ and\ \citenamefont {Paul}}]{Bernal:2018hjm}%
  \BibitemOpen
  \bibfield  {author} {\bibinfo {author} {\bibfnamefont {N.}~\bibnamefont {Bernal}}, \bibinfo {author} {\bibfnamefont {A.}~\bibnamefont {Chatterjee}}, \ and\ \bibinfo {author} {\bibfnamefont {A.}~\bibnamefont {Paul}},\ }\href {\doibase 10.1088/1475-7516/2018/12/020} {\bibfield  {journal} {\bibinfo  {journal} {JCAP}\ }\textbf {\bibinfo {volume} {12}},\ \bibinfo {pages} {020} (\bibinfo {year} {2018}{\natexlab{a}})},\ \Eprint {http://arxiv.org/abs/1809.02338} {arXiv:1809.02338 [hep-ph]} \BibitemShut {NoStop}%
\bibitem [{\citenamefont {Choudhury}\ \emph {et~al.}(2023{\natexlab{a}})\citenamefont {Choudhury}, \citenamefont {Panda},\ and\ \citenamefont {Sami}}]{Choudhury:2023hvf}%
  \BibitemOpen
  \bibfield  {author} {\bibinfo {author} {\bibfnamefont {S.}~\bibnamefont {Choudhury}}, \bibinfo {author} {\bibfnamefont {S.}~\bibnamefont {Panda}}, \ and\ \bibinfo {author} {\bibfnamefont {M.}~\bibnamefont {Sami}},\ }\href {\doibase 10.1088/1475-7516/2023/08/078} {\bibfield  {journal} {\bibinfo  {journal} {JCAP}\ }\textbf {\bibinfo {volume} {08}},\ \bibinfo {pages} {078} (\bibinfo {year} {2023}{\natexlab{a}})},\ \Eprint {http://arxiv.org/abs/2304.04065} {arXiv:2304.04065 [astro-ph.CO]} \BibitemShut {NoStop}%
\bibitem [{\citenamefont {Choudhury}\ \emph {et~al.}(2023{\natexlab{b}})\citenamefont {Choudhury}, \citenamefont {Panda},\ and\ \citenamefont {Sami}}]{Choudhury:2023rks}%
  \BibitemOpen
  \bibfield  {author} {\bibinfo {author} {\bibfnamefont {S.}~\bibnamefont {Choudhury}}, \bibinfo {author} {\bibfnamefont {S.}~\bibnamefont {Panda}}, \ and\ \bibinfo {author} {\bibfnamefont {M.}~\bibnamefont {Sami}},\ }\href {\doibase 10.1088/1475-7516/2023/11/066} {\bibfield  {journal} {\bibinfo  {journal} {JCAP}\ }\textbf {\bibinfo {volume} {11}},\ \bibinfo {pages} {066} (\bibinfo {year} {2023}{\natexlab{b}})},\ \Eprint {http://arxiv.org/abs/2303.06066} {arXiv:2303.06066 [astro-ph.CO]} \BibitemShut {NoStop}%
\bibitem [{\citenamefont {Choudhury}\ \emph {et~al.}(2023{\natexlab{c}})\citenamefont {Choudhury}, \citenamefont {Panda},\ and\ \citenamefont {Sami}}]{Choudhury:2023jlt}%
  \BibitemOpen
  \bibfield  {author} {\bibinfo {author} {\bibfnamefont {S.}~\bibnamefont {Choudhury}}, \bibinfo {author} {\bibfnamefont {S.}~\bibnamefont {Panda}}, \ and\ \bibinfo {author} {\bibfnamefont {M.}~\bibnamefont {Sami}},\ }\href {\doibase 10.1016/j.physletb.2023.138123} {\bibfield  {journal} {\bibinfo  {journal} {Phys. Lett. B}\ }\textbf {\bibinfo {volume} {845}},\ \bibinfo {pages} {138123} (\bibinfo {year} {2023}{\natexlab{c}})},\ \Eprint {http://arxiv.org/abs/2302.05655} {arXiv:2302.05655 [astro-ph.CO]} \BibitemShut {NoStop}%
\bibitem [{\citenamefont {Choudhury}\ \emph {et~al.}(2023{\natexlab{d}})\citenamefont {Choudhury}, \citenamefont {Gangopadhyay},\ and\ \citenamefont {Sami}}]{Choudhury:2023vuj}%
  \BibitemOpen
  \bibfield  {author} {\bibinfo {author} {\bibfnamefont {S.}~\bibnamefont {Choudhury}}, \bibinfo {author} {\bibfnamefont {M.~R.}\ \bibnamefont {Gangopadhyay}}, \ and\ \bibinfo {author} {\bibfnamefont {M.}~\bibnamefont {Sami}},\ }\href@noop {} {\  (\bibinfo {year} {2023}{\natexlab{d}})},\ \Eprint {http://arxiv.org/abs/2301.10000} {arXiv:2301.10000 [astro-ph.CO]} \BibitemShut {NoStop}%
\bibitem [{\citenamefont {Choudhury}(2014)}]{Choudhury:2014hua}%
  \BibitemOpen
  \bibfield  {author} {\bibinfo {author} {\bibfnamefont {S.}~\bibnamefont {Choudhury}},\ }\href {\doibase 10.1016/j.physletb.2014.06.029} {\bibfield  {journal} {\bibinfo  {journal} {Phys. Lett. B}\ }\textbf {\bibinfo {volume} {735}},\ \bibinfo {pages} {138} (\bibinfo {year} {2014})},\ \Eprint {http://arxiv.org/abs/1403.0676} {arXiv:1403.0676 [hep-th]} \BibitemShut {NoStop}%
\bibitem [{\citenamefont {Choudhury}\ \emph {et~al.}(2014)\citenamefont {Choudhury}, \citenamefont {Mazumdar},\ and\ \citenamefont {Pukartas}}]{Choudhury:2014sxa}%
  \BibitemOpen
  \bibfield  {author} {\bibinfo {author} {\bibfnamefont {S.}~\bibnamefont {Choudhury}}, \bibinfo {author} {\bibfnamefont {A.}~\bibnamefont {Mazumdar}}, \ and\ \bibinfo {author} {\bibfnamefont {E.}~\bibnamefont {Pukartas}},\ }\href {\doibase 10.1007/JHEP04(2014)077} {\bibfield  {journal} {\bibinfo  {journal} {JHEP}\ }\textbf {\bibinfo {volume} {04}},\ \bibinfo {pages} {077} (\bibinfo {year} {2014})},\ \Eprint {http://arxiv.org/abs/1402.1227} {arXiv:1402.1227 [hep-th]} \BibitemShut {NoStop}%
\bibitem [{\citenamefont {Choudhury}\ \emph {et~al.}(2013)\citenamefont {Choudhury}, \citenamefont {Mazumdar},\ and\ \citenamefont {Pal}}]{Choudhury:2013jya}%
  \BibitemOpen
  \bibfield  {author} {\bibinfo {author} {\bibfnamefont {S.}~\bibnamefont {Choudhury}}, \bibinfo {author} {\bibfnamefont {A.}~\bibnamefont {Mazumdar}}, \ and\ \bibinfo {author} {\bibfnamefont {S.}~\bibnamefont {Pal}},\ }\href {\doibase 10.1088/1475-7516/2013/07/041} {\bibfield  {journal} {\bibinfo  {journal} {JCAP}\ }\textbf {\bibinfo {volume} {07}},\ \bibinfo {pages} {041} (\bibinfo {year} {2013})},\ \Eprint {http://arxiv.org/abs/1305.6398} {arXiv:1305.6398 [hep-ph]} \BibitemShut {NoStop}%
\bibitem [{\citenamefont {Choudhury}\ and\ \citenamefont {Mazumdar}(2014)}]{Choudhury:2013iaa}%
  \BibitemOpen
  \bibfield  {author} {\bibinfo {author} {\bibfnamefont {S.}~\bibnamefont {Choudhury}}\ and\ \bibinfo {author} {\bibfnamefont {A.}~\bibnamefont {Mazumdar}},\ }\href {\doibase 10.1016/j.nuclphysb.2014.03.005} {\bibfield  {journal} {\bibinfo  {journal} {Nucl. Phys. B}\ }\textbf {\bibinfo {volume} {882}},\ \bibinfo {pages} {386} (\bibinfo {year} {2014})},\ \Eprint {http://arxiv.org/abs/1306.4496} {arXiv:1306.4496 [hep-ph]} \BibitemShut {NoStop}%
\bibitem [{\citenamefont {Choudhury}\ and\ \citenamefont {Pal}(2012{\natexlab{a}})}]{Choudhury:2011jt}%
  \BibitemOpen
  \bibfield  {author} {\bibinfo {author} {\bibfnamefont {S.}~\bibnamefont {Choudhury}}\ and\ \bibinfo {author} {\bibfnamefont {S.}~\bibnamefont {Pal}},\ }\href {\doibase 10.1088/1475-7516/2012/04/018} {\bibfield  {journal} {\bibinfo  {journal} {JCAP}\ }\textbf {\bibinfo {volume} {04}},\ \bibinfo {pages} {018} (\bibinfo {year} {2012}{\natexlab{a}})},\ \Eprint {http://arxiv.org/abs/1111.3441} {arXiv:1111.3441 [hep-ph]} \BibitemShut {NoStop}%
\bibitem [{\citenamefont {Choudhury}\ and\ \citenamefont {Pal}(2012{\natexlab{b}})}]{Choudhury:2011rz}%
  \BibitemOpen
  \bibfield  {author} {\bibinfo {author} {\bibfnamefont {S.}~\bibnamefont {Choudhury}}\ and\ \bibinfo {author} {\bibfnamefont {S.}~\bibnamefont {Pal}},\ }\href {\doibase 10.1016/j.nuclphysb.2011.12.006} {\bibfield  {journal} {\bibinfo  {journal} {Nucl. Phys. B}\ }\textbf {\bibinfo {volume} {857}},\ \bibinfo {pages} {85} (\bibinfo {year} {2012}{\natexlab{b}})},\ \Eprint {http://arxiv.org/abs/1108.5676} {arXiv:1108.5676 [hep-ph]} \BibitemShut {NoStop}%
\bibitem [{\citenamefont {McDonald}(2002)}]{McDonald:2001vt}%
  \BibitemOpen
  \bibfield  {author} {\bibinfo {author} {\bibfnamefont {J.}~\bibnamefont {McDonald}},\ }\href {\doibase 10.1103/PhysRevLett.88.091304} {\bibfield  {journal} {\bibinfo  {journal} {Phys. Rev. Lett.}\ }\textbf {\bibinfo {volume} {88}},\ \bibinfo {pages} {091304} (\bibinfo {year} {2002})},\ \Eprint {http://arxiv.org/abs/hep-ph/0106249} {arXiv:hep-ph/0106249} \BibitemShut {NoStop}%
\bibitem [{\citenamefont {Choi}\ and\ \citenamefont {Roszkowski}(2005)}]{Choi:2005vq}%
  \BibitemOpen
  \bibfield  {author} {\bibinfo {author} {\bibfnamefont {K.-Y.}\ \bibnamefont {Choi}}\ and\ \bibinfo {author} {\bibfnamefont {L.}~\bibnamefont {Roszkowski}},\ }\href {\doibase 10.1063/1.2149672} {\bibfield  {journal} {\bibinfo  {journal} {AIP Conf. Proc.}\ }\textbf {\bibinfo {volume} {805}},\ \bibinfo {pages} {30} (\bibinfo {year} {2005})},\ \Eprint {http://arxiv.org/abs/hep-ph/0511003} {arXiv:hep-ph/0511003} \BibitemShut {NoStop}%
\bibitem [{\citenamefont {Kusenko}(2006)}]{Kusenko:2006rh}%
  \BibitemOpen
  \bibfield  {author} {\bibinfo {author} {\bibfnamefont {A.}~\bibnamefont {Kusenko}},\ }\href {\doibase 10.1103/PhysRevLett.97.241301} {\bibfield  {journal} {\bibinfo  {journal} {Phys. Rev. Lett.}\ }\textbf {\bibinfo {volume} {97}},\ \bibinfo {pages} {241301} (\bibinfo {year} {2006})},\ \Eprint {http://arxiv.org/abs/hep-ph/0609081} {arXiv:hep-ph/0609081} \BibitemShut {NoStop}%
\bibitem [{\citenamefont {Petraki}\ and\ \citenamefont {Kusenko}(2008)}]{Petraki:2007gq}%
  \BibitemOpen
  \bibfield  {author} {\bibinfo {author} {\bibfnamefont {K.}~\bibnamefont {Petraki}}\ and\ \bibinfo {author} {\bibfnamefont {A.}~\bibnamefont {Kusenko}},\ }\href {\doibase 10.1103/PhysRevD.77.065014} {\bibfield  {journal} {\bibinfo  {journal} {Phys. Rev. D}\ }\textbf {\bibinfo {volume} {77}},\ \bibinfo {pages} {065014} (\bibinfo {year} {2008})},\ \Eprint {http://arxiv.org/abs/0711.4646} {arXiv:0711.4646 [hep-ph]} \BibitemShut {NoStop}%
\bibitem [{\citenamefont {Hall}\ \emph {et~al.}(2010)\citenamefont {Hall}, \citenamefont {Jedamzik}, \citenamefont {March-Russell},\ and\ \citenamefont {West}}]{Hall:2009bx}%
  \BibitemOpen
  \bibfield  {author} {\bibinfo {author} {\bibfnamefont {L.~J.}\ \bibnamefont {Hall}}, \bibinfo {author} {\bibfnamefont {K.}~\bibnamefont {Jedamzik}}, \bibinfo {author} {\bibfnamefont {J.}~\bibnamefont {March-Russell}}, \ and\ \bibinfo {author} {\bibfnamefont {S.~M.}\ \bibnamefont {West}},\ }\href {\doibase 10.1007/JHEP03(2010)080} {\bibfield  {journal} {\bibinfo  {journal} {JHEP}\ }\textbf {\bibinfo {volume} {03}},\ \bibinfo {pages} {080} (\bibinfo {year} {2010})},\ \Eprint {http://arxiv.org/abs/0911.1120} {arXiv:0911.1120 [hep-ph]} \BibitemShut {NoStop}%
\bibitem [{\citenamefont {Bernal}\ \emph {et~al.}(2017)\citenamefont {Bernal}, \citenamefont {Heikinheimo}, \citenamefont {Tenkanen}, \citenamefont {Tuominen},\ and\ \citenamefont {Vaskonen}}]{Bernal:2017kxu}%
  \BibitemOpen
  \bibfield  {author} {\bibinfo {author} {\bibfnamefont {N.}~\bibnamefont {Bernal}}, \bibinfo {author} {\bibfnamefont {M.}~\bibnamefont {Heikinheimo}}, \bibinfo {author} {\bibfnamefont {T.}~\bibnamefont {Tenkanen}}, \bibinfo {author} {\bibfnamefont {K.}~\bibnamefont {Tuominen}}, \ and\ \bibinfo {author} {\bibfnamefont {V.}~\bibnamefont {Vaskonen}},\ }\href {\doibase 10.1142/S0217751X1730023X} {\bibfield  {journal} {\bibinfo  {journal} {Int. J. Mod. Phys. A}\ }\textbf {\bibinfo {volume} {32}},\ \bibinfo {pages} {1730023} (\bibinfo {year} {2017})},\ \Eprint {http://arxiv.org/abs/1706.07442} {arXiv:1706.07442 [hep-ph]} \BibitemShut {NoStop}%
\bibitem [{\citenamefont {Haque}\ and\ \citenamefont {Maity}(2022)}]{Haque:2021mab}%
  \BibitemOpen
  \bibfield  {author} {\bibinfo {author} {\bibfnamefont {M.~R.}\ \bibnamefont {Haque}}\ and\ \bibinfo {author} {\bibfnamefont {D.}~\bibnamefont {Maity}},\ }\href {\doibase 10.1103/PhysRevD.106.023506} {\bibfield  {journal} {\bibinfo  {journal} {Phys. Rev. D}\ }\textbf {\bibinfo {volume} {106}},\ \bibinfo {pages} {023506} (\bibinfo {year} {2022})},\ \Eprint {http://arxiv.org/abs/2112.14668} {arXiv:2112.14668 [hep-ph]} \BibitemShut {NoStop}%
\bibitem [{\citenamefont {Haque}\ and\ \citenamefont {Maity}(2023)}]{Haque:2022kez}%
  \BibitemOpen
  \bibfield  {author} {\bibinfo {author} {\bibfnamefont {M.~R.}\ \bibnamefont {Haque}}\ and\ \bibinfo {author} {\bibfnamefont {D.}~\bibnamefont {Maity}},\ }\href {\doibase 10.1103/PhysRevD.107.043531} {\bibfield  {journal} {\bibinfo  {journal} {Phys. Rev. D}\ }\textbf {\bibinfo {volume} {107}},\ \bibinfo {pages} {043531} (\bibinfo {year} {2023})},\ \Eprint {http://arxiv.org/abs/2201.02348} {arXiv:2201.02348 [hep-ph]} \BibitemShut {NoStop}%
\bibitem [{\citenamefont {Haque}\ \emph {et~al.}(2023)\citenamefont {Haque}, \citenamefont {Maity},\ and\ \citenamefont {Mondal}}]{Haque:2023yra}%
  \BibitemOpen
  \bibfield  {author} {\bibinfo {author} {\bibfnamefont {M.~R.}\ \bibnamefont {Haque}}, \bibinfo {author} {\bibfnamefont {D.}~\bibnamefont {Maity}}, \ and\ \bibinfo {author} {\bibfnamefont {R.}~\bibnamefont {Mondal}},\ }\href {\doibase 10.1007/JHEP09(2023)012} {\bibfield  {journal} {\bibinfo  {journal} {JHEP}\ }\textbf {\bibinfo {volume} {09}},\ \bibinfo {pages} {012} (\bibinfo {year} {2023})},\ \Eprint {http://arxiv.org/abs/2301.01641} {arXiv:2301.01641 [hep-ph]} \BibitemShut {NoStop}%
\bibitem [{\citenamefont {Baer}\ \emph {et~al.}(2015)\citenamefont {Baer}, \citenamefont {Choi}, \citenamefont {Kim},\ and\ \citenamefont {Roszkowski}}]{Baer:2014eja}%
  \BibitemOpen
  \bibfield  {author} {\bibinfo {author} {\bibfnamefont {H.}~\bibnamefont {Baer}}, \bibinfo {author} {\bibfnamefont {K.-Y.}\ \bibnamefont {Choi}}, \bibinfo {author} {\bibfnamefont {J.~E.}\ \bibnamefont {Kim}}, \ and\ \bibinfo {author} {\bibfnamefont {L.}~\bibnamefont {Roszkowski}},\ }\href {\doibase 10.1016/j.physrep.2014.10.002} {\bibfield  {journal} {\bibinfo  {journal} {Phys. Rept.}\ }\textbf {\bibinfo {volume} {555}},\ \bibinfo {pages} {1} (\bibinfo {year} {2015})},\ \Eprint {http://arxiv.org/abs/1407.0017} {arXiv:1407.0017 [hep-ph]} \BibitemShut {NoStop}%
\bibitem [{\citenamefont {Garny}\ \emph {et~al.}(2016)\citenamefont {Garny}, \citenamefont {Sandora},\ and\ \citenamefont {Sloth}}]{Garny:2015sjg}%
  \BibitemOpen
  \bibfield  {author} {\bibinfo {author} {\bibfnamefont {M.}~\bibnamefont {Garny}}, \bibinfo {author} {\bibfnamefont {M.}~\bibnamefont {Sandora}}, \ and\ \bibinfo {author} {\bibfnamefont {M.~S.}\ \bibnamefont {Sloth}},\ }\href {\doibase 10.1103/PhysRevLett.116.101302} {\bibfield  {journal} {\bibinfo  {journal} {Phys. Rev. Lett.}\ }\textbf {\bibinfo {volume} {116}},\ \bibinfo {pages} {101302} (\bibinfo {year} {2016})},\ \Eprint {http://arxiv.org/abs/1511.03278} {arXiv:1511.03278 [hep-ph]} \BibitemShut {NoStop}%
\bibitem [{\citenamefont {Tang}\ and\ \citenamefont {Wu}(2016)}]{Tang:2016vch}%
  \BibitemOpen
  \bibfield  {author} {\bibinfo {author} {\bibfnamefont {Y.}~\bibnamefont {Tang}}\ and\ \bibinfo {author} {\bibfnamefont {Y.-L.}\ \bibnamefont {Wu}},\ }\href {\doibase 10.1016/j.physletb.2016.05.045} {\bibfield  {journal} {\bibinfo  {journal} {Phys. Lett. B}\ }\textbf {\bibinfo {volume} {758}},\ \bibinfo {pages} {402} (\bibinfo {year} {2016})},\ \Eprint {http://arxiv.org/abs/1604.04701} {arXiv:1604.04701 [hep-ph]} \BibitemShut {NoStop}%
\bibitem [{\citenamefont {Tang}\ and\ \citenamefont {Wu}(2017)}]{Tang:2017hvq}%
  \BibitemOpen
  \bibfield  {author} {\bibinfo {author} {\bibfnamefont {Y.}~\bibnamefont {Tang}}\ and\ \bibinfo {author} {\bibfnamefont {Y.-L.}\ \bibnamefont {Wu}},\ }\href {\doibase 10.1016/j.physletb.2017.10.034} {\bibfield  {journal} {\bibinfo  {journal} {Phys. Lett. B}\ }\textbf {\bibinfo {volume} {774}},\ \bibinfo {pages} {676} (\bibinfo {year} {2017})},\ \Eprint {http://arxiv.org/abs/1708.05138} {arXiv:1708.05138 [hep-ph]} \BibitemShut {NoStop}%
\bibitem [{\citenamefont {Garny}\ \emph {et~al.}(2018)\citenamefont {Garny}, \citenamefont {Palessandro}, \citenamefont {Sandora},\ and\ \citenamefont {Sloth}}]{Garny:2017kha}%
  \BibitemOpen
  \bibfield  {author} {\bibinfo {author} {\bibfnamefont {M.}~\bibnamefont {Garny}}, \bibinfo {author} {\bibfnamefont {A.}~\bibnamefont {Palessandro}}, \bibinfo {author} {\bibfnamefont {M.}~\bibnamefont {Sandora}}, \ and\ \bibinfo {author} {\bibfnamefont {M.~S.}\ \bibnamefont {Sloth}},\ }\href {\doibase 10.1088/1475-7516/2018/02/027} {\bibfield  {journal} {\bibinfo  {journal} {JCAP}\ }\textbf {\bibinfo {volume} {02}},\ \bibinfo {pages} {027} (\bibinfo {year} {2018})},\ \Eprint {http://arxiv.org/abs/1709.09688} {arXiv:1709.09688 [hep-ph]} \BibitemShut {NoStop}%
\bibitem [{\citenamefont {Bernal}\ \emph {et~al.}(2018{\natexlab{b}})\citenamefont {Bernal}, \citenamefont {Dutra}, \citenamefont {Mambrini}, \citenamefont {Olive}, \citenamefont {Peloso},\ and\ \citenamefont {Pierre}}]{Bernal:2018qlk}%
  \BibitemOpen
  \bibfield  {author} {\bibinfo {author} {\bibfnamefont {N.}~\bibnamefont {Bernal}}, \bibinfo {author} {\bibfnamefont {M.}~\bibnamefont {Dutra}}, \bibinfo {author} {\bibfnamefont {Y.}~\bibnamefont {Mambrini}}, \bibinfo {author} {\bibfnamefont {K.}~\bibnamefont {Olive}}, \bibinfo {author} {\bibfnamefont {M.}~\bibnamefont {Peloso}}, \ and\ \bibinfo {author} {\bibfnamefont {M.}~\bibnamefont {Pierre}},\ }\href {\doibase 10.1103/PhysRevD.97.115020} {\bibfield  {journal} {\bibinfo  {journal} {Phys. Rev. D}\ }\textbf {\bibinfo {volume} {97}},\ \bibinfo {pages} {115020} (\bibinfo {year} {2018}{\natexlab{b}})},\ \Eprint {http://arxiv.org/abs/1803.01866} {arXiv:1803.01866 [hep-ph]} \BibitemShut {NoStop}%
\bibitem [{\citenamefont {Barman}\ \emph {et~al.}(2023{\natexlab{a}})\citenamefont {Barman}, \citenamefont {Ghoshal}, \citenamefont {Grzadkowski},\ and\ \citenamefont {Socha}}]{Barman:2023ktz}%
  \BibitemOpen
  \bibfield  {author} {\bibinfo {author} {\bibfnamefont {B.}~\bibnamefont {Barman}}, \bibinfo {author} {\bibfnamefont {A.}~\bibnamefont {Ghoshal}}, \bibinfo {author} {\bibfnamefont {B.}~\bibnamefont {Grzadkowski}}, \ and\ \bibinfo {author} {\bibfnamefont {A.}~\bibnamefont {Socha}},\ }\href {\doibase 10.1007/JHEP07(2023)231} {\bibfield  {journal} {\bibinfo  {journal} {JHEP}\ }\textbf {\bibinfo {volume} {07}},\ \bibinfo {pages} {231} (\bibinfo {year} {2023}{\natexlab{a}})},\ \Eprint {http://arxiv.org/abs/2305.00027} {arXiv:2305.00027 [hep-ph]} \BibitemShut {NoStop}%
\bibitem [{\citenamefont {Barman}\ \emph {et~al.}(2023{\natexlab{b}})\citenamefont {Barman}, \citenamefont {Bhupal~Dev},\ and\ \citenamefont {Ghoshal}}]{Barman:2022scg}%
  \BibitemOpen
  \bibfield  {author} {\bibinfo {author} {\bibfnamefont {B.}~\bibnamefont {Barman}}, \bibinfo {author} {\bibfnamefont {P.~S.}\ \bibnamefont {Bhupal~Dev}}, \ and\ \bibinfo {author} {\bibfnamefont {A.}~\bibnamefont {Ghoshal}},\ }\href {\doibase 10.1103/PhysRevD.108.035037} {\bibfield  {journal} {\bibinfo  {journal} {Phys. Rev. D}\ }\textbf {\bibinfo {volume} {108}},\ \bibinfo {pages} {035037} (\bibinfo {year} {2023}{\natexlab{b}})},\ \Eprint {http://arxiv.org/abs/2210.07739} {arXiv:2210.07739 [hep-ph]} \BibitemShut {NoStop}%
\bibitem [{\citenamefont {Barman}\ and\ \citenamefont {Ghoshal}(2022{\natexlab{a}})}]{Barman:2022njh}%
  \BibitemOpen
  \bibfield  {author} {\bibinfo {author} {\bibfnamefont {B.}~\bibnamefont {Barman}}\ and\ \bibinfo {author} {\bibfnamefont {A.}~\bibnamefont {Ghoshal}},\ }\href {\doibase 10.1088/1475-7516/2022/10/082} {\bibfield  {journal} {\bibinfo  {journal} {JCAP}\ }\textbf {\bibinfo {volume} {10}},\ \bibinfo {pages} {082} (\bibinfo {year} {2022}{\natexlab{a}})},\ \Eprint {http://arxiv.org/abs/2203.13269} {arXiv:2203.13269 [hep-ph]} \BibitemShut {NoStop}%
\bibitem [{\citenamefont {Barman}\ \emph {et~al.}(2022)\citenamefont {Barman}, \citenamefont {Ghosh}, \citenamefont {Ghoshal},\ and\ \citenamefont {Mukherjee}}]{Barman:2021yaz}%
  \BibitemOpen
  \bibfield  {author} {\bibinfo {author} {\bibfnamefont {B.}~\bibnamefont {Barman}}, \bibinfo {author} {\bibfnamefont {P.}~\bibnamefont {Ghosh}}, \bibinfo {author} {\bibfnamefont {A.}~\bibnamefont {Ghoshal}}, \ and\ \bibinfo {author} {\bibfnamefont {L.}~\bibnamefont {Mukherjee}},\ }\href {\doibase 10.1088/1475-7516/2022/08/049} {\bibfield  {journal} {\bibinfo  {journal} {JCAP}\ }\textbf {\bibinfo {volume} {08}},\ \bibinfo {pages} {049} (\bibinfo {year} {2022})},\ \Eprint {http://arxiv.org/abs/2112.12798} {arXiv:2112.12798 [hep-ph]} \BibitemShut {NoStop}%
\bibitem [{\citenamefont {Barman}\ and\ \citenamefont {Ghoshal}(2022{\natexlab{b}})}]{Barman:2021lot}%
  \BibitemOpen
  \bibfield  {author} {\bibinfo {author} {\bibfnamefont {B.}~\bibnamefont {Barman}}\ and\ \bibinfo {author} {\bibfnamefont {A.}~\bibnamefont {Ghoshal}},\ }\href {\doibase 10.1088/1475-7516/2022/03/003} {\bibfield  {journal} {\bibinfo  {journal} {JCAP}\ }\textbf {\bibinfo {volume} {03}},\ \bibinfo {pages} {003} (\bibinfo {year} {2022}{\natexlab{b}})},\ \Eprint {http://arxiv.org/abs/2109.03259} {arXiv:2109.03259 [hep-ph]} \BibitemShut {NoStop}%
\bibitem [{\citenamefont {Ghosh}\ \emph {et~al.}(2024)\citenamefont {Ghosh}, \citenamefont {Ghoshal},\ and\ \citenamefont {Jeesun}}]{Ghosh:2023tyz}%
  \BibitemOpen
  \bibfield  {author} {\bibinfo {author} {\bibfnamefont {D.~K.}\ \bibnamefont {Ghosh}}, \bibinfo {author} {\bibfnamefont {A.}~\bibnamefont {Ghoshal}}, \ and\ \bibinfo {author} {\bibfnamefont {S.}~\bibnamefont {Jeesun}},\ }\href {\doibase 10.1007/JHEP01(2024)026} {\bibfield  {journal} {\bibinfo  {journal} {JHEP}\ }\textbf {\bibinfo {volume} {01}},\ \bibinfo {pages} {026} (\bibinfo {year} {2024})},\ \Eprint {http://arxiv.org/abs/2305.09188} {arXiv:2305.09188 [hep-ph]} \BibitemShut {NoStop}%
\bibitem [{\citenamefont {Elor}\ \emph {et~al.}(2023)\citenamefont {Elor}, \citenamefont {McGehee},\ and\ \citenamefont {Pierce}}]{Elor:2021swj}%
  \BibitemOpen
  \bibfield  {author} {\bibinfo {author} {\bibfnamefont {G.}~\bibnamefont {Elor}}, \bibinfo {author} {\bibfnamefont {R.}~\bibnamefont {McGehee}}, \ and\ \bibinfo {author} {\bibfnamefont {A.}~\bibnamefont {Pierce}},\ }\href {\doibase 10.1103/PhysRevLett.130.031803} {\bibfield  {journal} {\bibinfo  {journal} {Phys. Rev. Lett.}\ }\textbf {\bibinfo {volume} {130}},\ \bibinfo {pages} {031803} (\bibinfo {year} {2023})},\ \Eprint {http://arxiv.org/abs/2112.03920} {arXiv:2112.03920 [hep-ph]} \BibitemShut {NoStop}%
\bibitem [{\citenamefont {Bhattiprolu}\ \emph {et~al.}(2023)\citenamefont {Bhattiprolu}, \citenamefont {Elor}, \citenamefont {McGehee},\ and\ \citenamefont {Pierce}}]{Bhattiprolu:2022sdd}%
  \BibitemOpen
  \bibfield  {author} {\bibinfo {author} {\bibfnamefont {P.~N.}\ \bibnamefont {Bhattiprolu}}, \bibinfo {author} {\bibfnamefont {G.}~\bibnamefont {Elor}}, \bibinfo {author} {\bibfnamefont {R.}~\bibnamefont {McGehee}}, \ and\ \bibinfo {author} {\bibfnamefont {A.}~\bibnamefont {Pierce}},\ }\href {\doibase 10.1007/JHEP01(2023)128} {\bibfield  {journal} {\bibinfo  {journal} {JHEP}\ }\textbf {\bibinfo {volume} {01}},\ \bibinfo {pages} {128} (\bibinfo {year} {2023})},\ \Eprint {http://arxiv.org/abs/2210.15653} {arXiv:2210.15653 [hep-ph]} \BibitemShut {NoStop}%
\bibitem [{\citenamefont {Chakraborty}\ \emph {et~al.}(2023)\citenamefont {Chakraborty}, \citenamefont {Haque}, \citenamefont {Maity},\ and\ \citenamefont {Mondal}}]{Chakraborty:2023ocr}%
  \BibitemOpen
  \bibfield  {author} {\bibinfo {author} {\bibfnamefont {A.}~\bibnamefont {Chakraborty}}, \bibinfo {author} {\bibfnamefont {M.~R.}\ \bibnamefont {Haque}}, \bibinfo {author} {\bibfnamefont {D.}~\bibnamefont {Maity}}, \ and\ \bibinfo {author} {\bibfnamefont {R.}~\bibnamefont {Mondal}},\ }\href {\doibase 10.1103/PhysRevD.108.023515} {\bibfield  {journal} {\bibinfo  {journal} {Phys. Rev. D}\ }\textbf {\bibinfo {volume} {108}},\ \bibinfo {pages} {023515} (\bibinfo {year} {2023})},\ \Eprint {http://arxiv.org/abs/2304.13637} {arXiv:2304.13637 [astro-ph.CO]} \BibitemShut {NoStop}%
\bibitem [{\citenamefont {Ghoshal}\ \emph {et~al.}(2022{\natexlab{a}})\citenamefont {Ghoshal}, \citenamefont {Heurtier},\ and\ \citenamefont {Paul}}]{Ghoshal:2022ruy}%
  \BibitemOpen
  \bibfield  {author} {\bibinfo {author} {\bibfnamefont {A.}~\bibnamefont {Ghoshal}}, \bibinfo {author} {\bibfnamefont {L.}~\bibnamefont {Heurtier}}, \ and\ \bibinfo {author} {\bibfnamefont {A.}~\bibnamefont {Paul}},\ }\href {\doibase 10.1007/JHEP12(2022)105} {\bibfield  {journal} {\bibinfo  {journal} {JHEP}\ }\textbf {\bibinfo {volume} {12}},\ \bibinfo {pages} {105} (\bibinfo {year} {2022}{\natexlab{a}})},\ \Eprint {http://arxiv.org/abs/2208.01670} {arXiv:2208.01670 [hep-ph]} \BibitemShut {NoStop}%
\bibitem [{\citenamefont {Berbig}\ and\ \citenamefont {Ghoshal}(2023)}]{Berbig:2023yyy}%
  \BibitemOpen
  \bibfield  {author} {\bibinfo {author} {\bibfnamefont {M.}~\bibnamefont {Berbig}}\ and\ \bibinfo {author} {\bibfnamefont {A.}~\bibnamefont {Ghoshal}},\ }\href {\doibase 10.1007/JHEP05(2023)172} {\bibfield  {journal} {\bibinfo  {journal} {JHEP}\ }\textbf {\bibinfo {volume} {05}},\ \bibinfo {pages} {172} (\bibinfo {year} {2023})},\ \Eprint {http://arxiv.org/abs/2301.05672} {arXiv:2301.05672 [hep-ph]} \BibitemShut {NoStop}%
\bibitem [{\citenamefont {Paul}\ \emph {et~al.}(2019)\citenamefont {Paul}, \citenamefont {Ghoshal}, \citenamefont {Chatterjee},\ and\ \citenamefont {Pal}}]{Paul:2018njm}%
  \BibitemOpen
  \bibfield  {author} {\bibinfo {author} {\bibfnamefont {A.}~\bibnamefont {Paul}}, \bibinfo {author} {\bibfnamefont {A.}~\bibnamefont {Ghoshal}}, \bibinfo {author} {\bibfnamefont {A.}~\bibnamefont {Chatterjee}}, \ and\ \bibinfo {author} {\bibfnamefont {S.}~\bibnamefont {Pal}},\ }\href {\doibase 10.1140/epjc/s10052-019-7348-5} {\bibfield  {journal} {\bibinfo  {journal} {Eur. Phys. J. C}\ }\textbf {\bibinfo {volume} {79}},\ \bibinfo {pages} {818} (\bibinfo {year} {2019})},\ \Eprint {http://arxiv.org/abs/1808.09706} {arXiv:1808.09706 [astro-ph.CO]} \BibitemShut {NoStop}%
\bibitem [{\citenamefont {Ghoshal}\ and\ \citenamefont {Saha}(2024)}]{Ghoshal:2022jdt}%
  \BibitemOpen
  \bibfield  {author} {\bibinfo {author} {\bibfnamefont {A.}~\bibnamefont {Ghoshal}}\ and\ \bibinfo {author} {\bibfnamefont {P.}~\bibnamefont {Saha}},\ }\href {\doibase 10.1103/PhysRevD.109.023526} {\bibfield  {journal} {\bibinfo  {journal} {Phys. Rev. D}\ }\textbf {\bibinfo {volume} {109}},\ \bibinfo {pages} {023526} (\bibinfo {year} {2024})},\ \Eprint {http://arxiv.org/abs/2203.14424} {arXiv:2203.14424 [hep-ph]} \BibitemShut {NoStop}%
\bibitem [{\citenamefont {Ghoshal}\ \emph {et~al.}(2023{\natexlab{a}})\citenamefont {Ghoshal}, \citenamefont {Khlopov}, \citenamefont {Lalak},\ and\ \citenamefont {Porey}}]{Ghoshal:2023jhh}%
  \BibitemOpen
  \bibfield  {author} {\bibinfo {author} {\bibfnamefont {A.}~\bibnamefont {Ghoshal}}, \bibinfo {author} {\bibfnamefont {M.~Y.}\ \bibnamefont {Khlopov}}, \bibinfo {author} {\bibfnamefont {Z.}~\bibnamefont {Lalak}}, \ and\ \bibinfo {author} {\bibfnamefont {S.}~\bibnamefont {Porey}},\ }\href@noop {} {\  (\bibinfo {year} {2023}{\natexlab{a}})},\ \Eprint {http://arxiv.org/abs/2306.09409} {arXiv:2306.09409 [hep-ph]} \BibitemShut {NoStop}%
\bibitem [{\citenamefont {Nozari}\ and\ \citenamefont {Sadatian}(2008)}]{Nozari:2007eq}%
  \BibitemOpen
  \bibfield  {author} {\bibinfo {author} {\bibfnamefont {K.}~\bibnamefont {Nozari}}\ and\ \bibinfo {author} {\bibfnamefont {S.~D.}\ \bibnamefont {Sadatian}},\ }\href {\doibase 10.1142/S0217732308026698} {\bibfield  {journal} {\bibinfo  {journal} {Mod. Phys. Lett. A}\ }\textbf {\bibinfo {volume} {23}},\ \bibinfo {pages} {2933} (\bibinfo {year} {2008})},\ \Eprint {http://arxiv.org/abs/0710.0058} {arXiv:0710.0058 [astro-ph]} \BibitemShut {NoStop}%
\bibitem [{\citenamefont {Cheong}\ \emph {et~al.}(2022)\citenamefont {Cheong}, \citenamefont {Lee},\ and\ \citenamefont {Park}}]{Cheong:2021kyc}%
  \BibitemOpen
  \bibfield  {author} {\bibinfo {author} {\bibfnamefont {D.~Y.}\ \bibnamefont {Cheong}}, \bibinfo {author} {\bibfnamefont {S.~M.}\ \bibnamefont {Lee}}, \ and\ \bibinfo {author} {\bibfnamefont {S.~C.}\ \bibnamefont {Park}},\ }\href {\doibase 10.1088/1475-7516/2022/02/029} {\bibfield  {journal} {\bibinfo  {journal} {JCAP}\ }\textbf {\bibinfo {volume} {02}},\ \bibinfo {pages} {029} (\bibinfo {year} {2022})},\ \Eprint {http://arxiv.org/abs/2111.00825} {arXiv:2111.00825 [hep-ph]} \BibitemShut {NoStop}%
\bibitem [{\citenamefont {Kodama}\ and\ \citenamefont {Takahashi}(2022)}]{Kodama:2021yrm}%
  \BibitemOpen
  \bibfield  {author} {\bibinfo {author} {\bibfnamefont {T.}~\bibnamefont {Kodama}}\ and\ \bibinfo {author} {\bibfnamefont {T.}~\bibnamefont {Takahashi}},\ }\href {\doibase 10.1103/PhysRevD.105.063542} {\bibfield  {journal} {\bibinfo  {journal} {Phys. Rev. D}\ }\textbf {\bibinfo {volume} {105}},\ \bibinfo {pages} {063542} (\bibinfo {year} {2022})},\ \Eprint {http://arxiv.org/abs/2112.05283} {arXiv:2112.05283 [astro-ph.CO]} \BibitemShut {NoStop}%
\bibitem [{\citenamefont {Watanabe}\ and\ \citenamefont {Komatsu}(2007)}]{Watanabe:2006ku}%
  \BibitemOpen
  \bibfield  {author} {\bibinfo {author} {\bibfnamefont {Y.}~\bibnamefont {Watanabe}}\ and\ \bibinfo {author} {\bibfnamefont {E.}~\bibnamefont {Komatsu}},\ }\href {\doibase 10.1103/PhysRevD.75.061301} {\bibfield  {journal} {\bibinfo  {journal} {Phys. Rev. D}\ }\textbf {\bibinfo {volume} {75}},\ \bibinfo {pages} {061301} (\bibinfo {year} {2007})},\ \Eprint {http://arxiv.org/abs/gr-qc/0612120} {arXiv:gr-qc/0612120} \BibitemShut {NoStop}%
\bibitem [{\citenamefont {Kannike}\ \emph {et~al.}(2015)\citenamefont {Kannike}, \citenamefont {H\"utsi}, \citenamefont {Pizza}, \citenamefont {Racioppi}, \citenamefont {Raidal}, \citenamefont {Salvio},\ and\ \citenamefont {Strumia}}]{Kannike:2015fom}%
  \BibitemOpen
  \bibfield  {author} {\bibinfo {author} {\bibfnamefont {K.}~\bibnamefont {Kannike}}, \bibinfo {author} {\bibfnamefont {G.}~\bibnamefont {H\"utsi}}, \bibinfo {author} {\bibfnamefont {L.}~\bibnamefont {Pizza}}, \bibinfo {author} {\bibfnamefont {A.}~\bibnamefont {Racioppi}}, \bibinfo {author} {\bibfnamefont {M.}~\bibnamefont {Raidal}}, \bibinfo {author} {\bibfnamefont {A.}~\bibnamefont {Salvio}}, \ and\ \bibinfo {author} {\bibfnamefont {A.}~\bibnamefont {Strumia}},\ }\href {\doibase 10.22323/1.234.0379} {\bibfield  {journal} {\bibinfo  {journal} {PoS}\ }\textbf {\bibinfo {volume} {EPS-HEP2015}},\ \bibinfo {pages} {379} (\bibinfo {year} {2015})}\BibitemShut {NoStop}%
\bibitem [{\citenamefont {Shaposhnikov}\ \emph {et~al.}(2020)\citenamefont {Shaposhnikov}, \citenamefont {Shkerin},\ and\ \citenamefont {Zell}}]{Shaposhnikov:2020fdv}%
  \BibitemOpen
  \bibfield  {author} {\bibinfo {author} {\bibfnamefont {M.}~\bibnamefont {Shaposhnikov}}, \bibinfo {author} {\bibfnamefont {A.}~\bibnamefont {Shkerin}}, \ and\ \bibinfo {author} {\bibfnamefont {S.}~\bibnamefont {Zell}},\ }\href {\doibase 10.1088/1475-7516/2020/07/064} {\bibfield  {journal} {\bibinfo  {journal} {JCAP}\ }\textbf {\bibinfo {volume} {07}},\ \bibinfo {pages} {064} (\bibinfo {year} {2020})},\ \Eprint {http://arxiv.org/abs/2002.07105} {arXiv:2002.07105 [hep-ph]} \BibitemShut {NoStop}%
\bibitem [{\citenamefont {Park}\ and\ \citenamefont {Yamaguchi}(2008)}]{Park:2008hz}%
  \BibitemOpen
  \bibfield  {author} {\bibinfo {author} {\bibfnamefont {S.~C.}\ \bibnamefont {Park}}\ and\ \bibinfo {author} {\bibfnamefont {S.}~\bibnamefont {Yamaguchi}},\ }\href {\doibase 10.1088/1475-7516/2008/08/009} {\bibfield  {journal} {\bibinfo  {journal} {JCAP}\ }\textbf {\bibinfo {volume} {08}},\ \bibinfo {pages} {009} (\bibinfo {year} {2008})},\ \Eprint {http://arxiv.org/abs/0801.1722} {arXiv:0801.1722 [hep-ph]} \BibitemShut {NoStop}%
\bibitem [{\citenamefont {Oda}\ \emph {et~al.}(2018)\citenamefont {Oda}, \citenamefont {Okada}, \citenamefont {Raut},\ and\ \citenamefont {Takahashi}}]{Oda:2017zul}%
  \BibitemOpen
  \bibfield  {author} {\bibinfo {author} {\bibfnamefont {S.}~\bibnamefont {Oda}}, \bibinfo {author} {\bibfnamefont {N.}~\bibnamefont {Okada}}, \bibinfo {author} {\bibfnamefont {D.}~\bibnamefont {Raut}}, \ and\ \bibinfo {author} {\bibfnamefont {D.-s.}\ \bibnamefont {Takahashi}},\ }\href {\doibase 10.1103/PhysRevD.97.055001} {\bibfield  {journal} {\bibinfo  {journal} {Phys. Rev. D}\ }\textbf {\bibinfo {volume} {97}},\ \bibinfo {pages} {055001} (\bibinfo {year} {2018})},\ \Eprint {http://arxiv.org/abs/1711.09850} {arXiv:1711.09850 [hep-ph]} \BibitemShut {NoStop}%
\bibitem [{\citenamefont {J\"arv}\ \emph {et~al.}(2017)\citenamefont {J\"arv}, \citenamefont {Kannike}, \citenamefont {Marzola}, \citenamefont {Racioppi}, \citenamefont {Raidal}, \citenamefont {R\"unkla}, \citenamefont {Saal},\ and\ \citenamefont {Veerm\"ae}}]{Jarv:2016sow}%
  \BibitemOpen
  \bibfield  {author} {\bibinfo {author} {\bibfnamefont {L.}~\bibnamefont {J\"arv}}, \bibinfo {author} {\bibfnamefont {K.}~\bibnamefont {Kannike}}, \bibinfo {author} {\bibfnamefont {L.}~\bibnamefont {Marzola}}, \bibinfo {author} {\bibfnamefont {A.}~\bibnamefont {Racioppi}}, \bibinfo {author} {\bibfnamefont {M.}~\bibnamefont {Raidal}}, \bibinfo {author} {\bibfnamefont {M.}~\bibnamefont {R\"unkla}}, \bibinfo {author} {\bibfnamefont {M.}~\bibnamefont {Saal}}, \ and\ \bibinfo {author} {\bibfnamefont {H.}~\bibnamefont {Veerm\"ae}},\ }\href {\doibase 10.1103/PhysRevLett.118.151302} {\bibfield  {journal} {\bibinfo  {journal} {Phys. Rev. Lett.}\ }\textbf {\bibinfo {volume} {118}},\ \bibinfo {pages} {151302} (\bibinfo {year} {2017})},\ \Eprint {http://arxiv.org/abs/1612.06863} {arXiv:1612.06863 [hep-ph]} \BibitemShut {NoStop}%
\bibitem [{\citenamefont {Lyth}(1993)}]{Lyth:1993eu}%
  \BibitemOpen
  \bibfield  {author} {\bibinfo {author} {\bibfnamefont {D.~H.}\ \bibnamefont {Lyth}},\ }in\ \href@noop {} {\emph {\bibinfo {booktitle} {{Summer School in High-energy Physics and Cosmology (Includes Workshop on Strings, Gravity, and Related Topics 29-30 Jul 1993)}}}}\ (\bibinfo {year} {1993})\ pp.\ \bibinfo {pages} {0069--136},\ \Eprint {http://arxiv.org/abs/astro-ph/9312022} {arXiv:astro-ph/9312022} \BibitemShut {NoStop}%
\bibitem [{\citenamefont {Baumann}(2022)}]{Baumann:2022mni}%
  \BibitemOpen
  \bibfield  {author} {\bibinfo {author} {\bibfnamefont {D.}~\bibnamefont {Baumann}},\ }\href {\doibase 10.1017/9781108937092} {\emph {\bibinfo {title} {{Cosmology}}}}\ (\bibinfo  {publisher} {Cambridge University Press},\ \bibinfo {year} {2022})\BibitemShut {NoStop}%
\bibitem [{\citenamefont {Bambi}\ and\ \citenamefont {Dolgov}(2015)}]{Bambi:2015mba}%
  \BibitemOpen
  \bibfield  {author} {\bibinfo {author} {\bibfnamefont {C.}~\bibnamefont {Bambi}}\ and\ \bibinfo {author} {\bibfnamefont {A.~D.}\ \bibnamefont {Dolgov}},\ }\href {\doibase 10.1007/978-3-662-48078-6} {\emph {\bibinfo {title} {{Introduction to Particle Cosmology}}}},\ UNITEXT for Physics\ (\bibinfo  {publisher} {Springer},\ \bibinfo {year} {2015})\BibitemShut {NoStop}%
\bibitem [{\citenamefont {Aghanim}\ \emph {et~al.}(2020)\citenamefont {Aghanim} \emph {et~al.}}]{Aghanim:2018eyx}%
  \BibitemOpen
  \bibfield  {author} {\bibinfo {author} {\bibfnamefont {N.}~\bibnamefont {Aghanim}} \emph {et~al.} (\bibinfo {collaboration} {Planck}),\ }\href {\doibase 10.1051/0004-6361/201833910} {\bibfield  {journal} {\bibinfo  {journal} {Astron. Astrophys.}\ }\textbf {\bibinfo {volume} {641}},\ \bibinfo {pages} {A6} (\bibinfo {year} {2020})},\ \bibinfo {note} {[Erratum: Astron.Astrophys. 652, C4 (2021)]},\ \Eprint {http://arxiv.org/abs/1807.06209} {arXiv:1807.06209 [astro-ph.CO]} \BibitemShut {NoStop}%
\bibitem [{\citenamefont {Workman}\ \emph {et~al.}(2022)\citenamefont {Workman} \emph {et~al.}}]{ParticleDataGroup:2022pth}%
  \BibitemOpen
  \bibfield  {author} {\bibinfo {author} {\bibfnamefont {R.~L.}\ \bibnamefont {Workman}} \emph {et~al.} (\bibinfo {collaboration} {Particle Data Group}),\ }\href {\doibase 10.1093/ptep/ptac097} {\bibfield  {journal} {\bibinfo  {journal} {PTEP}\ }\textbf {\bibinfo {volume} {2022}},\ \bibinfo {pages} {083C01} (\bibinfo {year} {2022})}\BibitemShut {NoStop}%
\bibitem [{\citenamefont {Ade}\ \emph {et~al.}(2022{\natexlab{a}})\citenamefont {Ade} \emph {et~al.}}]{BICEPKeck:2022mhb}%
  \BibitemOpen
  \bibfield  {author} {\bibinfo {author} {\bibfnamefont {P.~A.~R.}\ \bibnamefont {Ade}} \emph {et~al.} (\bibinfo {collaboration} {BICEP/Keck}),\ }in\ \href@noop {} {\emph {\bibinfo {booktitle} {{56th Rencontres de Moriond on Cosmology}}}}\ (\bibinfo {year} {2022})\ \Eprint {http://arxiv.org/abs/2203.16556} {arXiv:2203.16556 [astro-ph.CO]} \BibitemShut {NoStop}%
\bibitem [{\citenamefont {Ade}\ \emph {et~al.}(2021)\citenamefont {Ade} \emph {et~al.}}]{BICEP:2021xfz}%
  \BibitemOpen
  \bibfield  {author} {\bibinfo {author} {\bibfnamefont {P.~A.~R.}\ \bibnamefont {Ade}} \emph {et~al.} (\bibinfo {collaboration} {BICEP, Keck}),\ }\href {\doibase 10.1103/PhysRevLett.127.151301} {\bibfield  {journal} {\bibinfo  {journal} {Phys. Rev. Lett.}\ }\textbf {\bibinfo {volume} {127}},\ \bibinfo {pages} {151301} (\bibinfo {year} {2021})},\ \Eprint {http://arxiv.org/abs/2110.00483} {arXiv:2110.00483 [astro-ph.CO]} \BibitemShut {NoStop}%
\bibitem [{\citenamefont {Campeti}\ and\ \citenamefont {Komatsu}(2022)}]{Campeti:2022vom}%
  \BibitemOpen
  \bibfield  {author} {\bibinfo {author} {\bibfnamefont {P.}~\bibnamefont {Campeti}}\ and\ \bibinfo {author} {\bibfnamefont {E.}~\bibnamefont {Komatsu}},\ }\href {\doibase 10.3847/1538-4357/ac9ea3} {\bibfield  {journal} {\bibinfo  {journal} {Astrophys. J.}\ }\textbf {\bibinfo {volume} {941}},\ \bibinfo {pages} {110} (\bibinfo {year} {2022})},\ \Eprint {http://arxiv.org/abs/2205.05617} {arXiv:2205.05617 [astro-ph.CO]} \BibitemShut {NoStop}%
\bibitem [{\citenamefont {Bernal}\ and\ \citenamefont {Xu}(2021)}]{Bernal:2021qrl}%
  \BibitemOpen
  \bibfield  {author} {\bibinfo {author} {\bibfnamefont {N.}~\bibnamefont {Bernal}}\ and\ \bibinfo {author} {\bibfnamefont {Y.}~\bibnamefont {Xu}},\ }\href {\doibase 10.1140/epjc/s10052-021-09694-5} {\bibfield  {journal} {\bibinfo  {journal} {Eur. Phys. J. C}\ }\textbf {\bibinfo {volume} {81}},\ \bibinfo {pages} {877} (\bibinfo {year} {2021})},\ \Eprint {http://arxiv.org/abs/2106.03950} {arXiv:2106.03950 [hep-ph]} \BibitemShut {NoStop}%
\bibitem [{\citenamefont {Ghoshal}\ \emph {et~al.}(2022{\natexlab{b}})\citenamefont {Ghoshal}, \citenamefont {Lambiase}, \citenamefont {Pal}, \citenamefont {Paul},\ and\ \citenamefont {Porey}}]{Ghoshal:2022jeo}%
  \BibitemOpen
  \bibfield  {author} {\bibinfo {author} {\bibfnamefont {A.}~\bibnamefont {Ghoshal}}, \bibinfo {author} {\bibfnamefont {G.}~\bibnamefont {Lambiase}}, \bibinfo {author} {\bibfnamefont {S.}~\bibnamefont {Pal}}, \bibinfo {author} {\bibfnamefont {A.}~\bibnamefont {Paul}}, \ and\ \bibinfo {author} {\bibfnamefont {S.}~\bibnamefont {Porey}},\ }\href {\doibase 10.1007/JHEP09(2022)231} {\bibfield  {journal} {\bibinfo  {journal} {JHEP}\ }\textbf {\bibinfo {volume} {09}},\ \bibinfo {pages} {231} (\bibinfo {year} {2022}{\natexlab{b}})},\ \Eprint {http://arxiv.org/abs/2206.10648} {arXiv:2206.10648 [hep-ph]} \BibitemShut {NoStop}%
\bibitem [{\citenamefont {Ghoshal}\ \emph {et~al.}(2023{\natexlab{b}})\citenamefont {Ghoshal}, \citenamefont {Lambiase}, \citenamefont {Pal}, \citenamefont {Paul},\ and\ \citenamefont {Porey}}]{Ghoshal:2023noe}%
  \BibitemOpen
  \bibfield  {author} {\bibinfo {author} {\bibfnamefont {A.}~\bibnamefont {Ghoshal}}, \bibinfo {author} {\bibfnamefont {G.}~\bibnamefont {Lambiase}}, \bibinfo {author} {\bibfnamefont {S.}~\bibnamefont {Pal}}, \bibinfo {author} {\bibfnamefont {A.}~\bibnamefont {Paul}}, \ and\ \bibinfo {author} {\bibfnamefont {S.}~\bibnamefont {Porey}},\ }\href {\doibase 10.3390/sym15020543} {\bibfield  {journal} {\bibinfo  {journal} {Symmetry}\ }\textbf {\bibinfo {volume} {15}},\ \bibinfo {pages} {543} (\bibinfo {year} {2023}{\natexlab{b}})}\BibitemShut {NoStop}%
\bibitem [{\citenamefont {Ghoshal}\ \emph {et~al.}(2022{\natexlab{c}})\citenamefont {Ghoshal}, \citenamefont {Lambiase}, \citenamefont {Pal}, \citenamefont {Paul},\ and\ \citenamefont {Porey}}]{Ghoshal:2022aoh}%
  \BibitemOpen
  \bibfield  {author} {\bibinfo {author} {\bibfnamefont {A.}~\bibnamefont {Ghoshal}}, \bibinfo {author} {\bibfnamefont {G.}~\bibnamefont {Lambiase}}, \bibinfo {author} {\bibfnamefont {S.}~\bibnamefont {Pal}}, \bibinfo {author} {\bibfnamefont {A.}~\bibnamefont {Paul}}, \ and\ \bibinfo {author} {\bibfnamefont {S.}~\bibnamefont {Porey}},\ }in\ \href@noop {} {\emph {\bibinfo {booktitle} {{25th Workshop on What Comes Beyond the Standard Models?}}}}\ (\bibinfo {year} {2022})\ \Eprint {http://arxiv.org/abs/2211.15061} {arXiv:2211.15061 [astro-ph.CO]} \BibitemShut {NoStop}%
\bibitem [{\citenamefont {Ghoshal}\ \emph {et~al.}(2024)\citenamefont {Ghoshal}, \citenamefont {Lalak}, \citenamefont {Pal},\ and\ \citenamefont {Porey}}]{Ghoshal:2024ycp}%
  \BibitemOpen
  \bibfield  {author} {\bibinfo {author} {\bibfnamefont {A.}~\bibnamefont {Ghoshal}}, \bibinfo {author} {\bibfnamefont {Z.}~\bibnamefont {Lalak}}, \bibinfo {author} {\bibfnamefont {S.}~\bibnamefont {Pal}}, \ and\ \bibinfo {author} {\bibfnamefont {S.}~\bibnamefont {Porey}},\ }\href@noop {} {\  (\bibinfo {year} {2024})},\ \Eprint {http://arxiv.org/abs/2401.17262} {arXiv:2401.17262 [astro-ph.CO]} \BibitemShut {NoStop}%
\bibitem [{\citenamefont {Drees}\ and\ \citenamefont {Xu}(2021)}]{Drees:2021wgd}%
  \BibitemOpen
  \bibfield  {author} {\bibinfo {author} {\bibfnamefont {M.}~\bibnamefont {Drees}}\ and\ \bibinfo {author} {\bibfnamefont {Y.}~\bibnamefont {Xu}},\ }\href {\doibase 10.1088/1475-7516/2021/09/012} {\bibfield  {journal} {\bibinfo  {journal} {JCAP}\ }\textbf {\bibinfo {volume} {09}},\ \bibinfo {pages} {012} (\bibinfo {year} {2021})},\ \Eprint {http://arxiv.org/abs/2104.03977} {arXiv:2104.03977 [hep-ph]} \BibitemShut {NoStop}%
\bibitem [{\citenamefont {Ghoshal}\ \emph {et~al.}(2023{\natexlab{c}})\citenamefont {Ghoshal}, \citenamefont {Lalak},\ and\ \citenamefont {Porey}}]{Ghoshal:2023phi}%
  \BibitemOpen
  \bibfield  {author} {\bibinfo {author} {\bibfnamefont {A.}~\bibnamefont {Ghoshal}}, \bibinfo {author} {\bibfnamefont {Z.}~\bibnamefont {Lalak}}, \ and\ \bibinfo {author} {\bibfnamefont {S.}~\bibnamefont {Porey}},\ }\href {\doibase 10.1103/PhysRevD.108.063030} {\bibfield  {journal} {\bibinfo  {journal} {Phys. Rev. D}\ }\textbf {\bibinfo {volume} {108}},\ \bibinfo {pages} {063030} (\bibinfo {year} {2023}{\natexlab{c}})},\ \Eprint {http://arxiv.org/abs/2302.03268} {arXiv:2302.03268 [hep-ph]} \BibitemShut {NoStop}%
\bibitem [{\citenamefont {Peccei}(2008)}]{Peccei:2006as}%
  \BibitemOpen
  \bibfield  {author} {\bibinfo {author} {\bibfnamefont {R.~D.}\ \bibnamefont {Peccei}},\ }\href {\doibase 10.1007/978-3-540-73518-2_1} {\bibfield  {journal} {\bibinfo  {journal} {Lect. Notes Phys.}\ }\textbf {\bibinfo {volume} {741}},\ \bibinfo {pages} {3} (\bibinfo {year} {2008})},\ \Eprint {http://arxiv.org/abs/hep-ph/0607268} {arXiv:hep-ph/0607268} \BibitemShut {NoStop}%
\bibitem [{\citenamefont {Hook}(2019)}]{Hook:2018dlk}%
  \BibitemOpen
  \bibfield  {author} {\bibinfo {author} {\bibfnamefont {A.}~\bibnamefont {Hook}},\ }\href@noop {} {\bibfield  {journal} {\bibinfo  {journal} {PoS}\ }\textbf {\bibinfo {volume} {TASI2018}},\ \bibinfo {pages} {004} (\bibinfo {year} {2019})},\ \Eprint {http://arxiv.org/abs/1812.02669} {arXiv:1812.02669 [hep-ph]} \BibitemShut {NoStop}%
\bibitem [{\citenamefont {Kim}\ and\ \citenamefont {Carosi}(2010)}]{Kim:2008hd}%
  \BibitemOpen
  \bibfield  {author} {\bibinfo {author} {\bibfnamefont {J.~E.}\ \bibnamefont {Kim}}\ and\ \bibinfo {author} {\bibfnamefont {G.}~\bibnamefont {Carosi}},\ }\href {\doibase 10.1103/RevModPhys.82.557} {\bibfield  {journal} {\bibinfo  {journal} {Rev. Mod. Phys.}\ }\textbf {\bibinfo {volume} {82}},\ \bibinfo {pages} {557} (\bibinfo {year} {2010})},\ \bibinfo {note} {[Erratum: Rev.Mod.Phys. 91, 049902 (2019)]},\ \Eprint {http://arxiv.org/abs/0807.3125} {arXiv:0807.3125 [hep-ph]} \BibitemShut {NoStop}%
\bibitem [{\citenamefont {Kim}\ \emph {et~al.}(2005)\citenamefont {Kim}, \citenamefont {Nilles},\ and\ \citenamefont {Peloso}}]{Kim:2004rp}%
  \BibitemOpen
  \bibfield  {author} {\bibinfo {author} {\bibfnamefont {J.~E.}\ \bibnamefont {Kim}}, \bibinfo {author} {\bibfnamefont {H.~P.}\ \bibnamefont {Nilles}}, \ and\ \bibinfo {author} {\bibfnamefont {M.}~\bibnamefont {Peloso}},\ }\href {\doibase 10.1088/1475-7516/2005/01/005} {\bibfield  {journal} {\bibinfo  {journal} {JCAP}\ }\textbf {\bibinfo {volume} {01}},\ \bibinfo {pages} {005} (\bibinfo {year} {2005})},\ \Eprint {http://arxiv.org/abs/hep-ph/0409138} {arXiv:hep-ph/0409138} \BibitemShut {NoStop}%
\bibitem [{\citenamefont {Freese}\ \emph {et~al.}(1990)\citenamefont {Freese}, \citenamefont {Frieman},\ and\ \citenamefont {Olinto}}]{Freese:1990rb}%
  \BibitemOpen
  \bibfield  {author} {\bibinfo {author} {\bibfnamefont {K.}~\bibnamefont {Freese}}, \bibinfo {author} {\bibfnamefont {J.~A.}\ \bibnamefont {Frieman}}, \ and\ \bibinfo {author} {\bibfnamefont {A.~V.}\ \bibnamefont {Olinto}},\ }\href {\doibase 10.1103/PhysRevLett.65.3233} {\bibfield  {journal} {\bibinfo  {journal} {Phys. Rev. Lett.}\ }\textbf {\bibinfo {volume} {65}},\ \bibinfo {pages} {3233} (\bibinfo {year} {1990})}\BibitemShut {NoStop}%
\bibitem [{\citenamefont {Savage}\ \emph {et~al.}(2006)\citenamefont {Savage}, \citenamefont {Freese},\ and\ \citenamefont {Kinney}}]{Savage:2006tr}%
  \BibitemOpen
  \bibfield  {author} {\bibinfo {author} {\bibfnamefont {C.}~\bibnamefont {Savage}}, \bibinfo {author} {\bibfnamefont {K.}~\bibnamefont {Freese}}, \ and\ \bibinfo {author} {\bibfnamefont {W.~H.}\ \bibnamefont {Kinney}},\ }\href {\doibase 10.1103/PhysRevD.74.123511} {\bibfield  {journal} {\bibinfo  {journal} {Phys. Rev. D}\ }\textbf {\bibinfo {volume} {74}},\ \bibinfo {pages} {123511} (\bibinfo {year} {2006})},\ \Eprint {http://arxiv.org/abs/hep-ph/0609144} {arXiv:hep-ph/0609144} \BibitemShut {NoStop}%
\bibitem [{\citenamefont {Cheng}\ \emph {et~al.}(2021)\citenamefont {Cheng}, \citenamefont {Bian},\ and\ \citenamefont {Zhou}}]{Cheng:2021qmc}%
  \BibitemOpen
  \bibfield  {author} {\bibinfo {author} {\bibfnamefont {W.}~\bibnamefont {Cheng}}, \bibinfo {author} {\bibfnamefont {L.}~\bibnamefont {Bian}}, \ and\ \bibinfo {author} {\bibfnamefont {Y.-F.}\ \bibnamefont {Zhou}},\ }\href {\doibase 10.1103/PhysRevD.104.063010} {\bibfield  {journal} {\bibinfo  {journal} {Phys. Rev. D}\ }\textbf {\bibinfo {volume} {104}},\ \bibinfo {pages} {063010} (\bibinfo {year} {2021})},\ \Eprint {http://arxiv.org/abs/2104.06602} {arXiv:2104.06602 [hep-ph]} \BibitemShut {NoStop}%
\bibitem [{\citenamefont {Daido}\ \emph {et~al.}(2017)\citenamefont {Daido}, \citenamefont {Takahashi},\ and\ \citenamefont {Yin}}]{Daido:2017wwb}%
  \BibitemOpen
  \bibfield  {author} {\bibinfo {author} {\bibfnamefont {R.}~\bibnamefont {Daido}}, \bibinfo {author} {\bibfnamefont {F.}~\bibnamefont {Takahashi}}, \ and\ \bibinfo {author} {\bibfnamefont {W.}~\bibnamefont {Yin}},\ }\href {\doibase 10.1088/1475-7516/2017/05/044} {\bibfield  {journal} {\bibinfo  {journal} {JCAP}\ }\textbf {\bibinfo {volume} {05}},\ \bibinfo {pages} {044} (\bibinfo {year} {2017})},\ \Eprint {http://arxiv.org/abs/1702.03284} {arXiv:1702.03284 [hep-ph]} \BibitemShut {NoStop}%
\bibitem [{\citenamefont {Reyimuaji}\ and\ \citenamefont {Zhang}(2021)}]{Reyimuaji:2020goi}%
  \BibitemOpen
  \bibfield  {author} {\bibinfo {author} {\bibfnamefont {Y.}~\bibnamefont {Reyimuaji}}\ and\ \bibinfo {author} {\bibfnamefont {X.}~\bibnamefont {Zhang}},\ }\href {\doibase 10.1088/1475-7516/2021/03/059} {\bibfield  {journal} {\bibinfo  {journal} {JCAP}\ }\textbf {\bibinfo {volume} {03}},\ \bibinfo {pages} {059} (\bibinfo {year} {2021})},\ \Eprint {http://arxiv.org/abs/2012.14248} {arXiv:2012.14248 [astro-ph.CO]} \BibitemShut {NoStop}%
\bibitem [{\citenamefont {Birrell}\ and\ \citenamefont {Davies}(1984)}]{Birrell:1982ix}%
  \BibitemOpen
  \bibfield  {author} {\bibinfo {author} {\bibfnamefont {N.~D.}\ \bibnamefont {Birrell}}\ and\ \bibinfo {author} {\bibfnamefont {P.~C.~W.}\ \bibnamefont {Davies}},\ }\href {\doibase 10.1017/CBO9780511622632} {\emph {\bibinfo {title} {{Quantum Fields in Curved Space}}}},\ Cambridge Monographs on Mathematical Physics\ (\bibinfo  {publisher} {Cambridge Univ. Press},\ \bibinfo {address} {Cambridge, UK},\ \bibinfo {year} {1984})\BibitemShut {NoStop}%
\bibitem [{\citenamefont {Faraoni}(1996)}]{Faraoni:1996rf}%
  \BibitemOpen
  \bibfield  {author} {\bibinfo {author} {\bibfnamefont {V.}~\bibnamefont {Faraoni}},\ }\href {\doibase 10.1103/PhysRevD.53.6813} {\bibfield  {journal} {\bibinfo  {journal} {Phys. Rev. D}\ }\textbf {\bibinfo {volume} {53}},\ \bibinfo {pages} {6813} (\bibinfo {year} {1996})},\ \Eprint {http://arxiv.org/abs/astro-ph/9602111} {arXiv:astro-ph/9602111} \BibitemShut {NoStop}%
\bibitem [{\citenamefont {Hrycyna}(2017)}]{Hrycyna:2015vvs}%
  \BibitemOpen
  \bibfield  {author} {\bibinfo {author} {\bibfnamefont {O.}~\bibnamefont {Hrycyna}},\ }\href {\doibase 10.1016/j.physletb.2017.02.062} {\bibfield  {journal} {\bibinfo  {journal} {Phys. Lett. B}\ }\textbf {\bibinfo {volume} {768}},\ \bibinfo {pages} {218} (\bibinfo {year} {2017})},\ \Eprint {http://arxiv.org/abs/1511.08736} {arXiv:1511.08736 [astro-ph.CO]} \BibitemShut {NoStop}%
\bibitem [{\citenamefont {Hazumi}\ \emph {et~al.}(2020)\citenamefont {Hazumi} \emph {et~al.}}]{LiteBIRD:2020khw}%
  \BibitemOpen
  \bibfield  {author} {\bibinfo {author} {\bibfnamefont {M.}~\bibnamefont {Hazumi}} \emph {et~al.} (\bibinfo {collaboration} {LiteBIRD}),\ }\href {\doibase 10.1117/12.2563050} {\bibfield  {journal} {\bibinfo  {journal} {Proc. SPIE Int. Soc. Opt. Eng.}\ }\textbf {\bibinfo {volume} {11443}},\ \bibinfo {pages} {114432F} (\bibinfo {year} {2020})},\ \Eprint {http://arxiv.org/abs/2101.12449} {arXiv:2101.12449 [astro-ph.IM]} \BibitemShut {NoStop}%
\bibitem [{\citenamefont {Abazajian}\ \emph {et~al.}(2016)\citenamefont {Abazajian} \emph {et~al.}}]{CMB-S4:2016ple}%
  \BibitemOpen
  \bibfield  {author} {\bibinfo {author} {\bibfnamefont {K.~N.}\ \bibnamefont {Abazajian}} \emph {et~al.} (\bibinfo {collaboration} {CMB-S4}),\ }\href@noop {} {\  (\bibinfo {year} {2016})},\ \Eprint {http://arxiv.org/abs/1610.02743} {arXiv:1610.02743 [astro-ph.CO]} \BibitemShut {NoStop}%
\bibitem [{\citenamefont {Ade}\ \emph {et~al.}(2019)\citenamefont {Ade} \emph {et~al.}}]{SimonsObservatory:2018koc}%
  \BibitemOpen
  \bibfield  {author} {\bibinfo {author} {\bibfnamefont {P.}~\bibnamefont {Ade}} \emph {et~al.} (\bibinfo {collaboration} {Simons Observatory}),\ }\href {\doibase 10.1088/1475-7516/2019/02/056} {\bibfield  {journal} {\bibinfo  {journal} {JCAP}\ }\textbf {\bibinfo {volume} {02}},\ \bibinfo {pages} {056} (\bibinfo {year} {2019})},\ \Eprint {http://arxiv.org/abs/1808.07445} {arXiv:1808.07445 [astro-ph.CO]} \BibitemShut {NoStop}%
\bibitem [{\citenamefont {Enqvist}(2012)}]{Enqvist:2012qc}%
  \BibitemOpen
  \bibfield  {author} {\bibinfo {author} {\bibfnamefont {K.}~\bibnamefont {Enqvist}},\ }in\ \href@noop {} {\emph {\bibinfo {booktitle} {{2010 European School of High Energy Physics}}}}\ (\bibinfo {year} {2012})\ \Eprint {http://arxiv.org/abs/1201.6164} {arXiv:1201.6164 [gr-qc]} \BibitemShut {NoStop}%
\bibitem [{\citenamefont {Garcia}\ \emph {et~al.}(2020)\citenamefont {Garcia}, \citenamefont {Kaneta}, \citenamefont {Mambrini},\ and\ \citenamefont {Olive}}]{Garcia:2020eof}%
  \BibitemOpen
  \bibfield  {author} {\bibinfo {author} {\bibfnamefont {M.~A.~G.}\ \bibnamefont {Garcia}}, \bibinfo {author} {\bibfnamefont {K.}~\bibnamefont {Kaneta}}, \bibinfo {author} {\bibfnamefont {Y.}~\bibnamefont {Mambrini}}, \ and\ \bibinfo {author} {\bibfnamefont {K.~A.}\ \bibnamefont {Olive}},\ }\href {\doibase 10.1103/PhysRevD.101.123507} {\bibfield  {journal} {\bibinfo  {journal} {Phys. Rev. D}\ }\textbf {\bibinfo {volume} {101}},\ \bibinfo {pages} {123507} (\bibinfo {year} {2020})},\ \Eprint {http://arxiv.org/abs/2004.08404} {arXiv:2004.08404 [hep-ph]} \BibitemShut {NoStop}%
\bibitem [{\citenamefont {Kannike}\ \emph {et~al.}(2017)\citenamefont {Kannike}, \citenamefont {Racioppi},\ and\ \citenamefont {Raidal}}]{Kannike:2016jfs}%
  \BibitemOpen
  \bibfield  {author} {\bibinfo {author} {\bibfnamefont {K.}~\bibnamefont {Kannike}}, \bibinfo {author} {\bibfnamefont {A.}~\bibnamefont {Racioppi}}, \ and\ \bibinfo {author} {\bibfnamefont {M.}~\bibnamefont {Raidal}},\ }\href {\doibase 10.1016/j.nuclphysb.2017.02.019} {\bibfield  {journal} {\bibinfo  {journal} {Nucl. Phys. B}\ }\textbf {\bibinfo {volume} {918}},\ \bibinfo {pages} {162} (\bibinfo {year} {2017})},\ \Eprint {http://arxiv.org/abs/1605.09378} {arXiv:1605.09378 [hep-ph]} \BibitemShut {NoStop}%
\bibitem [{\citenamefont {Babichev}\ \emph {et~al.}(2020)\citenamefont {Babichev}, \citenamefont {Gorbunov}, \citenamefont {Ramazanov},\ and\ \citenamefont {Reverberi}}]{Babichev:2020yeo}%
  \BibitemOpen
  \bibfield  {author} {\bibinfo {author} {\bibfnamefont {E.}~\bibnamefont {Babichev}}, \bibinfo {author} {\bibfnamefont {D.}~\bibnamefont {Gorbunov}}, \bibinfo {author} {\bibfnamefont {S.}~\bibnamefont {Ramazanov}}, \ and\ \bibinfo {author} {\bibfnamefont {L.}~\bibnamefont {Reverberi}},\ }\href {\doibase 10.1088/1475-7516/2020/09/059} {\bibfield  {journal} {\bibinfo  {journal} {JCAP}\ }\textbf {\bibinfo {volume} {09}},\ \bibinfo {pages} {059} (\bibinfo {year} {2020})},\ \Eprint {http://arxiv.org/abs/2006.02225} {arXiv:2006.02225 [hep-ph]} \BibitemShut {NoStop}%
\bibitem [{\citenamefont {Cembranos}\ \emph {et~al.}(2020)\citenamefont {Cembranos}, \citenamefont {Garay},\ and\ \citenamefont {S\'anchez~Vel\'azquez}}]{Cembranos:2019qlm}%
  \BibitemOpen
  \bibfield  {author} {\bibinfo {author} {\bibfnamefont {J.~A.~R.}\ \bibnamefont {Cembranos}}, \bibinfo {author} {\bibfnamefont {L.~J.}\ \bibnamefont {Garay}}, \ and\ \bibinfo {author} {\bibfnamefont {J.~M.}\ \bibnamefont {S\'anchez~Vel\'azquez}},\ }\href {\doibase 10.1007/JHEP06(2020)084} {\bibfield  {journal} {\bibinfo  {journal} {JHEP}\ }\textbf {\bibinfo {volume} {06}},\ \bibinfo {pages} {084} (\bibinfo {year} {2020})},\ \Eprint {http://arxiv.org/abs/1910.13937} {arXiv:1910.13937 [hep-ph]} \BibitemShut {NoStop}%
\bibitem [{\citenamefont {Kolb}\ \emph {et~al.}(2003)\citenamefont {Kolb}, \citenamefont {Notari},\ and\ \citenamefont {Riotto}}]{Kolb:2003ke}%
  \BibitemOpen
  \bibfield  {author} {\bibinfo {author} {\bibfnamefont {E.~W.}\ \bibnamefont {Kolb}}, \bibinfo {author} {\bibfnamefont {A.}~\bibnamefont {Notari}}, \ and\ \bibinfo {author} {\bibfnamefont {A.}~\bibnamefont {Riotto}},\ }\href {\doibase 10.1103/PhysRevD.68.123505} {\bibfield  {journal} {\bibinfo  {journal} {Phys. Rev. D}\ }\textbf {\bibinfo {volume} {68}},\ \bibinfo {pages} {123505} (\bibinfo {year} {2003})},\ \Eprint {http://arxiv.org/abs/hep-ph/0307241} {arXiv:hep-ph/0307241} \BibitemShut {NoStop}%
\bibitem [{\citenamefont {Cat\`a}\ \emph {et~al.}(2017)\citenamefont {Cat\`a}, \citenamefont {Ibarra},\ and\ \citenamefont {Ingenh\"utt}}]{Cata:2016epa}%
  \BibitemOpen
  \bibfield  {author} {\bibinfo {author} {\bibfnamefont {O.}~\bibnamefont {Cat\`a}}, \bibinfo {author} {\bibfnamefont {A.}~\bibnamefont {Ibarra}}, \ and\ \bibinfo {author} {\bibfnamefont {S.}~\bibnamefont {Ingenh\"utt}},\ }\href {\doibase 10.1103/PhysRevD.95.035011} {\bibfield  {journal} {\bibinfo  {journal} {Phys. Rev. D}\ }\textbf {\bibinfo {volume} {95}},\ \bibinfo {pages} {035011} (\bibinfo {year} {2017})},\ \Eprint {http://arxiv.org/abs/1611.00725} {arXiv:1611.00725 [hep-ph]} \BibitemShut {NoStop}%
\bibitem [{\citenamefont {Cat\`a}\ \emph {et~al.}(2016)\citenamefont {Cat\`a}, \citenamefont {Ibarra},\ and\ \citenamefont {Ingenh\"utt}}]{Cata:2016dsg}%
  \BibitemOpen
  \bibfield  {author} {\bibinfo {author} {\bibfnamefont {O.}~\bibnamefont {Cat\`a}}, \bibinfo {author} {\bibfnamefont {A.}~\bibnamefont {Ibarra}}, \ and\ \bibinfo {author} {\bibfnamefont {S.}~\bibnamefont {Ingenh\"utt}},\ }\href {\doibase 10.1103/PhysRevLett.117.021302} {\bibfield  {journal} {\bibinfo  {journal} {Phys. Rev. Lett.}\ }\textbf {\bibinfo {volume} {117}},\ \bibinfo {pages} {021302} (\bibinfo {year} {2016})},\ \Eprint {http://arxiv.org/abs/1603.03696} {arXiv:1603.03696 [hep-ph]} \BibitemShut {NoStop}%
\bibitem [{\citenamefont {Barman}\ \emph {et~al.}(2023{\natexlab{c}})\citenamefont {Barman}, \citenamefont {Bernal},\ and\ \citenamefont {Rubio}}]{Barman:2023opy}%
  \BibitemOpen
  \bibfield  {author} {\bibinfo {author} {\bibfnamefont {B.}~\bibnamefont {Barman}}, \bibinfo {author} {\bibfnamefont {N.}~\bibnamefont {Bernal}}, \ and\ \bibinfo {author} {\bibfnamefont {J.}~\bibnamefont {Rubio}},\ }\href@noop {} {\  (\bibinfo {year} {2023}{\natexlab{c}})},\ \Eprint {http://arxiv.org/abs/2310.06039} {arXiv:2310.06039 [hep-ph]} \BibitemShut {NoStop}%
\bibitem [{\citenamefont {Kannike}\ \emph {et~al.}(2016)\citenamefont {Kannike}, \citenamefont {Racioppi},\ and\ \citenamefont {Raidal}}]{Kannike:2015kda}%
  \BibitemOpen
  \bibfield  {author} {\bibinfo {author} {\bibfnamefont {K.}~\bibnamefont {Kannike}}, \bibinfo {author} {\bibfnamefont {A.}~\bibnamefont {Racioppi}}, \ and\ \bibinfo {author} {\bibfnamefont {M.}~\bibnamefont {Raidal}},\ }\href {\doibase 10.1007/JHEP01(2016)035} {\bibfield  {journal} {\bibinfo  {journal} {JHEP}\ }\textbf {\bibinfo {volume} {01}},\ \bibinfo {pages} {035} (\bibinfo {year} {2016})},\ \Eprint {http://arxiv.org/abs/1509.05423} {arXiv:1509.05423 [hep-ph]} \BibitemShut {NoStop}%
\bibitem [{\citenamefont {Giudice}\ \emph {et~al.}(2001)\citenamefont {Giudice}, \citenamefont {Kolb},\ and\ \citenamefont {Riotto}}]{Giudice:2000ex}%
  \BibitemOpen
  \bibfield  {author} {\bibinfo {author} {\bibfnamefont {G.~F.}\ \bibnamefont {Giudice}}, \bibinfo {author} {\bibfnamefont {E.~W.}\ \bibnamefont {Kolb}}, \ and\ \bibinfo {author} {\bibfnamefont {A.}~\bibnamefont {Riotto}},\ }\href {\doibase 10.1103/PhysRevD.64.023508} {\bibfield  {journal} {\bibinfo  {journal} {Phys. Rev. D}\ }\textbf {\bibinfo {volume} {64}},\ \bibinfo {pages} {023508} (\bibinfo {year} {2001})},\ \Eprint {http://arxiv.org/abs/hep-ph/0005123} {arXiv:hep-ph/0005123} \BibitemShut {NoStop}%
\bibitem [{\citenamefont {Chung}\ \emph {et~al.}(1999)\citenamefont {Chung}, \citenamefont {Kolb},\ and\ \citenamefont {Riotto}}]{Chung:1998rq}%
  \BibitemOpen
  \bibfield  {author} {\bibinfo {author} {\bibfnamefont {D.~J.~H.}\ \bibnamefont {Chung}}, \bibinfo {author} {\bibfnamefont {E.~W.}\ \bibnamefont {Kolb}}, \ and\ \bibinfo {author} {\bibfnamefont {A.}~\bibnamefont {Riotto}},\ }\href {\doibase 10.1103/PhysRevD.60.063504} {\bibfield  {journal} {\bibinfo  {journal} {Phys. Rev. D}\ }\textbf {\bibinfo {volume} {60}},\ \bibinfo {pages} {063504} (\bibinfo {year} {1999})},\ \Eprint {http://arxiv.org/abs/hep-ph/9809453} {arXiv:hep-ph/9809453} \BibitemShut {NoStop}%
\bibitem [{\citenamefont {Drewes}\ \emph {et~al.}(2017)\citenamefont {Drewes}, \citenamefont {Kang},\ and\ \citenamefont {Mun}}]{Drewes:2017fmn}%
  \BibitemOpen
  \bibfield  {author} {\bibinfo {author} {\bibfnamefont {M.}~\bibnamefont {Drewes}}, \bibinfo {author} {\bibfnamefont {J.~U.}\ \bibnamefont {Kang}}, \ and\ \bibinfo {author} {\bibfnamefont {U.~R.}\ \bibnamefont {Mun}},\ }\href {\doibase 10.1007/JHEP11(2017)072} {\bibfield  {journal} {\bibinfo  {journal} {JHEP}\ }\textbf {\bibinfo {volume} {11}},\ \bibinfo {pages} {072} (\bibinfo {year} {2017})},\ \Eprint {http://arxiv.org/abs/1708.01197} {arXiv:1708.01197 [astro-ph.CO]} \BibitemShut {NoStop}%
\bibitem [{\citenamefont {Drewes}(2022)}]{Drewes:2019rxn}%
  \BibitemOpen
  \bibfield  {author} {\bibinfo {author} {\bibfnamefont {M.}~\bibnamefont {Drewes}},\ }\href {\doibase 10.1088/1475-7516/2022/09/069} {\bibfield  {journal} {\bibinfo  {journal} {JCAP}\ }\textbf {\bibinfo {volume} {09}},\ \bibinfo {pages} {069} (\bibinfo {year} {2022})},\ \Eprint {http://arxiv.org/abs/1903.09599} {arXiv:1903.09599 [astro-ph.CO]} \BibitemShut {NoStop}%
\bibitem [{\citenamefont {Hasegawa}\ \emph {et~al.}(2019)\citenamefont {Hasegawa}, \citenamefont {Hiroshima}, \citenamefont {Kohri}, \citenamefont {Hansen}, \citenamefont {Tram},\ and\ \citenamefont {Hannestad}}]{Hasegawa:2019jsa}%
  \BibitemOpen
  \bibfield  {author} {\bibinfo {author} {\bibfnamefont {T.}~\bibnamefont {Hasegawa}}, \bibinfo {author} {\bibfnamefont {N.}~\bibnamefont {Hiroshima}}, \bibinfo {author} {\bibfnamefont {K.}~\bibnamefont {Kohri}}, \bibinfo {author} {\bibfnamefont {R.~S.~L.}\ \bibnamefont {Hansen}}, \bibinfo {author} {\bibfnamefont {T.}~\bibnamefont {Tram}}, \ and\ \bibinfo {author} {\bibfnamefont {S.}~\bibnamefont {Hannestad}},\ }\href {\doibase 10.1088/1475-7516/2019/12/012} {\bibfield  {journal} {\bibinfo  {journal} {JCAP}\ }\textbf {\bibinfo {volume} {12}},\ \bibinfo {pages} {012} (\bibinfo {year} {2019})},\ \Eprint {http://arxiv.org/abs/1908.10189} {arXiv:1908.10189 [hep-ph]} \BibitemShut {NoStop}%
\bibitem [{\citenamefont {Kawasaki}\ \emph {et~al.}(1999)\citenamefont {Kawasaki}, \citenamefont {Kohri},\ and\ \citenamefont {Sugiyama}}]{Kawasaki:1999na}%
  \BibitemOpen
  \bibfield  {author} {\bibinfo {author} {\bibfnamefont {M.}~\bibnamefont {Kawasaki}}, \bibinfo {author} {\bibfnamefont {K.}~\bibnamefont {Kohri}}, \ and\ \bibinfo {author} {\bibfnamefont {N.}~\bibnamefont {Sugiyama}},\ }\href {\doibase 10.1103/PhysRevLett.82.4168} {\bibfield  {journal} {\bibinfo  {journal} {Phys. Rev. Lett.}\ }\textbf {\bibinfo {volume} {82}},\ \bibinfo {pages} {4168} (\bibinfo {year} {1999})},\ \Eprint {http://arxiv.org/abs/astro-ph/9811437} {arXiv:astro-ph/9811437} \BibitemShut {NoStop}%
\bibitem [{\citenamefont {Kawasaki}\ \emph {et~al.}(2000)\citenamefont {Kawasaki}, \citenamefont {Kohri},\ and\ \citenamefont {Sugiyama}}]{Kawasaki:2000en}%
  \BibitemOpen
  \bibfield  {author} {\bibinfo {author} {\bibfnamefont {M.}~\bibnamefont {Kawasaki}}, \bibinfo {author} {\bibfnamefont {K.}~\bibnamefont {Kohri}}, \ and\ \bibinfo {author} {\bibfnamefont {N.}~\bibnamefont {Sugiyama}},\ }\href {\doibase 10.1103/PhysRevD.62.023506} {\bibfield  {journal} {\bibinfo  {journal} {Phys. Rev. D}\ }\textbf {\bibinfo {volume} {62}},\ \bibinfo {pages} {023506} (\bibinfo {year} {2000})},\ \Eprint {http://arxiv.org/abs/astro-ph/0002127} {arXiv:astro-ph/0002127} \BibitemShut {NoStop}%
\bibitem [{\citenamefont {Ferreira}\ \emph {et~al.}(2018)\citenamefont {Ferreira}, \citenamefont {Notari},\ and\ \citenamefont {Simeon}}]{Ferreira:2018nav}%
  \BibitemOpen
  \bibfield  {author} {\bibinfo {author} {\bibfnamefont {R.~Z.}\ \bibnamefont {Ferreira}}, \bibinfo {author} {\bibfnamefont {A.}~\bibnamefont {Notari}}, \ and\ \bibinfo {author} {\bibfnamefont {G.}~\bibnamefont {Simeon}},\ }\href {\doibase 10.1088/1475-7516/2018/11/021} {\bibfield  {journal} {\bibinfo  {journal} {JCAP}\ }\textbf {\bibinfo {volume} {11}},\ \bibinfo {pages} {021} (\bibinfo {year} {2018})},\ \Eprint {http://arxiv.org/abs/1806.05511} {arXiv:1806.05511 [astro-ph.CO]} \BibitemShut {NoStop}%
\bibitem [{\citenamefont {Racioppi}(2017)}]{Racioppi:2017spw}%
  \BibitemOpen
  \bibfield  {author} {\bibinfo {author} {\bibfnamefont {A.}~\bibnamefont {Racioppi}},\ }\href {\doibase 10.1088/1475-7516/2017/12/041} {\bibfield  {journal} {\bibinfo  {journal} {JCAP}\ }\textbf {\bibinfo {volume} {12}},\ \bibinfo {pages} {041} (\bibinfo {year} {2017})},\ \Eprint {http://arxiv.org/abs/1710.04853} {arXiv:1710.04853 [astro-ph.CO]} \BibitemShut {NoStop}%
\bibitem [{\citenamefont {Kannike}\ \emph {et~al.}(2014)\citenamefont {Kannike}, \citenamefont {Racioppi},\ and\ \citenamefont {Raidal}}]{Kannike:2014mia}%
  \BibitemOpen
  \bibfield  {author} {\bibinfo {author} {\bibfnamefont {K.}~\bibnamefont {Kannike}}, \bibinfo {author} {\bibfnamefont {A.}~\bibnamefont {Racioppi}}, \ and\ \bibinfo {author} {\bibfnamefont {M.}~\bibnamefont {Raidal}},\ }\href {\doibase 10.1007/JHEP06(2014)154} {\bibfield  {journal} {\bibinfo  {journal} {JHEP}\ }\textbf {\bibinfo {volume} {06}},\ \bibinfo {pages} {154} (\bibinfo {year} {2014})},\ \Eprint {http://arxiv.org/abs/1405.3987} {arXiv:1405.3987 [hep-ph]} \BibitemShut {NoStop}%
\bibitem [{\citenamefont {Bostan}(2020)}]{Bostan:2019uvv}%
  \BibitemOpen
  \bibfield  {author} {\bibinfo {author} {\bibfnamefont {N.}~\bibnamefont {Bostan}},\ }\href {\doibase 10.1016/j.physletb.2020.135954} {\bibfield  {journal} {\bibinfo  {journal} {Phys. Lett. B}\ }\textbf {\bibinfo {volume} {811}},\ \bibinfo {pages} {135954} (\bibinfo {year} {2020})},\ \Eprint {http://arxiv.org/abs/1907.13235} {arXiv:1907.13235 [gr-qc]} \BibitemShut {NoStop}%
\bibitem [{\citenamefont {Maji}\ and\ \citenamefont {Shafi}(2023)}]{Maji:2022jzu}%
  \BibitemOpen
  \bibfield  {author} {\bibinfo {author} {\bibfnamefont {R.}~\bibnamefont {Maji}}\ and\ \bibinfo {author} {\bibfnamefont {Q.}~\bibnamefont {Shafi}},\ }\href {\doibase 10.1088/1475-7516/2023/03/007} {\bibfield  {journal} {\bibinfo  {journal} {JCAP}\ }\textbf {\bibinfo {volume} {03}},\ \bibinfo {pages} {007} (\bibinfo {year} {2023})},\ \Eprint {http://arxiv.org/abs/2208.08137} {arXiv:2208.08137 [hep-ph]} \BibitemShut {NoStop}%
\bibitem [{\citenamefont {Coleman}\ and\ \citenamefont {Weinberg}(1973)}]{Coleman:1973jx}%
  \BibitemOpen
  \bibfield  {author} {\bibinfo {author} {\bibfnamefont {S.~R.}\ \bibnamefont {Coleman}}\ and\ \bibinfo {author} {\bibfnamefont {E.~J.}\ \bibnamefont {Weinberg}},\ }\href {\doibase 10.1103/PhysRevD.7.1888} {\bibfield  {journal} {\bibinfo  {journal} {Phys. Rev. D}\ }\textbf {\bibinfo {volume} {7}},\ \bibinfo {pages} {1888} (\bibinfo {year} {1973})}\BibitemShut {NoStop}%
\bibitem [{\citenamefont {Abazajian}\ \emph {et~al.}(2022)\citenamefont {Abazajian} \emph {et~al.}}]{CMB-S4:2020lpa}%
  \BibitemOpen
  \bibfield  {author} {\bibinfo {author} {\bibfnamefont {K.}~\bibnamefont {Abazajian}} \emph {et~al.} (\bibinfo {collaboration} {CMB-S4}),\ }\href {\doibase 10.3847/1538-4357/ac1596} {\bibfield  {journal} {\bibinfo  {journal} {Astrophys. J.}\ }\textbf {\bibinfo {volume} {926}},\ \bibinfo {pages} {54} (\bibinfo {year} {2022})},\ \Eprint {http://arxiv.org/abs/2008.12619} {arXiv:2008.12619 [astro-ph.CO]} \BibitemShut {NoStop}%
\bibitem [{\citenamefont {Allys}\ \emph {et~al.}(2023)\citenamefont {Allys} \emph {et~al.}}]{LiteBIRD:2022cnt}%
  \BibitemOpen
  \bibfield  {author} {\bibinfo {author} {\bibfnamefont {E.}~\bibnamefont {Allys}} \emph {et~al.} (\bibinfo {collaboration} {LiteBIRD}),\ }\href {\doibase 10.1093/ptep/ptac150} {\bibfield  {journal} {\bibinfo  {journal} {PTEP}\ }\textbf {\bibinfo {volume} {2023}},\ \bibinfo {pages} {042F01} (\bibinfo {year} {2023})},\ \Eprint {http://arxiv.org/abs/2202.02773} {arXiv:2202.02773 [astro-ph.IM]} \BibitemShut {NoStop}%
\bibitem [{\citenamefont {Hazumi}\ \emph {et~al.}(2019)\citenamefont {Hazumi} \emph {et~al.}}]{Hazumi:2019lys}%
  \BibitemOpen
  \bibfield  {author} {\bibinfo {author} {\bibfnamefont {M.}~\bibnamefont {Hazumi}} \emph {et~al.},\ }\href {\doibase 10.1007/s10909-019-02150-5} {\bibfield  {journal} {\bibinfo  {journal} {J. Low Temp. Phys.}\ }\textbf {\bibinfo {volume} {194}},\ \bibinfo {pages} {443} (\bibinfo {year} {2019})}\BibitemShut {NoStop}%
\bibitem [{\citenamefont {Adak}\ \emph {et~al.}(2022)\citenamefont {Adak}, \citenamefont {Sen}, \citenamefont {Basak}, \citenamefont {Delabrouille}, \citenamefont {Ghosh}, \citenamefont {Rotti}, \citenamefont {Mart\'\i{}nez-Solaeche},\ and\ \citenamefont {Souradeep}}]{Adak:2021lbu}%
  \BibitemOpen
  \bibfield  {author} {\bibinfo {author} {\bibfnamefont {D.}~\bibnamefont {Adak}}, \bibinfo {author} {\bibfnamefont {A.}~\bibnamefont {Sen}}, \bibinfo {author} {\bibfnamefont {S.}~\bibnamefont {Basak}}, \bibinfo {author} {\bibfnamefont {J.}~\bibnamefont {Delabrouille}}, \bibinfo {author} {\bibfnamefont {T.}~\bibnamefont {Ghosh}}, \bibinfo {author} {\bibfnamefont {A.}~\bibnamefont {Rotti}}, \bibinfo {author} {\bibfnamefont {G.}~\bibnamefont {Mart\'\i{}nez-Solaeche}}, \ and\ \bibinfo {author} {\bibfnamefont {T.}~\bibnamefont {Souradeep}},\ }\href {\doibase 10.1093/mnras/stac1474} {\bibfield  {journal} {\bibinfo  {journal} {Mon. Not. Roy. Astron. Soc.}\ }\textbf {\bibinfo {volume} {514}},\ \bibinfo {pages} {3002} (\bibinfo {year} {2022})},\ \Eprint {http://arxiv.org/abs/2110.12362} {arXiv:2110.12362 [astro-ph.CO]} \BibitemShut {NoStop}%
\bibitem [{\citenamefont {Sayre}\ \emph {et~al.}(2020)\citenamefont {Sayre} \emph {et~al.}}]{SPT:2019nip}%
  \BibitemOpen
  \bibfield  {author} {\bibinfo {author} {\bibfnamefont {J.~T.}\ \bibnamefont {Sayre}} \emph {et~al.} (\bibinfo {collaboration} {SPT}),\ }\href {\doibase 10.1103/PhysRevD.101.122003} {\bibfield  {journal} {\bibinfo  {journal} {Phys. Rev. D}\ }\textbf {\bibinfo {volume} {101}},\ \bibinfo {pages} {122003} (\bibinfo {year} {2020})},\ \Eprint {http://arxiv.org/abs/1910.05748} {arXiv:1910.05748 [astro-ph.CO]} \BibitemShut {NoStop}%
\bibitem [{\citenamefont {Suzuki}\ \emph {et~al.}(2016)\citenamefont {Suzuki} \emph {et~al.}}]{POLARBEAR:2015ixw}%
  \BibitemOpen
  \bibfield  {author} {\bibinfo {author} {\bibfnamefont {A.}~\bibnamefont {Suzuki}} \emph {et~al.} (\bibinfo {collaboration} {POLARBEAR}),\ }\href {\doibase 10.1007/s10909-015-1425-4} {\bibfield  {journal} {\bibinfo  {journal} {J. Low Temp. Phys.}\ }\textbf {\bibinfo {volume} {184}},\ \bibinfo {pages} {805} (\bibinfo {year} {2016})},\ \Eprint {http://arxiv.org/abs/1512.07299} {arXiv:1512.07299 [astro-ph.IM]} \BibitemShut {NoStop}%
\bibitem [{\citenamefont {Aiola}\ \emph {et~al.}(2020)\citenamefont {Aiola} \emph {et~al.}}]{ACT:2020gnv}%
  \BibitemOpen
  \bibfield  {author} {\bibinfo {author} {\bibfnamefont {S.}~\bibnamefont {Aiola}} \emph {et~al.} (\bibinfo {collaboration} {ACT}),\ }\href {\doibase 10.1088/1475-7516/2020/12/047} {\bibfield  {journal} {\bibinfo  {journal} {JCAP}\ }\textbf {\bibinfo {volume} {12}},\ \bibinfo {pages} {047} (\bibinfo {year} {2020})},\ \Eprint {http://arxiv.org/abs/2007.07288} {arXiv:2007.07288 [astro-ph.CO]} \BibitemShut {NoStop}%
\bibitem [{\citenamefont {Harrington}\ \emph {et~al.}(2016)\citenamefont {Harrington} \emph {et~al.}}]{Harrington:2016jrz}%
  \BibitemOpen
  \bibfield  {author} {\bibinfo {author} {\bibfnamefont {K.}~\bibnamefont {Harrington}} \emph {et~al.},\ }\href {\doibase 10.1117/12.2233125} {\bibfield  {journal} {\bibinfo  {journal} {Proc. SPIE Int. Soc. Opt. Eng.}\ }\textbf {\bibinfo {volume} {9914}},\ \bibinfo {pages} {99141K} (\bibinfo {year} {2016})},\ \Eprint {http://arxiv.org/abs/1608.08234} {arXiv:1608.08234 [astro-ph.IM]} \BibitemShut {NoStop}%
\bibitem [{\citenamefont {Addamo}\ \emph {et~al.}(2021)\citenamefont {Addamo} \emph {et~al.}}]{LSPE:2020uos}%
  \BibitemOpen
  \bibfield  {author} {\bibinfo {author} {\bibfnamefont {G.}~\bibnamefont {Addamo}} \emph {et~al.} (\bibinfo {collaboration} {LSPE}),\ }\href {\doibase 10.1088/1475-7516/2021/08/008} {\bibfield  {journal} {\bibinfo  {journal} {JCAP}\ }\textbf {\bibinfo {volume} {08}},\ \bibinfo {pages} {008} (\bibinfo {year} {2021})},\ \Eprint {http://arxiv.org/abs/2008.11049} {arXiv:2008.11049 [astro-ph.IM]} \BibitemShut {NoStop}%
\bibitem [{\citenamefont {Mennella}\ \emph {et~al.}(2019)\citenamefont {Mennella} \emph {et~al.}}]{Mennella:2019cwk}%
  \BibitemOpen
  \bibfield  {author} {\bibinfo {author} {\bibfnamefont {A.}~\bibnamefont {Mennella}} \emph {et~al.},\ }\href {\doibase 10.3390/universe5020042} {\bibfield  {journal} {\bibinfo  {journal} {Universe}\ }\textbf {\bibinfo {volume} {5}},\ \bibinfo {pages} {42} (\bibinfo {year} {2019})}\BibitemShut {NoStop}%
\bibitem [{\citenamefont {Ade}\ \emph {et~al.}(2022{\natexlab{b}})\citenamefont {Ade} \emph {et~al.}}]{SPIDER:2021ncy}%
  \BibitemOpen
  \bibfield  {author} {\bibinfo {author} {\bibfnamefont {P.~A.~R.}\ \bibnamefont {Ade}} \emph {et~al.} (\bibinfo {collaboration} {SPIDER}),\ }\href {\doibase 10.3847/1538-4357/ac20df} {\bibfield  {journal} {\bibinfo  {journal} {Astrophys. J.}\ }\textbf {\bibinfo {volume} {927}},\ \bibinfo {pages} {174} (\bibinfo {year} {2022}{\natexlab{b}})},\ \Eprint {http://arxiv.org/abs/2103.13334} {arXiv:2103.13334 [astro-ph.CO]} \BibitemShut {NoStop}%
\bibitem [{\citenamefont {Adshead}\ \emph {et~al.}(2016)\citenamefont {Adshead}, \citenamefont {Cui},\ and\ \citenamefont {Shelton}}]{Adshead:2016xxj}%
  \BibitemOpen
  \bibfield  {author} {\bibinfo {author} {\bibfnamefont {P.}~\bibnamefont {Adshead}}, \bibinfo {author} {\bibfnamefont {Y.}~\bibnamefont {Cui}}, \ and\ \bibinfo {author} {\bibfnamefont {J.}~\bibnamefont {Shelton}},\ }\href {\doibase 10.1007/JHEP06(2016)016} {\bibfield  {journal} {\bibinfo  {journal} {JHEP}\ }\textbf {\bibinfo {volume} {06}},\ \bibinfo {pages} {016} (\bibinfo {year} {2016})},\ \Eprint {http://arxiv.org/abs/1604.02458} {arXiv:1604.02458 [hep-ph]} \BibitemShut {NoStop}%
\bibitem [{\citenamefont {Adshead}\ \emph {et~al.}(2019)\citenamefont {Adshead}, \citenamefont {Ralegankar},\ and\ \citenamefont {Shelton}}]{Adshead:2019uwj}%
  \BibitemOpen
  \bibfield  {author} {\bibinfo {author} {\bibfnamefont {P.}~\bibnamefont {Adshead}}, \bibinfo {author} {\bibfnamefont {P.}~\bibnamefont {Ralegankar}}, \ and\ \bibinfo {author} {\bibfnamefont {J.}~\bibnamefont {Shelton}},\ }\href {\doibase 10.1007/JHEP08(2019)151} {\bibfield  {journal} {\bibinfo  {journal} {JHEP}\ }\textbf {\bibinfo {volume} {08}},\ \bibinfo {pages} {151} (\bibinfo {year} {2019})},\ \Eprint {http://arxiv.org/abs/1906.02755} {arXiv:1906.02755 [hep-ph]} \BibitemShut {NoStop}%
\end{thebibliography}

\end{document}